\newcommand\papertitle{Dust moments: towards a new modeling of the galactic dust emission for CMB B-modes analysis}

\documentclass[twocolumn]{aa}
\makeatletter
\renewcommand*\aa@pageof{, page \thepage{} of \pageref*{LastPage}}
\makeatother

\usepackage{graphicx}
\usepackage{txfonts}
\usepackage{url}
\usepackage{natbib}
\bibpunct{(}{)}{;}{a}{}{,}
\usepackage{color}
\usepackage{colortbl} 
\usepackage[dvipsnames]{xcolor}
\usepackage{enumitem}

\usepackage[breaklinks, colorlinks, citecolor=blue]{hyperref}
\usepackage{subfigure}
\usepackage{epsfig}
\usepackage{multirow}
\usepackage{array}

\usepackage{url}
\usepackage{psfrag}
\usepackage[T1]{fontenc}
\usepackage{rotating}
\def\be{\begin{equation}}
\def\ee{\end{equation}}
\def\bea{\begin{eqnarray}}
\def\eea{\end{eqnarray}}

\newcolumntype{L}[1]{>{\raggedright\let\newline\\\arraybackslash\hspace{0pt}}m{#1}}
\newcolumntype{C}[1]{>{\centering\let\newline\\\arraybackslash\hspace{0pt}}m{#1}}
\newcolumntype{R}[1]{>{\raggedleft\let\newline\\\arraybackslash\hspace{0pt}}m{#1}}

 \def\be{\begin{equation}}
 \def\ee{\end{equation}}
 \def\bea{\begin{eqnarray}}
 \def\eea{\end{eqnarray}}
\newcommand{\sect}[1]{Sect.~\ref{#1}}
\newcommand{\app}[1]{Appendix~\ref{#1}}

\newcommand{\fig}[1]{Fig.~\ref{#1}}

\newcommand{\eq}[1]{Eq.~(\ref{#1})}

\newcommand{\Planck}{{\it Planck}}
\newcommand{\PlanckHFI}{\Planck{}-HFI}

\definecolor{Blue}{rgb}{0.,0.,1.}
\definecolor{LightSkyBlue}{rgb}{0.691,0.827,1.}
\definecolor{Red}{rgb}{1.,0.,0.}
\definecolor{Green}{rgb}{0.,1.,0.}
\definecolor{Purple}{rgb}{0.5, 0., 0.5}
\definecolor{Try}{rgb}{0.15,0.,1}
\definecolor{Black}{rgb}{0., 0., 0.}

\newcommand{\Tbar}{T_0}
\newcommand{\betabar}{\beta_0}
\newcommand{\betaellb}{{\beta}_{0}(\ell)}
\newcommand{\betaellbcorr}{{\beta}^{\rm corr}_{0}(\ell)}

\newcommand{\ID}{{I_{\rm D}}}
\newcommand{\AD}{{A_{\rm D}}}

\newcommand{\momi}[1]{{\omega_{#1}}}
\newcommand{\momilm}[1]{{(\omega_{#1})_{\ell m}}}

\def\cl2dl{\frac{\ell (\ell+1)}{2 \pi} }
\newcommand{\colmat}[2][.8]{
  \scalebox{#1}{
    \renewcommand{\arraystretch}{.8}
    $\begin{bmatrix}#2\end{bmatrix}$
  }
}

\makeatletter
\newcommand\footnoteref[1]{\protected@xdef\@thefnmark{\ref{#1}}\@footnotemark}
\makeatother

\begin{document}

%

\title{\papertitle}

\offprints{jonathan.aumont@irap.omp.eu}
\authorrunning{Mangilli et al.}
\titlerunning{Dust moments}


\author{A. Mangilli \inst{\ref{inst1}} 
\and J. Aumont  \inst{\ref{inst1}} 
\and A. Rotti  \inst{\ref{inst3}}
\and F. Boulanger   \inst{\ref{inst2}} 
\and J. Chluba \inst{\ref{inst3}}
\and T. Ghosh  \inst{\ref{inst4}}
\and L. Montier  \inst{\ref{inst1}} 
} 

\institute{
IRAP, Universit\'e de Toulouse, CNRS, CNES, UPS, Toulouse, France
\label{inst1}
\and
Laboratoire de Physique de l'ENS, ENS, Universit\'e PSL, CNRS, Sorbonne Universit\'e, Universit\'e Paris-Diderot, Paris, France
\label{inst2}
\and Jodrell Bank Centre for Astrophysics, School of Physics and Astronomy, University of Manchester, Manchester M13 9PL, U.K. \label{inst3}
\and
School of Physical Sciences, National Institute of Science Education and Research, HBNI, Jatni, 752050, Odisha, India \label{inst4}
}


\abstract{
The characterization of the spectral energy distribution (SED) of dust emission has become a critical issue in the quest for primordial $B$-modes. The dust SED is often approximated by a modified black body (MBB) emission law but the extent to which this is accurate is unclear.  This paper addresses this question, expanding the dust SED at the power spectrum level. The expansion is performed by means of moments around the MBB law, related to derivatives with respect to the dust spectral index.
We present the mathematical formalism and apply it to simulations and  \Planck{}{} total intensity data, from 143 to 857\,GHz, because no polarized data are yet available that provide the required sensitivity to perform this analysis. With simulations, we demonstrate the
ability of high-order moments to account for spatial variations in MBB parameters. Neglecting these moments leads to poor fits and a bias in the recovered dust spectral index. 
We identify the main moments that are required to fit the \Planck{} data. The comparison with simulations helps us to disentangle the respective contributions from dust and the cosmic infrared background to the high-order moments, but the simulations give an insufficient description of the actual \Planck{} data. Extending our model to cosmic microwave background $B$-mode analyses within a simplified framework, we find that ignoring the dust SED distortions, or trying to model them with a single decorrelation parameter, could lead to biases that are  larger than the targeted sensitivity for the next generation of CMB $B$-mode experiments.}


\maketitle


\section{Introduction}\label{sec:introduction}

The precise characterization of the properties of the polarized dust emission from our Galaxy is a crucial issue in the quest for the primordial $B$-modes of the cosmic microwave background (CMB). If measured, this faint cosmological signal imprinted by the primordial gravitational wave background would be evidence of the inflation epoch and could be used to quantify its energy scale, providing a rigorous test of fundamental physics far beyond the reach of accelerators \citep{polnarev1985polarization,kamionkowski,seljak}. However, accurate determination of diffuse CMB $B$-mode foregrounds ---among which the polarized Galactic dust emission dominates at observing frequencies $\gtrsim$ 70 GHz \citep[see e.g.][]{krachmalnicoff,planck2018XI}--- is required to get an unbiased estimate of the ratio $r$ between tensor and scalar primordial perturbations, a parameter of unknown amplitude scaling the CMB $B$-mode power on the sky and directly linked to the energy scale at which inflation occurred. 

The frequency dependence of the dust emission, assessed through its spectral energy distribution (SED) is one of the characteristics that needs to be determined with the highest accuracy. The \Planck{} data have shown that the SED of the dust emission for total intensity and polarization can be fitted by a modified black body (MBB) law \citep[][]{pipxvii,2014A&A...571A..30P,2015A&A...576A.107P,planck2018XI}. Maps of the dust MBB spectral indices and temperatures have been fitted to the total intensity \Planck{} data \citep[see e.g.,][]{planck2013xi,planck2015x}. These provide evidence that dust emission properties vary across the sky. The data do not provide comparable observational evidence for polarization due to insufficient signal to noise ratio (S/N). However, analyzing total intensity data is 
 directly relevant to polarization.  Indeed, \Planck{} and the Balloon-borne Large Aperture Submillimeter Telescope for Polarimetry (BLASTPol) observations have shown that the polarization fraction is remarkably constant from the far-infrared to millimeter wavelengths, suggesting that the polarized and total dust emission arise predominantly from a single grain population \citep{2015A&A...576A.107P,Gandilo2016,Ashton2017,Guillet18,Shariff2019,planck2018XI}.  

In the inference of cosmological parameters from CMB data, the spectral frequency dependence of dust polarization and its angular structure on the sky are most often assumed to be separable. This is a simplifying assumption that needs to be improved upon. When spatial variations of the dust emission law are present but ignored, biases that compromise the cosmological analysis can arise and bias the sought-after CMB $B$-mode signal \citep{Tassis15,pipl,Poh17}. A bias can also be introduced by additional Galactic emission components or if the dust emission is not perfectly fitted by a MBB emission law. Even if the MBB is observed to be a good fit to the data, dust models do anticipate such departures \citep[e.g.,][]{Draine13,Hensley2018}.

Spatial SED variations induce a \emph{decorrelation} between dust maps in different frequency bands, causing a loss of power in the map cross-correlation compared to the geometrical mean of power spectra, which can generate a bias in CMB spectra if it is not properly accounted for. A tentative detection of this effect with the \Planck{} polarization dust data \citep{pipl} was later dismissed  \citep{sheehy}. This analysis was further extended by \citet{planck2018XI}, performing a global multi-frequency fit of the polarized \Planck{} HFI channels with a spectral model that includes  decorrelation. These latter authors derive an upper limit on frequency decorrelation that depends  on an ad-hoc, possibly misleading model. This  shortcoming also applies to the analysis of the latest BICEP2/Keck data \citep{bicep2018}. Although current data analyses do not provide conclusive evidence for frequency decorrelation, this is an effect that must be present at some level. The spectral modeling of polarized dust emission has thus become one of the main challenges in the quest of primordial $B$-modes.

An additional complexity is due to the fact that in sky maps, local variations in dust emission properties across the interstellar medium are averaged along the line of sight and within the beam. When computing power spectra, further averaging occurs through the computation of the  expansion of spherical harmonics. Even if the emission law of the dust in any infinitesimal volume element of the Galaxy is perfectly described by a MBB law, after averaging, the dust spectral frequency dependence is no longer a MBB and SED distortions arise. The dust frequency dependence and its angular structure on the sky become interdependent in ways that are difficult to model and generally involve nonlinear transformations.

\cite{2017MNRAS.472.1195C} proposed a way to describe the variations of the spectral properties along the line of sight, inside the beam, and across the sky using the \emph{moment} decomposition around the MBB in the map pixel space.
The moments can capture SED distortions due to variations in the dust temperature and spectral index, along the line of sight or between lines of sight, and also in the potential contribution of minor dust emission components. The present paper  
extends the \emph{moment} formalism from the map to the angular cross-power spectra, which are highly relevant to CMB $B$-mode analyses, and to getting rid of noise bias and uncorrelated systematic effects. Similar extensions could also play an important role in the extraction of primordial CMB spectral distortions that are caused by energy release \citep{Zeldovich1969, Sunyaev1970SPEC, Sunyaev1970mu, Chluba2011therm} and which could be targeted in the future \citep{Kogut2019BASS, Chluba2019Voyage}.

We present the formalism and assess its ability to fit simulations of increasing complexity before using it to analyze \Planck{} High Frequency Instrument (\PlanckHFI{}) total intensity data and build a spectral model in terms of harmonic space moments of the dust spectral index. While the polarization data available today are not sensitive enough to perform such an analysis, \PlanckHFI{} total intensity data offer the required sensitivity and frequency coverage to build a direct spectral model based on astrophysical grounds. Here, we consider \Planck{} intensity data as a proxy to data from future CMB $B$-modes experiments from Space \citep{litebird,2019arXiv190210541H}, in terms of S/N and frequency coverage. 

The paper is organized as follows. In \sect{sec:formalism}, we describe the formalism of the dust moment expansion in harmonic space including angular cross-power spectra. \sect{sec:method_and_implementation} details our methodology and  implementation of the dust moments analysis. We present the simulations and the actual \Planck{} data to which we fit our spectral model in \sect{sec:data_and_sims}; the results of the fits are presented in \sect{sec:results}.  In \sect{sec:impact_on_r}, within a simplified framework, we relate our spectral model to the analysis of CMB $B$-modes data. We summarize the main results and present our conclusions in \sect{sec:summary_and_conclusions}. The paper has five Appendices. The first three detail the simulations (\app{sec:templates}), the cross-spectra covariance matrix (\app{sec:correlations}),
and additional fit results (\app{sec:multi_order}). In  \app{sec:sync_vs_CIB} we asses the impact of synchrotron emission on our analysis and in \app{sec:dust_beta_tconst} we assess the effect of the spatial variations of dust temperatures. 

\section{Formalism}\label{sec:formalism}

In this section we present the formalism to describe the \emph{moment expansion} of the dust intensity SED built from angular cross-power spectra of spherical sky maps. As presented in \citet{2017MNRAS.472.1195C}, this formalism is a powerful tool with which  to account for SED distortions arising from the various averaging effects. We first reiterate the usual and the moment-expansion dust SED parameterizations in \sect{sec:standard_formalism_map_space}.  In \sect{sec:dust_moments_formalism_harmonics} and \sect{sec:dust_moments_formalism_spectra} we present the spherical harmonics and the cross-power spectra moment expansion that is used in the analysis we present. 

We present the formalism to describe the spectral departures from the MBB associated with derivatives with respect to the dust spectral index $\beta$, which is known to vary across the sky \citep[e.g.,][]{planck2015x}. This can be easily generalized to include derivatives with respect to the dust temperature \citep{2017MNRAS.472.1195C}. However, in the following analysis, we only use the spectral index moment expansion in the Rayleigh-Jeans regime, as temperature and $\beta$-variations can have similar effects on the dust SED built from microwave and submillimeter data.

\vspace{-0mm}
\subsection{Dust SED parametrization in the map domain}\label{sec:standard_formalism_map_space}

\subsubsection{Modified black body formalism}\label{sec:standard_formalism}

The commonly used parametrization for a single-temperature dust spectrum is a MBB emission law. We first consider the MBB parametrization without any spatial variations of the SED, that is, with a constant  spectral index $\betabar$ and a constant temperature $\Tbar$ across the sky. At a given frequency $\nu$ the dust intensity map $\ID(\nu,{\bf \hat{n}\textbf{}})$ takes the form: 
\begin{align}
\ID(\nu, {\bf \hat{n}}) = \frac{I_\nu(\Tbar,\betabar)}{I_{\nu_0}(\Tbar,\betabar)} \, \AD({\bf \hat{n}})= \left(\frac{\nu}{\nu_0}\right)^{\betabar} \frac{B_\nu(\Tbar)}{B_{\nu_0}(\Tbar)}\,\AD({\bf \hat{n}})\,, 
\label{eq:MBB}
\end{align}
where $\AD({\bf \hat{n}})$ is in this case a frequency-independent dust intensity template map, $\nu_0$ a reference frequency, and $B_\nu(\Tbar)$ is the Planck law for the temperature $T_0$.
As long as the MBB with constant temperature and spectral index is the correct emission law for all lines of sight, the spatial and the frequency dependence are separable. Nevertheless, different lines of sight probe Galactic regions with very different physical conditions (temperature, dust composition) in the three dimensions of space (we note here that \emph{spatial variations} between lines of sight and 3D variations along one line of sight lead to similar effects on the final SED). Emission from these regions are \emph{averaged} along the line of sight and within instrumental beams so that even if the SED of every infinitesimal volume element were to accurately describe by Eq.~\eqref{eq:MBB}, the MBB would no longer  accurately describe the 
effective emission law. In the following, we formally consider only the variations from one line of sight to another, so that the effective spectral index $\beta$ or temperature $T$ in Eq.~\eqref{eq:MBB} can vary spatially. 


\vspace{-0mm}
\subsubsection{Modified black body with spatially varying spectral index}
One general attempt to describe the spatial variations of the dust SED in the Rayleigh-Jeans regime of the dust SED is to allow the MBB spectral index to vary spatially. As a consequence, the frequency and spatial dependence of the dust emission are no longer trivially separable. Therefore, the standard MBB formalism of \sect{sec:standard_formalism} must be extended. In this case, the dust intensity map can be written as follows (assuming that the dust temperature remains constant across the sky):
\be
\ID(\nu, {\bf \hat{n}}) = \frac{I_\nu\big(\Tbar,{\bf \beta({\bf \hat{n}})}\big)}{I_{\nu_0}\big(\Tbar,{\bf \beta({\bf \hat{n}})}\big)} \, \AD({\bf \hat{n}})~, \ee
where now $\beta({\bf \hat{n}})$ accounts for the fact that the spectral index varies from pixel to pixel over the sky. We note that for the sake of clarity, in writing this equation, we ignore the variations along the line of sight. 

\subsubsection{Modified black body spectral index moment expansion}
A more general and powerful parametrization of the dust SED was proposed by \citet{2017MNRAS.472.1195C}. It introduces the \emph{moment expansion} of the dust SED, a perturbative expansion of the SED by means of derivatives of the MBB. In our frequency regime, we use derivatives with respect to the dust spectral index $\beta$ so that the dust map $\ID(\nu, {\bf \hat{n}})$ at a frequency $\nu$ now reads:
\begin{align}
\label{eq:moments_definition}
\ID(\nu, {\bf \hat{n}}) 
&\simeq \frac{I_{\nu}(T_0,\beta_0)}{I_{\nu_0}(T_0,\beta_0)}\left[\AD({\bf \hat{n}})+ \momi{1}({\bf \hat{n}}) \ln{\left(\frac{\nu}{\nu_0}\right)}+\frac{1}{2}  \momi{2}({\bf \hat{n}})  \ln^2{\left(\frac{\nu}{\nu_0}\right)} \right. \nonumber\\[1mm]
&\qquad\qquad\qquad\qquad +\left. \frac{1}{6}  \momi{3}({\bf \hat{n}})  \ln^3{\left(\frac{\nu}{\nu_0}\right)}+\dots \right] \,,
\end{align}
where $\momi{i}({\bf \hat{n}})=\AD({\bf \hat{n}}) \Delta \beta({\bf \hat{n}})^i$ is the $i^{\rm th}$ order moment map associated to the $i^{\rm th}$ derivative of the MBB with respect to $\beta$ (here in an expansion up to third order\footnote{Ultimately, the required maximal moment order is driven by the sensitivity and spectral coverage of the experiment under consideration and the target signal level.}), and $\Delta \beta({\bf \hat{n}})=\beta({\bf \hat{n}})-\beta_0$.

\subsection{Dust SED parametrization in spherical harmonics}\label{sec:dust_moments_formalism_harmonics}

By definition, decomposition of the dust map into spherical harmonics coefficients implies averaging over various lines of sight over the sky, which is mathematically equivalent to averaging pixels with different SEDs along the line of sight or in an instrumental beam \citep{2017MNRAS.472.1195C}. Therefore, we can use the spectral moment decomposition described in Eq.~\eqref{eq:moments_definition}. This leads to the following expansion in the spherical harmonics space: 
\begin{align}
(\ID)^{\nu}_{\ell m} 
&= \frac{I_\nu(\Tbar,\betaellb)}{I{\nu_0}(\Tbar,\betaellb)} \times \left[ (\AD)_{\ell m}  + \momilm{1} \ln \left(\frac{\nu}{\nu_0}\right)\right.\nonumber 
\\  
&\qquad +  \left.\frac{1}{2} \momilm{2} \ln^2 \left(\frac{\nu}{\nu_0}\right) + \frac{1}{6} \momilm{3} \ln^3 \left(\frac{\nu}{\nu_0}\right) + .... \right] \, , 
\label{eq:harmonics_moments_3}
\end{align}
where\footnote{More generally, one could introduce $\beta_0(\ell,m)$ for each multipole. However, we are mainly concerned with the power spectra, and so $\betaellb$ is a better starting point.}
$\betaellb$ refers to the effective value of the dust spectral index $\betabar$ for each multipole, as we see in the following. 
The variables $\momilm{i}$ ($i\in\{1,\dots,N\}$) are the spherical harmonic coefficient of the $i^{\rm th}$ order moment map $\momi{i}({\bf \hat{n}})$.
We note that the spherical harmonics moments $(\omega_i)_{\ell m}$ are not the same as the spatial moments $\omega_i({\bf \hat{n}})$, as they involve different averaging.

We stress that the formalism in \eq{eq:harmonics_moments_3} accounts for SED distortions in the most general case and not only the ones associated to the averages due to the spherical harmonics decomposition. Therefore, no further extension is required to capture the effects of line-of-sight or beam averaging. We also stress that the line-of-sight and beam-averaging effects, for real-world experiments, can never be avoided, such that $\beta({\bf \hat{n}})$ should already be interpreted as an averaged dust-amplitude-weighted quantity. 

\subsection{Dust SED parametrization in the cross-power spectra}\label{sec:dust_moments_formalism_spectra}

Based on \eq{eq:harmonics_moments_3}, we can compute the cross-spectrum between the maps observed in the frequency bands $\nu_1$ and $\nu_2$. Up to third order in terms of $\beta$ derivatives, this takes the form:

\begin{align}
\label{eq:dust_moment_cross_3}
\mathcal{D}_\ell^{\nu_1\times\nu_2} &= \frac{I_{\nu_1}(\Tbar,\betaellb) I_{\nu_2}(\Tbar,\betaellb)}{I^2_{\nu_0}(\Tbar,\betaellb)} 
\times  \bigg\{ \mathcal{D}_{\ell}^{\AD\AD}   
\nonumber \\[2mm]
\text{1. order}\;&
\begin{cases}
&+\Big[ \ln\left(\frac{\nu_1}{\nu_0}\right) 
+ \ln \left(\frac{\nu_2}{\nu_0}\right)  \Big] \,\mathcal{D}_{\ell}^{\AD \momi{1}}
\\[2mm]
&+\Big[ \ln\left(\frac{\nu_1}{\nu_0}\right)  \ln\left(\frac{\nu_2}{\nu_0}\right) \Big] \,\mathcal{D}_{\ell}^{\momi{1}\momi{1}}
\end{cases}
\nonumber \\[2mm]
\text{2. order}\;&
\begin{cases}
&+ \frac{1}{2} \left[ \ln^2\left(\frac{\nu_1}{\nu_0}\right) 
+ \ln^2\left(\frac{\nu_2}{\nu_0}\right) 
\right] \mathcal{D}_{\ell}^{\AD\momi{2}}
\\[2mm]
&+ \frac{1}{2} \Big[ \ln \left(\frac{\nu_1}{\nu_0}\right)  \ln^2\left(\frac{\nu_2}{\nu_0}\right)  
+
\ln\left(\frac{\nu_2}{\nu_0}\right)
\ln^2 \left(\frac{\nu_1}{\nu_0}\right) \Big] \mathcal{D}_{\ell}^{\momi{1} \momi{2}}
\\[2mm]
&+\frac{1}{4} \left[\ln^2 \left(\frac{\nu_1}{\nu_0}\right)  \ln^2 \left(\frac{\nu_2}{\nu_0}\right) \right]
\,\mathcal{D}_{\ell}^{\momi{2} \momi{2}}
\end{cases}
\nonumber \\[2mm]
\text{3. order}\;&
\begin{cases}
&+ \frac{1}{6} \left[ \ln^3\left(\frac{\nu_1}{\nu_0}\right) 
+ \ln^3\left(\frac{\nu_2}{\nu_0}\right) 
\right] \mathcal{D}_{\ell}^{\AD\momi{3}}
\\[2mm]
&+ \frac{1}{6} \Big[ \ln \left(\frac{\nu_1}{\nu_0}\right)  \ln^3\left(\frac{\nu_2}{\nu_0}\right) 
+
\ln\left(\frac{\nu_2}{\nu_0}\right)
\ln^3 \left(\frac{\nu_1}{\nu_0}\right) \Big] \mathcal{D}_{\ell}^{\momi{1} \momi{3}}
\\[2mm]
&+ \frac{1}{12} \Big[ \ln^2 \left(\frac{\nu_1}{\nu_0}\right)  \ln^3\left(\frac{\nu_2}{\nu_0}\right) 
+
\ln^2\left(\frac{\nu_2}{\nu_0}\right)
\ln^3 \left(\frac{\nu_1}{\nu_0}\right) \Big] \mathcal{D}_{\ell}^{\momi{2} \momi{3}}
\\[2mm]
&+\frac{1}{36} \left[
\ln^3 \left(\frac{\nu_1}{\nu_0}\right)  
\ln^3 \left(\frac{\nu_2}{\nu_0}\right) 
\right]
\,\mathcal{D}_{\ell}^{\momi{3} \momi{3}}
\end{cases}
\nonumber \\[2mm]
&\quad\,\, + ... \quad\bigg\}~,
\end{align}
where the \emph{moment} cross-power spectra, $\mathcal{D}_{\ell}^{ab}$, $\{a,b\}\in\{\AD,\momi{1},\momi{2},\momi{3}\}$ are defined as\footnote{In the rest of this work we always use $\mathcal{D}_\ell$ angular power spectra ($\mathcal{D}_\ell=\ell(\ell+1)\mathcal{C}_\ell/2\pi$), where $\mathcal{C}_\ell$ is the original angular power spectrum.}:
\be
\mathcal{D}_{\ell}^{ab} = \cl2dl \sum_{-\ell \le m,m'\le \ell} (a)_{\ell m} (b)_{\ell m'}  \label{eq:dust_moment_cross_ab}.
\ee
In Eq.~\eqref{eq:dust_moment_cross_3}, we group terms according to the maximal derivative order in $\beta$ that appears. These spectral moment truncations are motivated by first constructing the moment maps and then computing all the corresponding cross power spectra, though in this work we measure these cross moment spectra directly from the cross frequency data power spectra. Using this ordering, truncating after the first moment order means including the first three terms, truncating after the second moment order means including the first six terms, and so on.
As it is useful in the following, we define the \emph{moment functions} for a given cross-spectrum, $\mathcal{M}_\ell^{ab}(\nu_i,\nu_j)$, as the moments $\mathcal{D}_{\ell}^{ab}$ normalized to the dust amplitude spectrum $\mathcal{D}_{\ell}^{\AD\AD}$ and re-scaled by the corresponding frequency-dependent numerical factors for the $\nu_i \times \nu_j$ cross-spectrum, as: 
\be
\label{eq:moment_vivj}
\mathcal{M}_\ell^{ab}(\nu_i,\nu_j)=c^{ab}(\nu_i,\nu_j,\nu_0) \; \mathcal{D}_{\ell}^{ab}/\mathcal{D}_{\ell}^{\AD\AD},
\ee
where $c^{ab}(\nu_i,\nu_j,\nu_0)$ are the numerical coefficients which involve the sum and/or the product of the $\ln(\nu_i/\nu_0)$ and $\ln(\nu_j/\nu_0)$ terms, as appearing in \eq{eq:dust_moment_cross_3}. In this way, the $\mathcal{M}_\ell^{ab}$ functions of \eq{eq:moment_vivj} show, for a given cross-spectrum, the effective contribution of each moment to the departure of the standard MBB SED. We stress that, in \eq{eq:dust_moment_cross_3}, the dust spectral index $\betaellb$ is fixed and assigned with optimized values (these steps are described further) for each multipole $\ell$. 

The model for the dust SED in \eq{eq:dust_moment_cross_3} can be truncated at any order, depending on the complexity one wants to capture or the available degrees of freedom in the data to be modeled. At zero order (keeping only the first element of the sum, $\mathcal{D}_{\ell}^{\AD\AD}$), \eq{eq:dust_moment_cross_3} becomes the cross-power spectrum with a $\ell$-dependent spectral index.
The higher order terms account for the dust SED distortions to the MBB, meaning that \eq{eq:dust_moment_cross_3} provides us with a consistent and robust description of the dust SED distortions with respect to the MBB emission law. 

Furthermore, as this formalism describes the corrections to the MBB in the angular power spectrum domain ---as a function of the multipole $\ell$--- it naturally provides an efficient framework with which to characterize the frequency \emph{decorrelation} of the cross-power spectra due to spatial variations of the SED. In contrast to previous attempts \citep[e.g.,][]{pipl,2017A&A...603A..62V,bicep2018}, the frequency decorrelation is not introduced by an ad-hoc choice.

We also note that the normalized first-order term can be interpreted as a correction to the MBB spectral index $\betaellb$ needed to recover the ``true'' $\beta(\ell)$ when spatial and/or line-of-sight variations are present. We therefore define: 
\be
\Delta \betaellb \equiv \mathcal{D}_{\ell}^{\AD \momi{1}}/\mathcal{D}_{\ell}^{\AD\AD},
\label{eq:deltabetaell}
\ee

where the $\Delta \betaellb$ pivot parameter solution depends on the total number of moments that are included and becomes unbiased in the limit of many moments.
In fact, this term quantifies the scale-dependent bias that arises from neglecting SED corrections. Adding the moment expansion allows us to eliminate the bias while having nonzero moments at first and higher orders that include the full covariance between the parameters. This connection can be used to introduce an iterative scheme to the analysis, as discussed in \sect{sec:method_and_implementation}. 

We stress that $\betaellb$ in Eq.~\eqref{eq:deltabetaell} is related to a power-spectrum weighted average of $\beta({\bf \hat{n}})$ (which, strictly-speaking, itself is a line-of-sight-averaged quantity). A similar averaging process was discussed for the temperature power spectrum stemming from the relativistic Sunyaev-Zeldovich effect \citep{Remazeilles2019}.
The higher order moment terms in Eq.~\eqref{eq:dust_moment_cross_3}  quantify (3D) cross-correlations between the spectral indices \citep{2017MNRAS.472.1195C}. In general, they each come with their own spatial dependence and maps of these moments can help identify parts of the sky with substantial SED variations.

In summary, Eq.~\eqref{eq:dust_moment_cross_3} provides, for the first time, a physically motivated model for dealing with spatial and line-of-sight variations of the dust spectral index $\beta$ at the power-spectrum level.


\section{Methodology and implementation}\label{sec:method_and_implementation}
In this section, we detail the methodology and the implementation of the dust moments analysis. The analysis consists of an $\ell$-by-$\ell$ SED fit of a cross-spectra data vector $\vec{\mathcal{D}_\ell}$, gathering all the cross spectra that can be computed from a set of maps $M_{\nu_i}$ at frequencies $\nu_i,\ i\in\{1,\dots,N\}$:

\be 
\label{eq:dl_vector_general}
\vec{\mathcal{D}_\ell}\equiv\left[\begin{array}{@{}c@{}}
           \mathcal{D}_\ell^{\nu_1 \times \nu_1} \\
           \mathcal{D}_\ell^{\nu_1 \times \nu_2} \\
           \mathcal{D}_\ell^{\nu_1 \times \nu_3} \\
           \vdots \\
           \mathcal{D}_\ell^{\nu_1 \times \nu_N}\\
 \vdots \\

           \mathcal{D}_\ell^{\nu_N \times \nu_N}
           \end{array} \right]
           \equiv\left[\begin{array}{@{}c@{}}
           \mathcal{D}_\ell\left(M_{\nu_1} \times M_{\nu_1}\right) \\
           \mathcal{D}_\ell\left(M_{\nu_1} \times M_{\nu_2}\right) \\
           \mathcal{D}_\ell\left(M_{\nu_1} \times M_{\nu_3}\right) \\
           \vdots \\
           \mathcal{D}_\ell\left(M_{\nu_1} \times M_{\nu_N}\right)\\
 \vdots \\

           \mathcal{D}_\ell\left(M_{\nu_N} \times M_{\nu_N}\right)
           \end{array} \right]
           . 
\ee
The analysis is divided in three steps: 

\begin{itemize}
    \item Step 1: We fit for each multipole $\ell$ the two parameters $\betaellb$ and $\mathcal{D}_\ell^{\AD\AD}$ ($\Tbar$ is fixed). Zero-order or MBB fits, in the following, refer to this first step 
    \item Step 2: We fix $\betaellb$ and then fit \eq{eq:dust_moment_cross_3}. Three parameters are fitted at the first order, six at the second order and ten at the third order. 
    \item Step 3: We update the value of the spectral index to $\betaellbcorr=\betaellb+\Delta\betaellb$, fix it and redo the fits as in step 2. In practice, step 3 may need to be repeated in order to ensure that $\Delta\betaellb$ and thus $\mathcal{D}_\ell^{A_{\rm D}\omega_1}$ become compatible with zero, indicating the convergence.  
\end{itemize}

For instance, in the case of the dust moment expansion up to the third order, the fitted parameters in step 3, for each multipole bin, are: $\mathcal{D}_\ell^{\AD\AD}$, $\mathcal{D}_\ell^{\AD \momi1}$, $\mathcal{D}_\ell^{\momi1\momi1 }$, $\mathcal{D}_\ell^{\AD \momi2}$, $\mathcal{D}_\ell^{\momi1 \momi2}$, $\mathcal{D}_\ell^{\momi2 \momi2}$, $\mathcal{D}_\ell^{\AD \momi3}$, $\mathcal{D}_\ell^{\momi1 \momi3}$, $\mathcal{D}_\ell^{\momi2 \momi3}$,  $\mathcal{D}_\ell^{\momi3 \momi3}$. 
This ``three-step'' method for the dust moments fit allows us to consistently search for spectral departures from the standard MBB, quantifying, at each multipole, the corrections to the $\betaellb$ due to SED averaging effects, and therefore providing a description of the scale dependence of the frequency decorrelation.

In practice, we perform a Levenberg-Marquardt least-squares minimization of ${\bf r}={\bf y} - {\bf{m}_i}$, where, at each multipole bin, ${\bf y}\equiv \vec{\mathcal{D}_\ell}$ is the input cross-spectra data vector of \eq{eq:dl_vector_general} and ${\bf{m}_i} \equiv \vec{\mathcal{D}^D_\ell}$ is the model vector:
\be 
\label{eq:dl_model_vector}
\vec{\mathcal{D}^D_\ell}=\left[\begin{array}{@{}c@{}}
           \mathcal{D}_\ell^{{\rm D},\nu_1 \times \nu_1} \\
           \mathcal{D}_\ell^{{\rm D},\nu_1 \times \nu_2} \\
           \mathcal{D}_\ell^{{\rm D},\nu_1 \times \nu_3} \\
           \vdots \\
           \mathcal{D}_\ell^{{\rm D},\nu_1 \times \nu_N}\\
 \vdots \\

           \mathcal{D}_\ell^{{\rm D},\nu_N \times \nu_N}
           \end{array} \right].
\ee
For step 1, the model refers to the dust cross-spectra function truncated at zero order, ${\bf{m}_i\equiv {\bf{m}_0}}$, that is, the MBB emission with constant temperature and spectral index $\betaellb$. For steps 2 and 3, the model refers to the dust moment cross-spectra function of \eq{eq:dust_moment_cross_3}, truncated at orders 1, 2, and 3. 

In order to take into account the correlations between the cross-spectra $\mathcal{D}^{\nu_i\times\nu_j}$ and $\mathcal{D}^{\nu_k\times\nu_l}$, the cross-spectra covariance matrix $\mathbb{C}$ is included in the fit:

\begin{align}
\label{eq:covariance}
\mathbb{C}&\equiv \mathcal{C}_{ijkl}\left(\ell\right) \equiv {\rm cov}\left(\mathcal{D}_\ell^{\nu_i\times\nu_j},\mathcal{D}_\ell^{\nu_k\times\nu_l}\right)\nonumber
\\[2mm] 
&=\colmat[0.65]{
{\rm var}\left(\mathcal{D}_\ell^{\nu_1\times\nu_1},\mathcal{D}_\ell^{\nu_1\times\nu_1}\right)&\cdots&{\rm cov}\left(\mathcal{D}_\ell^{\nu_1\times\nu_1},\mathcal{D}_\ell^{\nu_1\times\nu_N}\right)&\cdots&{\rm cov}\left(\mathcal{D}_\ell^{\nu_1\times\nu_1},\mathcal{D}_\ell^{\nu_N\times\nu_N}\right)\\
\vdots&\ddots&\vdots&&\vdots\\
{\rm cov}\left(\mathcal{D}_\ell^{\nu_1\times\nu_N},\mathcal{D}_\ell^{\nu_1\times\nu_1}\right)&\cdots&{\rm var}\left(\mathcal{D}_\ell^{\nu_1\times\nu_N},\mathcal{D}_\ell^{\nu_1\times\nu_N}\right)&\cdots&{\rm cov}\left(\mathcal{D}_\ell^{\nu_1\times\nu_N},\mathcal{D}_\ell^{\nu_N\times\nu_N}\right)\\
\vdots&&\vdots&\ddots&\vdots\\
{\rm cov}\left(\mathcal{D}_\ell^{\nu_N\times\nu_N},\mathcal{D}_\ell^{\nu_1\times\nu_1}\right)&\cdots&{\rm cov}\left(\mathcal{D}_\ell^{\nu_N\times\nu_N},\mathcal{D}_\ell^{\nu_1\times\nu_N}\right)&\cdots&{\rm var}\left(\mathcal{D}_\ell^{\nu_N\times\nu_N},\mathcal{D}_\ell^{\nu_N\times\nu_N}\right)\\
}.
\end{align}
The computation of this cross-spectra covariance matrix from simulations is described in detail, for our applications, in \app{sec:correlations}. 
The reduced $\chi^2$ to be minimized is then defined as:
\be
\label{eq:chi2_definition}
\chi^2=\frac{{\bf r}^T \mathbb{C}^{-1} {\bf r}}{N_{\rm dof}},
\ee
where $N_{\rm dof}$ is the number of degrees of freedom.
%

\section{Data and simulation settings}\label{sec:data_and_sims}

In this section we present the \Planck{} intensity data and the simulations used in this paper. We only use the five highest-frequency \PlanckHFI{} channels (143, 217, 353, 545 and 857\,GHz). We discard lower frequencies in order to minimize the impact of emission components other than the thermal dust emission which are significant at frequencies lower than 143 GHz but may be neglected at higher frequencies at high Galactic latitude, such as the synchrotron emission, the anomalous microwave emission (AME), and the free-free emission. The CO emission lines at 115, 230, and 345\,GHz are significant in the \Planck{} 217 and  353\,GHz  channels, but their impact can be strongly reduced by applying a tailored mask, which is what we do as detailed later in this section and in \app{sec:templates}. The cosmic infrared background (CIB) emission is a significant emission component to the total intensity in all frequency bands from 143 to 857\,GHz, with an amplitude relative to dust that increases towards small angular scales \citep{2014A&A...571A..30P}. 
 
 In the following analysis, we consider five different data sets (full sky intensity maps at 143, 217, 353, 545 and 857 GHz): four types of simulations of the \Planck{} intensity data (labeled SIM1, SIM2, SIM3 and SIM4) and the actual \Planck{} intensity data (labeled PR3). All the maps used in this work use the {\tt HEALPix}\footnote{\url{http://healpix.jpl.nasa.gov}} pixelization \citep{healpix}.
 
\subsection{Simulated map data sets}

For each simulation type, the frequency channel maps $M^{\rm SIM}_{\nu_i}$ are the sum of a noise component $N$ and a sky template $S$. 
The sky template $S$ includes a dust component of increasing complexity from SIM1 to SIM4 which has been implemented in order to explore the impact of the spatial variations of the dust spectral index ---and eventually other components, such as the CIB--- to assess its potential contribution to the moment expansion analysis of the dust.

\subsubsection{Noise component}
\label{sec:noise_component}

The noise component $N$ is the same for each simulation type. We use 300 \Planck{} End-to-End noise maps ($N_{\nu_i}^\eta,\ \eta\in\{1,300\}$, 300 realizations per frequency band $\nu_i$) obtained from the {\tt FFP10} \Planck{} simulations (which include the contributions of both the noise and the residual systematic errors). These maps are publicly available in the \Planck{} Legacy Archive (PLA\footnote{\url{http://pla.esac.esa.int/pla/}\label{plafootnote}}).  

\subsubsection{Dust component}
\label{sec:dust_component}

The dust component is built from a dust intensity map template at 353 GHz, defined, in MJy$\,\rm{sr}^{-1}$ units, as: 
\be\label{eq:dust_353}
S^{\rm D}_{353}({\bf \hat{n}})=10^{20}\cdot M^{\rm D}_{\tau_{353}}({\bf \hat{n}})\cdot B_{\nu=353\,{\rm GHz}}(T_0=19.6\,{\rm K}),
\ee
where $B_{\nu=353\,{\rm GHz}}(T_0=19.6\,{\rm K})$ is the Planck function at 353\,GHz for the temperature $T_0=19.6\,{\rm K}$ in W\,m$^2$\,sr$^{-1}$, and $M^{\rm D}_{\tau_{353}}({\bf \hat{n}})$ is the the dust optical depth map at 353\,GHz. For all the simulations, we use the $M^{\rm D}_{\tau_{353}}$ map derived from an MBB fit of 
\Planck{} dust total intensity maps, obtained with the GNILC component separation method designed to separate dust from CIB anisotropies  \citep{2016A&A...596A.109P}\footnote{The maps are available in the PLA}. 
The GNILC analysis produced all-sky maps of Planck thermal dust emission  at 353, 545, and 857\,GHz with reduced CIB contamination. Reducing the CIB contamination of the thermal dust maps is crucial to building maps of dust optical depth, temperature, and dust spectral index that are accurate at high Galactic latitudes.
The dust map template $S^{\rm D}_{353}({\bf \hat{n}})$ at 353\,GHz as defined in \eq{eq:dust_353} is shown in the top panel of \fig{Fig:templates}.

The coefficients used to re-scale the dust template at 353\,GHz to a different frequency $\nu_i$ are defined as:
\be 
\label{eq:extrapolation1}
\alpha_{\nu_i}({\bf \hat{n}})=\frac{\nu_i^{\beta({\bf \hat{n}})} B_{\nu_i}(T({\bf \hat{n}}))}{(353\,{\rm GHz)}^{\beta({\bf \hat{n}})} B_{353\,{\rm GHz}}(T({\bf \hat{n}}))},
\ee
meaning that the dust map at the frequency $\nu_i$ is:
\be
\label{eq:extrapolation2}
S^{\rm D}_{\nu_i}({\bf \hat{n}})=S^{\rm D}_{353}({\bf \hat{n}}) \, \alpha_{\nu_i}({\bf \hat{n}}).
\ee

We define three types of dust simulation, with different levels of complexity in the frequency scaling of the dust intensity in \eq{eq:extrapolation1} and \eq{eq:extrapolation2}:

\begin{enumerate}

\item $S^{{\rm D}_1}_{\nu_i}$: Simulations with constant dust temperature ($T({\bf \hat{n}})=T_0=19.6$\,K) and spectral index ($\beta({\bf \hat{n}})=\beta_0=1.59$).

\item $S^{{\rm D}_2}_{\nu_i}$: Simulations with Gaussian variations of the dust spectral index and fixed temperature ($T({\bf \hat{n}})=T_0=19.6$\,K). The Gaussian spectral index map $\beta({\bf \hat{n}})=\mathcal{N}(\beta_0,\Delta\beta^2)$ is a single random realization (i.e., the same realization for all the simulations) of the normal distribution with a mean $\beta_0=1.59$ and a 1-$\sigma$ dispersion $\Delta\beta=0.1$. The $\Delta\beta$ value is chosen to roughly match the observed dispersion of the spectral index variations in the \Planck{} data \citep{2015A&A...576A.107P,2016A&A...596A.109P}. The corresponding $\beta({\bf \hat{n}})$ map is shown in the top panel of \fig{Fig:beta_maps}.

\item $S^{{\rm D}_3}_{\nu_i}$: Simulations using \Planck{} sky maps of the dust spectral index and temperature.  The dust spectral index map $\beta({\bf \hat{n}})$ and the temperature map $T({\bf \hat{n}})$ are those derived from the MBB fit of the \Planck{} GNILC dust total intensity maps \citep{2016A&A...596A.109P}. For illustration, the $\beta({\bf \hat{n}})$ map is shown in the bottom panel of \fig{Fig:beta_maps}.

\item $S^{{\rm D}_4}_{\nu_i}$: Simulations using the GNILC \Planck{} sky map of the dust spectral index $\beta({\bf \hat{n}})$ and a constant dust temperature $T_0=19.6$\,K. These simulations are not used in the main analysis but presented in \app{sec:dust_beta_tconst} to assess the effect of the dust temperature spatial variations.

\end{enumerate}
The simulations are computed at the \PlanckHFI{} reference frequencies. For comparison, the PR3 data will be color-corrected 
to account for the \PlanckHFI{} bandpasses \citep{bandpass}. 

\subsubsection{Cosmic infrared background and synchrotron}

In one of our simulation sets, detailed below, we include a CIB component $S^{\rm CIB}_{\nu_i}$. For this, we use the multi-frequency CIB simulation of the \Planck{} Sky Model (PSM) version 1.9 described in \cite{2013A&A...553A..96D}\footnote{\url{http://www.apc.univ-paris7.fr/~delabrou/PSM/psm.html}}. This CIB map is a random Gaussian realization matching the \Planck{} measured CIB power spectra of \citet{planckcib}. 
As an illustration, we show the CIB  map at 353 GHz in the middle panel of \fig{Fig:templates}.
Given that the synchrotron emission has a negligible impact on our analysis we do not detail the simulations of this component here, but comment on it in \app{sec:sync_vs_CIB}.

\subsubsection{Simulation types}

From the components above, we produce four batches of 100 simulated intensity maps at 143, 217, 353, 545, and 857\,GHz, numbered by the superscript $\eta\in\{1,100\}$:

\begin{enumerate}[label=(\alph*)]

\item {\bf SIM1}: $M^{\rm SIM1,\eta}_{\nu_i} =  N^\eta_{\nu_i}+S^{{\rm D}_1}_{\nu_i}$,
\item {\bf SIM2}: $M^{\rm SIM2,\eta}_{\nu_i} =  N^\eta_{\nu_i}+S^{{\rm D}_2}_{\nu_i}$,
\item {\bf SIM3}: $M^{\rm SIM3,\eta}_{\nu_i} =  N^\eta_{\nu_i}+S^{{\rm D}_3}_{\nu_i}$,
\item {\bf SIM4}: $M^{\rm SIM4,\eta}_{\nu_i} =  N^\eta_{\nu_i}+S^{{\rm D}_3}_{\nu_i}+S^{\rm CIB}_{\nu_i}$.

\end{enumerate}
We note that for a given simulation type and in a given frequency band, only the noise realization changes from one simulation to another while the sky component remains constant. 

\subsection{\Planck{} map data set}

For the actual \Planck{} map data set, we use the publicly available total intensity maps (observed at 143, 217, 353, 545 and 857\,GHz) from the third and latest \Planck{} release. When referring to the data we therefore use the "PR3" label.
The CMB component is subtracted from the PR3 intensity data set at the map level. We subtract the SMICA CMB map \citep{smica2018} from the five \PlanckHFI{} frequency channel maps we use in this work. As the SMICA map resolution is different from that of the \PlanckHFI{} frequency channel maps, we first deconvolve the SMICA CMB map from its beam function and then convolve it with the beam of individual \PlanckHFI{} frequency channel maps before subtraction. All the required information to do so is available in the PLA\footnoteref{plafootnote}.

\subsection{Cross-power spectra computation}
\label{sec:crosscomputation}

The cross-angular power spectra are computed from the data sets (SIM1, SIM2, SIM3, SIM4 and PR3) applying the LR42 sky mask defined in \cite{2016A&A...586A.133P} and used in several \Planck{} publications. This mask leaves $42\%$ of the sky for the analysis; it is apodized and includes a galactic mask, a point source mask, and a CO mask, as described in \cite{2016A&A...586A.133P}, and is shown in \fig{Fig:maskLR42}. 

To compute the cross-spectra we use the {\tt Xpol} code described in \cite{Tristram:2004if}. The {\tt Xpol} code is a pseudo-$C_\ell$ power spectrum estimator that corrects for the incomplete sky coverage, the filtering effects, and the pixel and beam window functions. 

For both the \Planck{} and the simulation data sets, we compute the 15 possible cross-spectra\footnote{$143\times143$, $143\times217$, $143\times353$, $143\times545$, $143\times857$, $217\times217$, $217\times353$, $217\times545$, $217\times857$, $353\times353$, $353\times545$, $353\times857$, $545\times545$, $545\times857$, and $857\times857$.}  between the five \PlanckHFI{} channels from 143 to 857\,GHz, as described in \sect{sec:method_and_implementation}, and store them in the data vector $\vec{\mathcal{D}_\ell}$ (see \eq{eq:dl_vector_general}). The cross-spectra vector is binned in 15 bins of multipoles\footnote{Centered on the multipoles $\ell=$ 23, 40, 60, 80, 100, 120, 140, 160, 180, 200, 220, 240, 260, 280 and 300, respectively.} of size $\Delta\ell=20$ in the range most relevant for CMB primordial $B$-modes analysis ($\ell\in\{20,300\}$), such that $\vec{\mathcal{D}_b}\equiv\langle\vec{\mathcal{D}_\ell}\rangle_{\ell\in b}$. As we work only with the binned version of the cross-spectra, in the remainder of this paper and for the sake of clarity, we do not make the distinction between $b$ and $\ell$, and  so $\vec{\mathcal{D}_\ell}$ refers  to $\vec{\mathcal{D}_b}$.

In order to avoid the noise auto-correlation bias and to reduce the level of correlated systematic errors, we compute the cross-spectra from data split maps (the \Planck{} half-mission maps, HM). We explain how we combine half-mission maps and compute the covariance matrix of the cross-spectra in \app{sec:correlations}.

After the computation of the cross-spectra $\vec{\mathcal{D}_\ell}$, we subtract, from every data set (PR3 and simulations), the averaged cross-spectrum computed from the 300 \Planck{} End-to-End simulations, which include instrumental noise and systematic effects (see \sect{sec:noise_component}). This allows us to correct for the small bias linked to residual systematic errors in the data and, by construction, in our simulations. 
Finally, we apply a color-correction to the PR3 data set cross-spectra to get rid of \PlanckHFI{}-specific calibration effects as described in \citet{bandpass}. The units of the total intensity cross-spectra for all the data sets are [(MJy\,sr$^{-1}$)$^2$].

\section{Results for simulations and \Planck{} data}\label{sec:results}

Here we present the results of the fits made for the
five data sets (SIM1, SIM2, SIM3, SIM4 and PR3) described in \sect{sec:data_and_sims}. 
Each data set is fitted independently for each multipole bin $\ell$ using the MBB moments expansion with the three-step method introduced in \sect{sec:method_and_implementation}. The maximum order of the moments expansion is limited to third order by the number of degrees of freedom of our data sets (see \app{sec:correlations}).
Here, we focus on our results for the MBB and the moment expansion to this maximum order, but additional plots in \app{sec:multi_order} include results for first and second orders.  

This section is organized as follows. We report on the goodness of  fits and the dust amplitude spectrum $\mathcal{D}_\ell^{\AD\AD}$ in 
\sect{sec:chi2}, on the dust spectral index $\betaellb$ and its leading order correction $\Delta \betaellb$ in \sect{sec:beta}, and  on higher order moments  in \sect{sec:higher_moments}.  \sect{sec:moments_discussion} 
summarizes these results. 
For each set of simulations, we present mean results averaged over 100 realizations, with error bars corresponding to the standard deviation among realizations. For the PR3 data, we estimate error bars propagating uncertainties through the fit. We checked that this approach, when applied to simulations, yields comparable error bars to those computed from the 100 realizations.

\subsection{Goodness of fits and dust amplitude spectra}\label{sec:chi2}
\begin{figure*}%
\centering
\subfigure{%
\includegraphics[width=0.65\columnwidth]{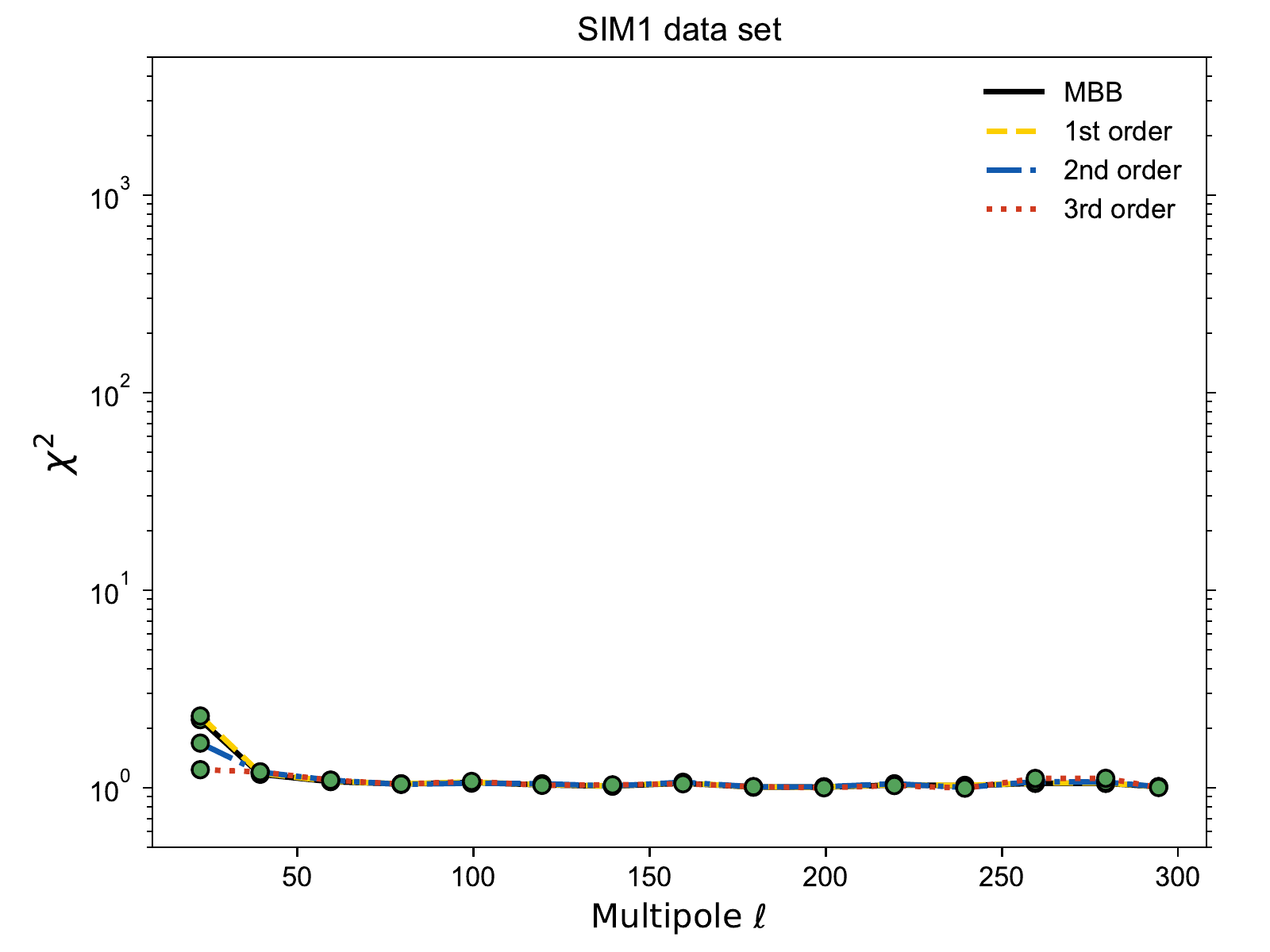}}
\subfigure{%
\includegraphics[width=0.65\columnwidth]{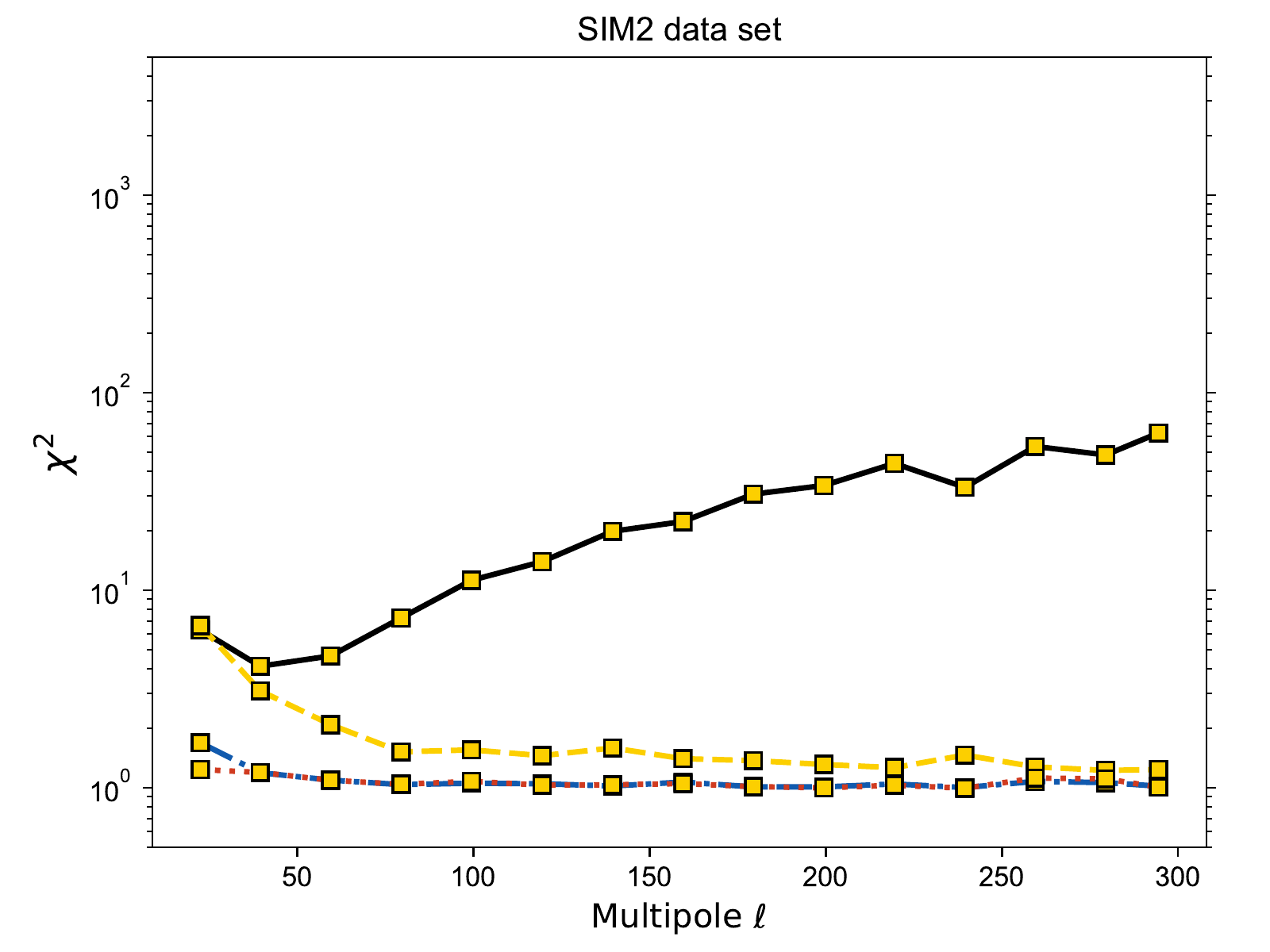}}
\subfigure{%
\includegraphics[width=0.65\columnwidth]{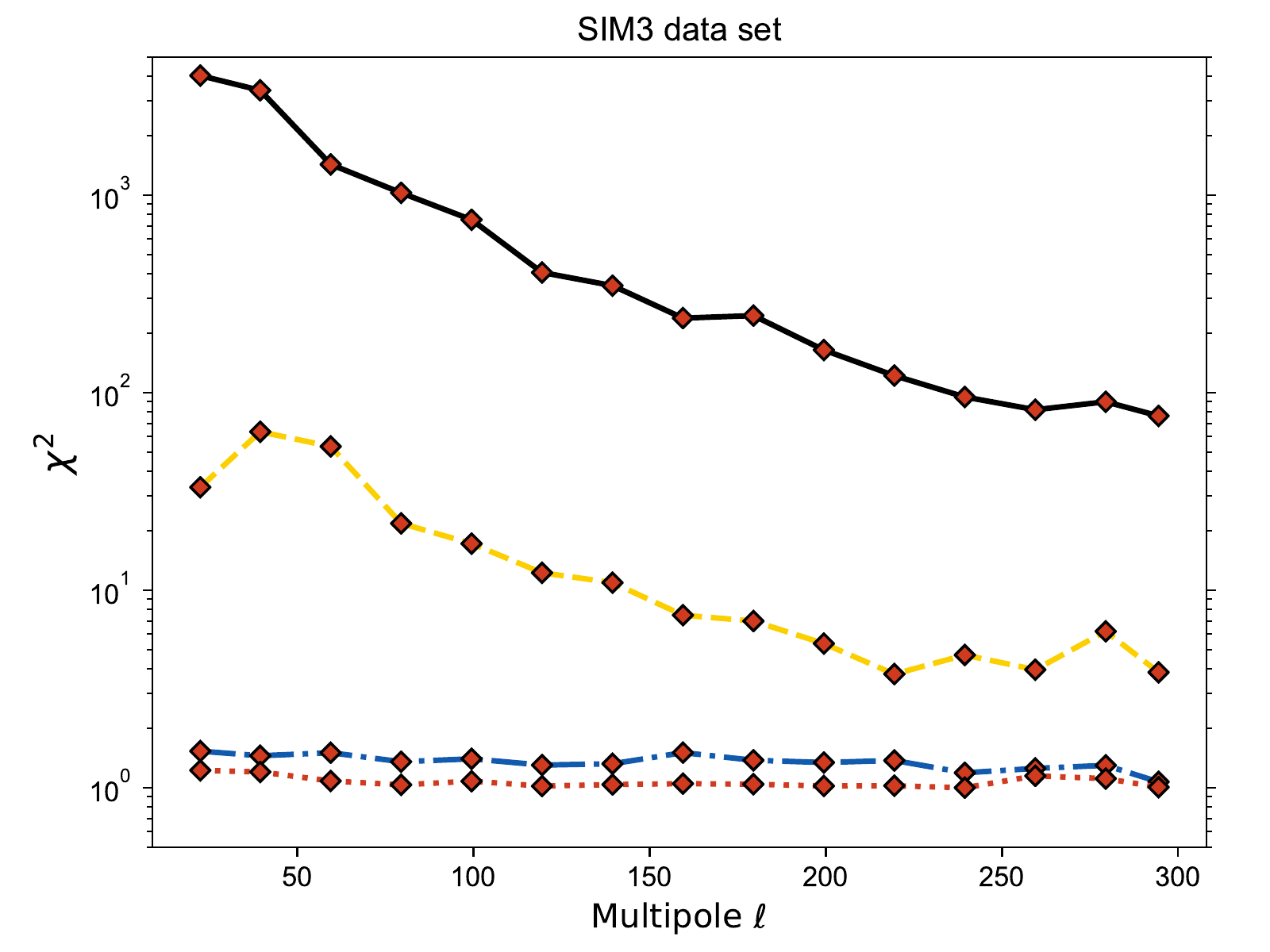}}
\subfigure{%
\includegraphics[width=0.65\columnwidth]{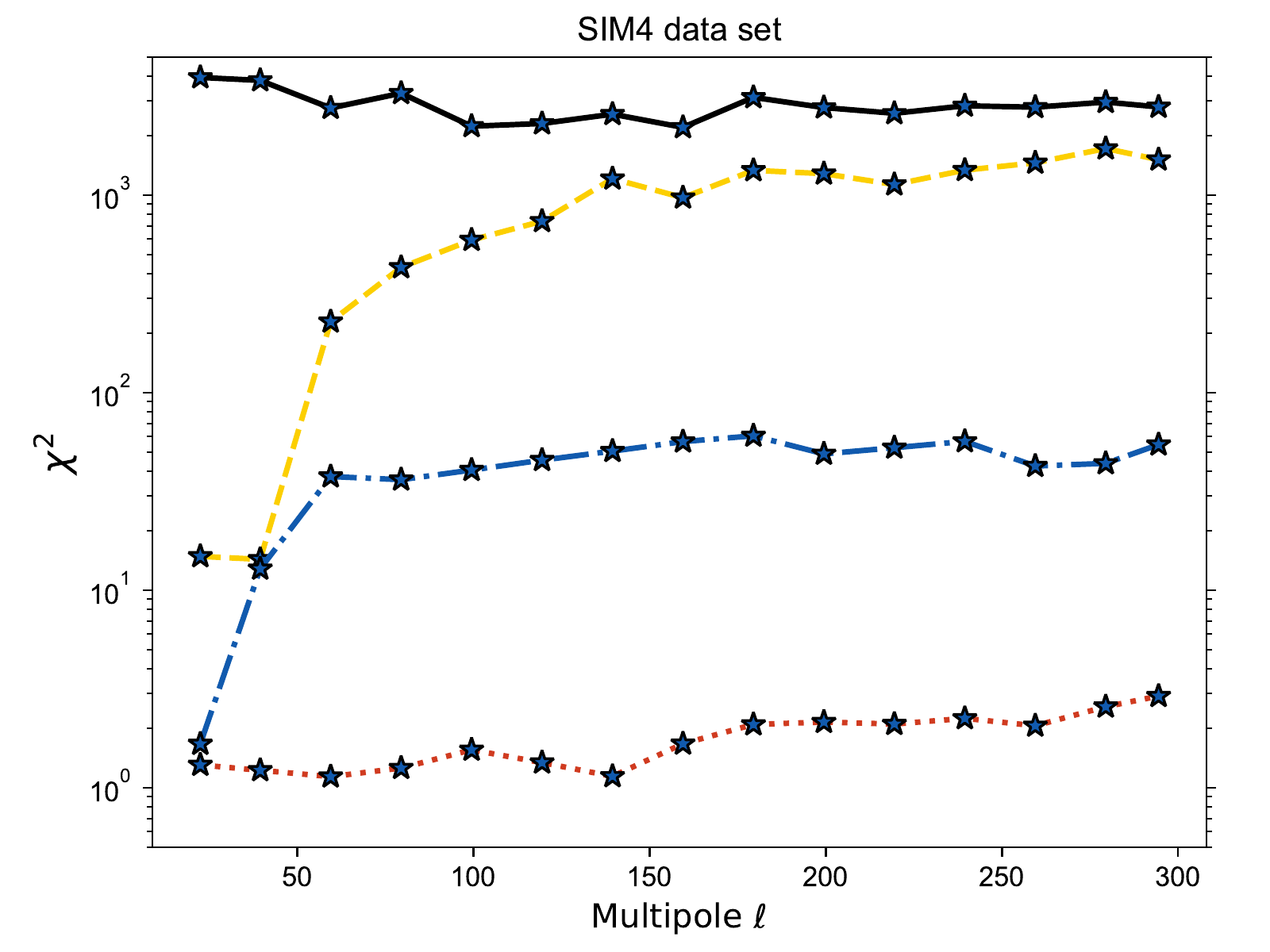}}
\subfigure{%
\includegraphics[width=0.65\columnwidth]{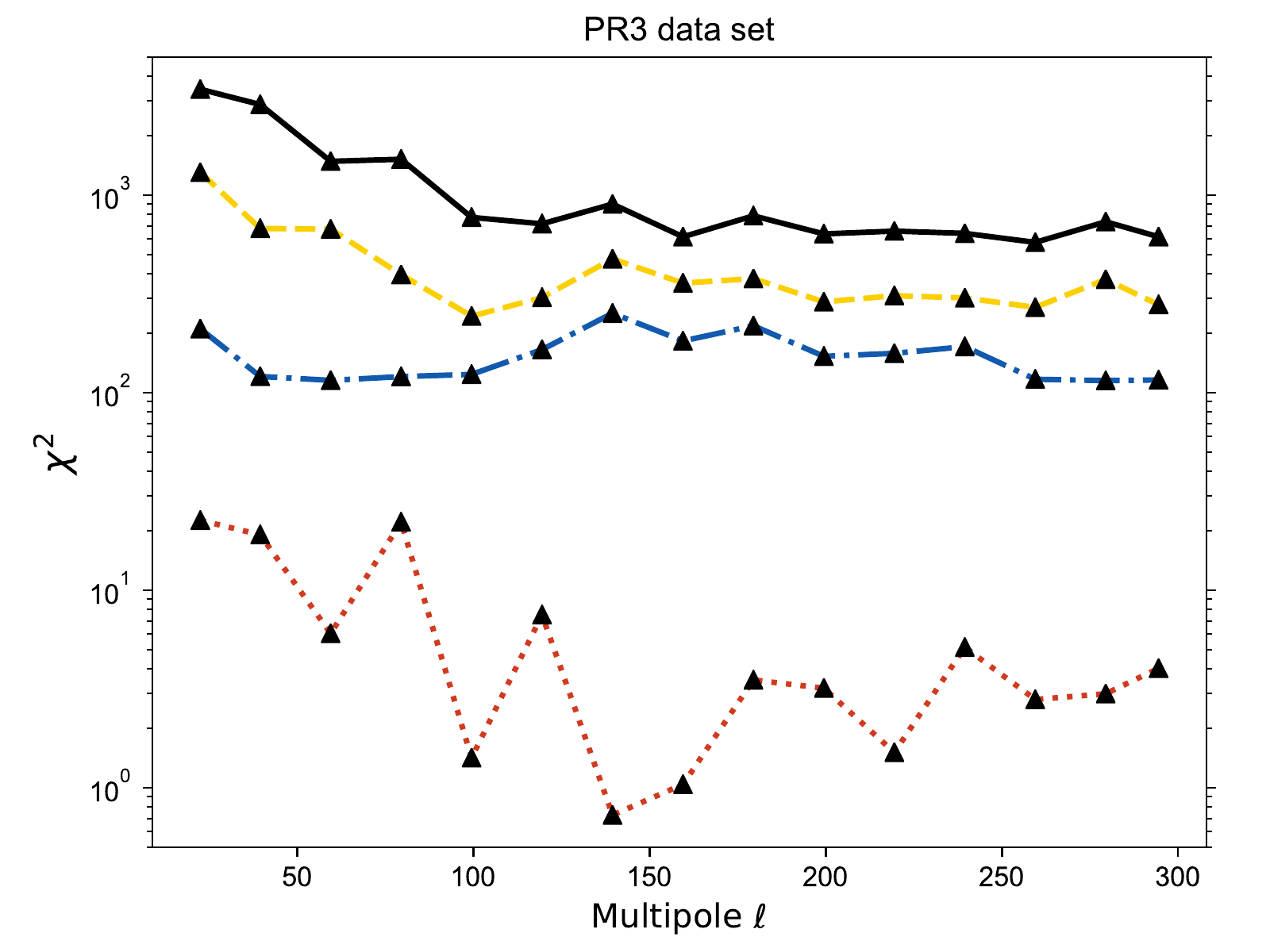}}
\caption{
Reduced $\chi^2$ of the moment expansion fits as a function of the multipole $\ell$. From the top left to the bottom right, the panels show the $\chi^2(\ell)$ results for the SIM1 (green circles), SIM2 (yellow squares), SIM3 (red diamonds), SIM4 (blue stars), and PR3 (black triangles) data sets. The $\chi^2(\ell)$ of the fits at zero (solid black), first (dashed yellow), second (dashed-dotted blue), and third order (dotted red) are displayed.}

\label{Fig:dust_moment_fit_vs_MBB_chi2_vs_ell}
\end{figure*}

The goodness of the fits is quantified with the reduced $\chi^2(\ell)$ plots, one per data set, shown in \fig{Fig:dust_moment_fit_vs_MBB_chi2_vs_ell}.
In each plot, we show the reduced $\chi^2(\ell)$  for four fits: the MBB law and the moments expansion in \eq{eq:dust_moment_cross_3} truncated to  first, second, and third order.

As expected, the SIM1 set is  well fitted by the MBB because these simulations do not include variations of the MBB parameters. The fact that the reduced $\chi^2$ of the  fit has a value close to 1 indicates that our correction for \Planck{} residual systematic errors described in \sect{sec:crosscomputation} is effective. There is some weak evidence of residual systematic errors in the lowest $\ell=23$ bin; indeed, for this bin a third-order fit is needed for the $\chi^2(\ell)$ to reach unity. For SIM2, fitting with higher order terms is required to get a reduced $\chi^2(\ell) \sim 1$. First order gives a fair $\chi^2(\ell)$ (except at very low $\ell$) and second order is needed to get a $\chi^2(\ell)$ close to unity. 

For the SIM3 and SIM4 simulations, as well as for the PR3 data, the MBB yields very poor fits. This illustrates 
the dual impact of multipole averaging, which introduces spectral complexity and increases the signal-to-noise ratio. 
For SIM3, a moment expansion up to second-order is required to get a fair fit and up to third order to reach $\chi^2(\ell)\sim 1$, while for the SIM4 set that includes the CIB, the third order is needed to have a fair reduced $\chi^2(\ell)$. 
For the PR3 data, the reduced $\chi^2(\ell)$ is somewhat worse than those for SIM4 (except for the MBB fit which is slightly better), indicating that the \Planck{} data have more spectral complexity than the SIM4 simulations.

\begin{figure}[ht!]%
\centering
\includegraphics[width=\columnwidth]{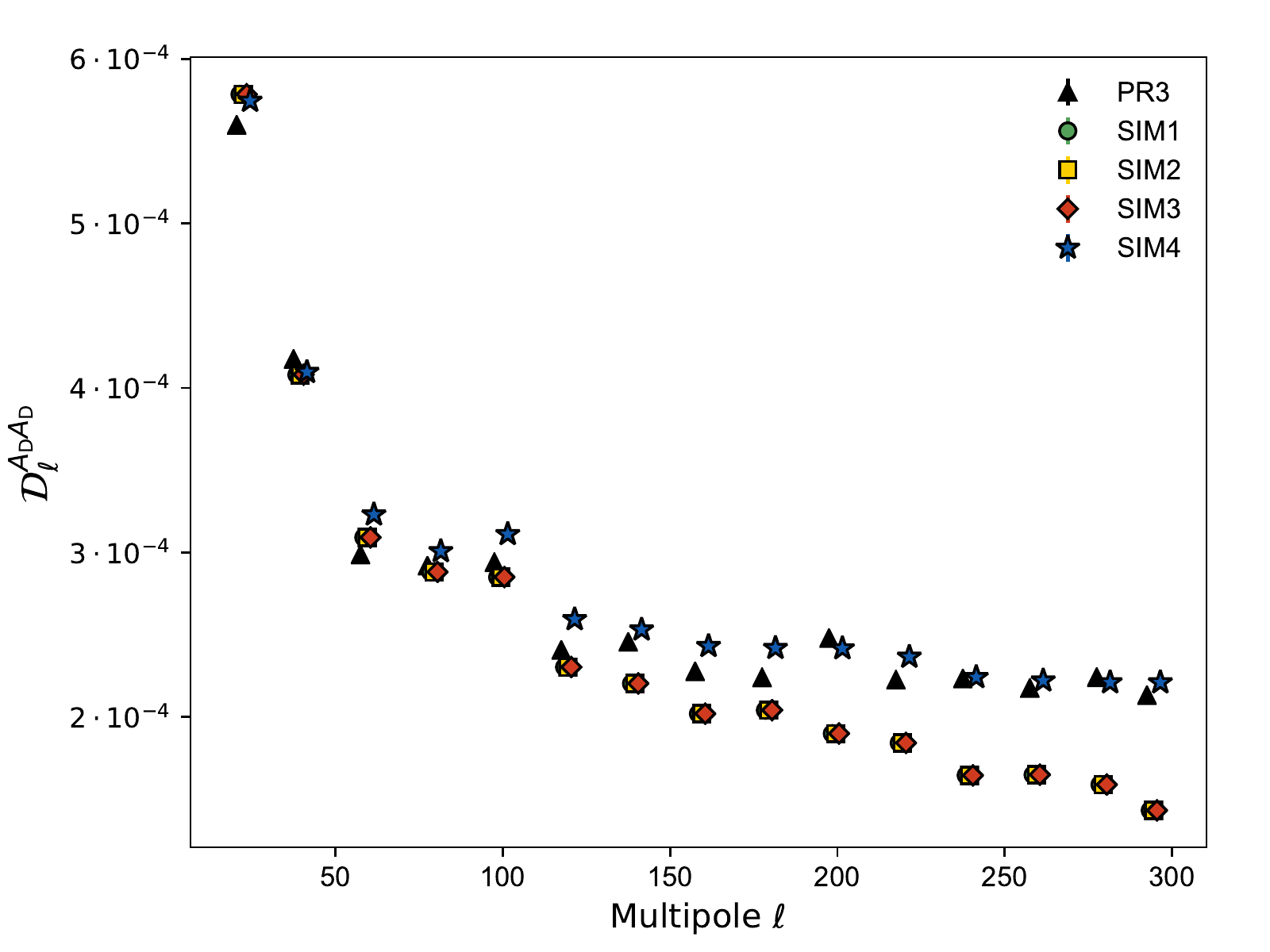}
\caption{Amplitude of the dust spectrum $\mathcal{D}_\ell^{\AD\AD}$ as a function of the multipole $\ell$. The 
symbols are green circles, yellow squares, red diamonds, and blue stars for the simulation sets SIM1 to SIM4,
and black triangles for the \Planck{} PR3 data. The error bars are smaller than the symbols. }
\label{Fig:DlAA}
\end{figure}

Figure~\ref{Fig:DlAA} shows the amplitude of the dust spectra $\mathcal{D}_\ell^{\AD\AD}$ for the third-order fit of
each data set. We note that these amplitudes do not significantly depend on the order of the fit (see \fig{fig:multi_order0}).
 As expected, the SIM1, SIM2, and SIM3 amplitudes are essentially indiscernible because they are built from the same dust spatial template  and they only differ in the modeling of the dust SED. The SIM4 and PR3 spectra are close to each other. Both depart from the dust-only simulations for multipoles $\ell\gtrsim100$, that is, for angular scales where the contribution from the CIB component is significant. We note that the power measured on the SIM4 simulations is somewhat larger than that measured for the \Planck{} data for multipoles ranging from about 100 to 220.
This may be due to the fact that the GNILC maps used as templates in the dust simulations are not fully free of CIB \citep{Chiang19}.

\subsection{Dust spectral index}
\label{sec:beta}

\begin{figure}%
\centering
\includegraphics[width=\columnwidth]{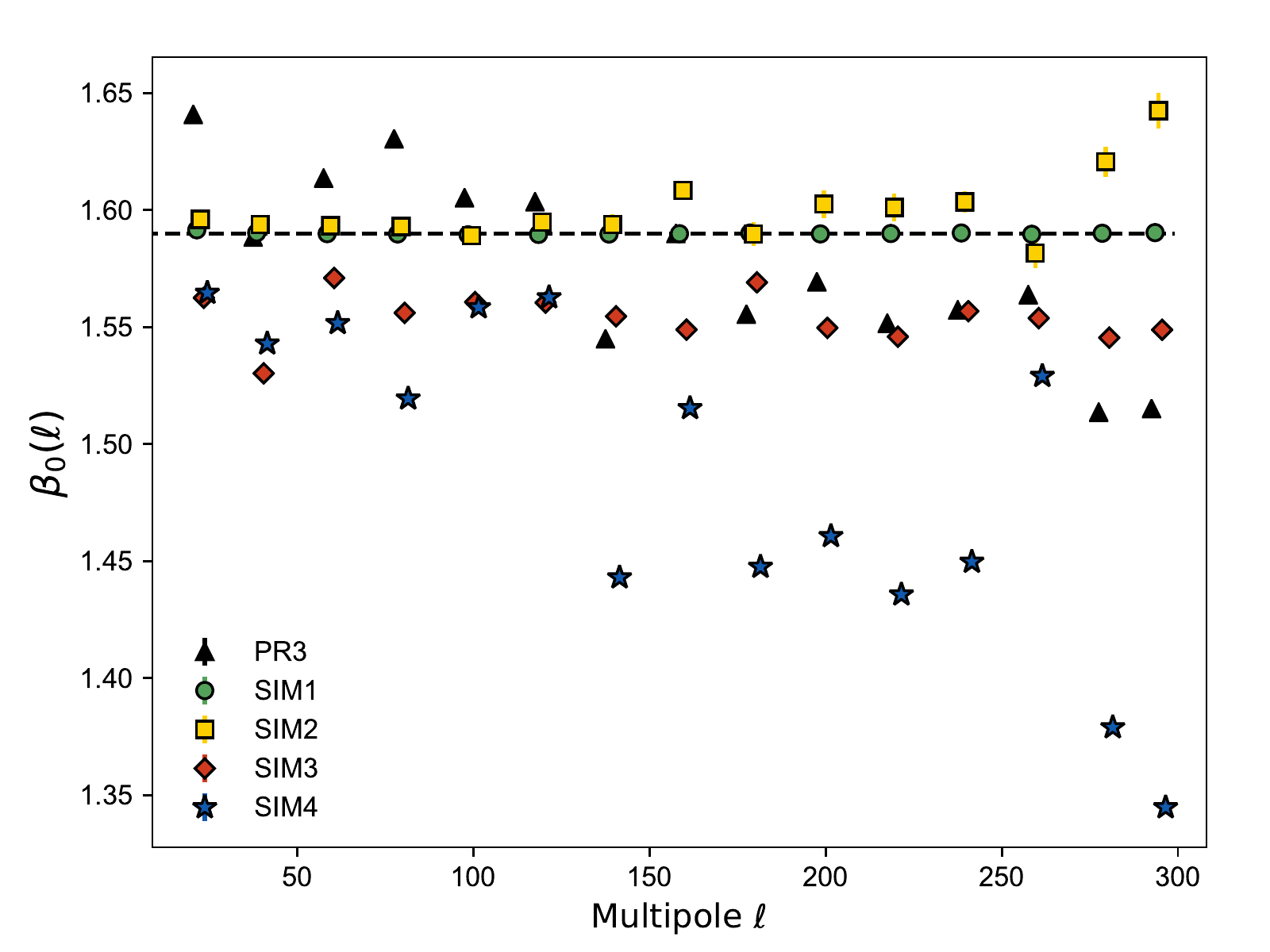}
\includegraphics[width=\columnwidth]{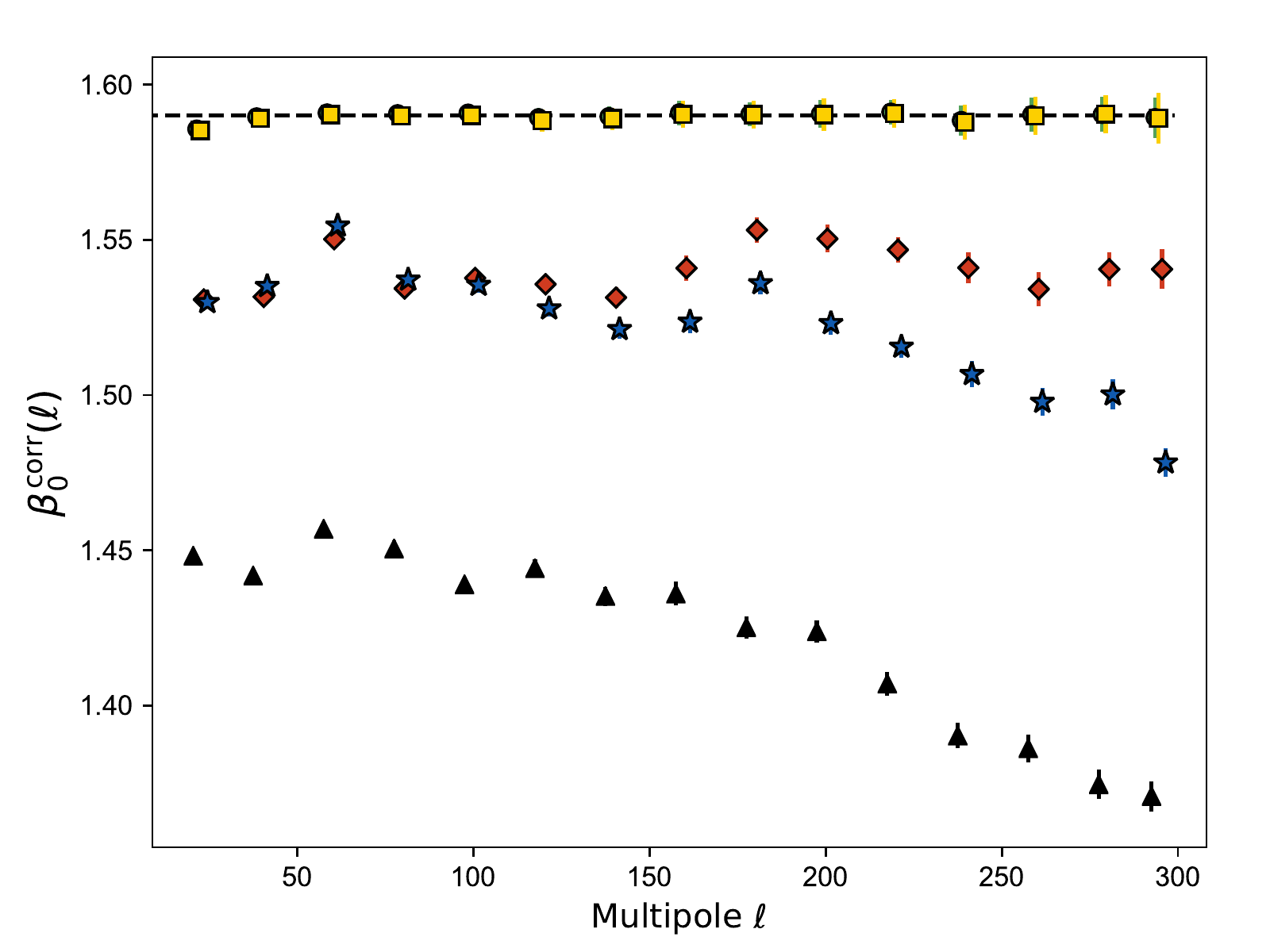}
\caption{{\it Upper panel.} 
Spectral index $\betaellb$ as a function of the multipole $\ell$ for the MBB fit (step 1). {\it Bottom panel.}  Corrected dust spectral index:  $\betaellbcorr=\betaellb+\Delta \betaellb$ from step 3 of the model fit. 
The symbols for the different data sets are as in \fig{Fig:DlAA}.}
\label{Fig:beta_deltabeta_betacorr}
\end{figure}

The dust spectral indices derived from our fitting are presented in \fig{Fig:beta_deltabeta_betacorr}. The top plot displays the spectral indices 
$\betaellb$ derived from the MBB fit, and 
the bottom plot the \emph{corrected}
dust spectral index $\betaellbcorr\equiv\betaellb+\Delta \betaellb$
in \eq{eq:deltabetaell}, obtained by fitting the spectral model in \eq{eq:dust_moment_cross_3} up to third order. The correction depends on the truncation order of the fit as illustrated in Fig.~\ref{fig:multi_order1}.

We find that $\betaellb$ 
matches the input spectral index of the simulation $\beta_0=1.59$ for the SIM1 set. For SIM2, $\betaellb$ values  are slightly larger than the input value at $\ell\gtrsim 100$ but the small difference is corrected when computing $\betaellbcorr$. 

Our results are less straightforward for the additional sets. 
For  SIM3,  we find $\langle\betaellbcorr\rangle_{\rm{SIM3}}=1.537\pm 0.003$
with no systematic dependence with $\ell$. 
For comparison, we computed the spectral index for a MBB
from the ratio between the cross-spectra of the 217 and 353\,GHz maps and the 353\,GHz power spectrum, as in \citet{planck2018XI}. We find values in the range 1.52 to 1.55 with no systematic dependence on $\ell$, in good 
agreement with the corrected index from our model. We also computed 
the median spectral index  of the GNILC input map (corrected to the reference temperature $T_0 =19.6\,$K we use in our model) over the L42 mask. 
The resulting value, 1.6, computed giving equal weights to each pixel, is slightly larger than $\langle\betaellbcorr\rangle_{\rm{SIM3}}$. 

The comparison of SIM3 and SIM4 in \fig{Fig:beta_deltabeta_betacorr}  shows that the CIB  lowers both $\betaellb$ and $\betaellbcorr$ for $\ell > 100$. Above this multipole, the difference between SIM3 and SIM4 values of $\betaellbcorr$  increases steadily with $\ell$, as does the CIB contribution. 
The values of $\betaellbcorr$ obtained on the PR3 data are systematically lower than the corresponding values for SIM4, but the dependence on $\ell$ is similar for both sets. We direct the reader to 
Fig.~\ref{fig:multi_order1}, which shows a systematic dependence of $\betaellbcorr$ on the order of the expansion, that is, on the spectral model. It is interesting to note that from second to third order the values of $\betaellbcorr$ move in the opposite direction for SIM4 and PR3. While the two sets of values roughly match for the second-order fits, they differ at third order.

To conclude this section, we compare our PR3 results with earlier determinations of the dust spectral index obtained by analyzing \Planck{} total intensity data.

\cite{planck2018XI} measured the dust spectral index from the ratio between the $217\times353$ and the $353\times353$ cross-spectra over the multipole range $\ell\in\{4,170\}$. For $\ell < 100$, where our analysis indicates that the measured spectral index is not biased by the CIB contribution, the spectral indices reported in their Table C.4 for the L42 mask are in the range 1.47 to 1.50, a little larger than our value $\langle\betaellbcorr\rangle_{\rm{PR3}}=1.45 \pm0.01$ averaged over the same range of multipoles. 
In \cite{planck2018XI}, the spectral index increases with $\ell$ for $\ell > 100$  up to 
$1.53\pm 0.01$ in their $\ell=150$ bin. In our analysis, we observe an opposite trend  with the spectral index decreasing for $\ell > 100$. It is not clear to us how to interpret this result. We are inclined to consider this  as evidence of a spectral mismatch between the SIM4 simulations and the PR3 data, rather than an outcome of the spectral model used to determine the spectral indices. This hypothesis is supported by the fit results in Fig.~\ref{fig:multi_order1}, which show that the order of the moments expansion does not have a significant effect on the $\ell$-dependence.

\subsection{High-order moments} 
\label{sec:higher_moments}
\begin{figure*}%
\centering
\includegraphics[width=0.65\columnwidth]{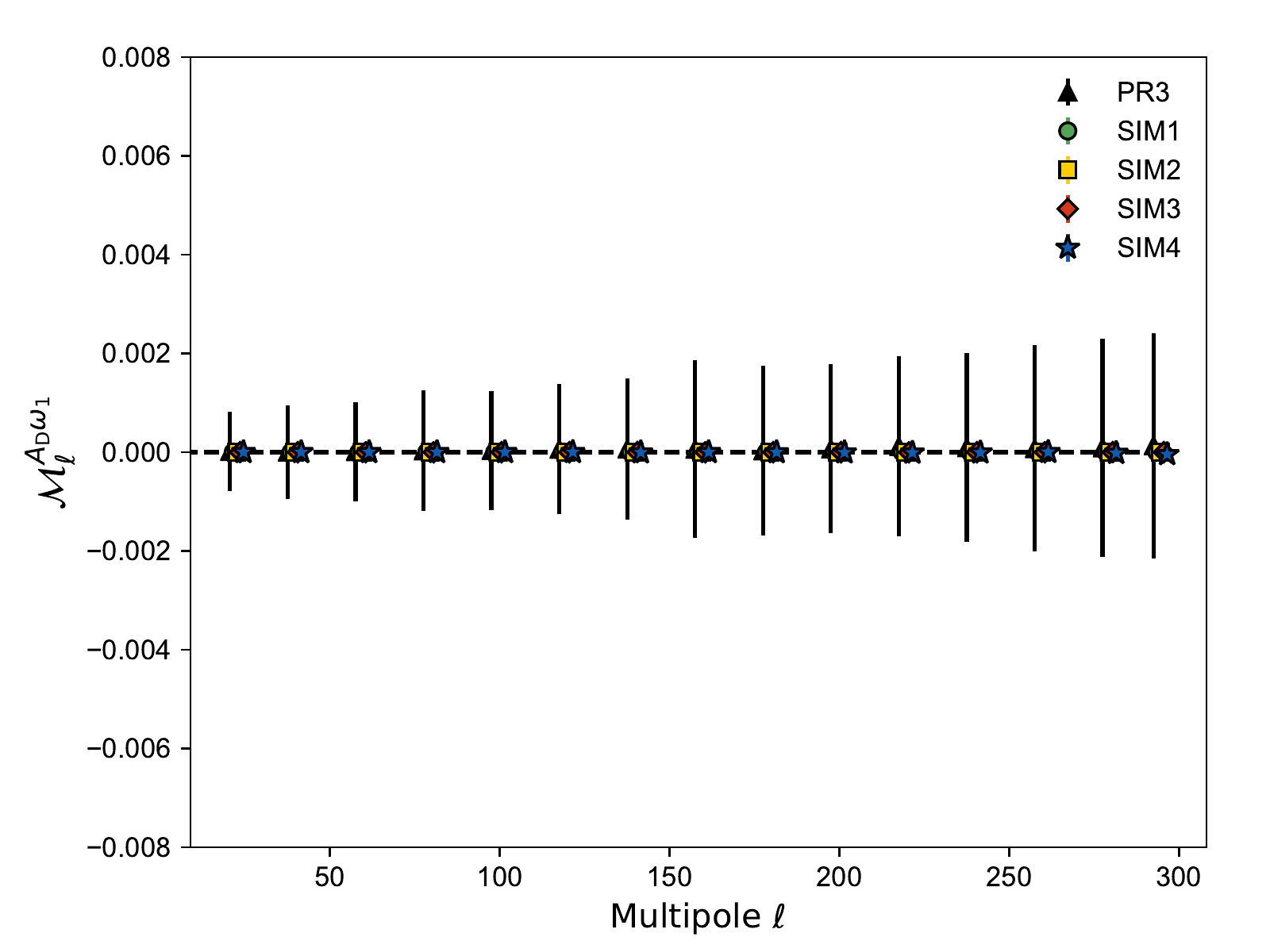}
\includegraphics[width=0.65\columnwidth]{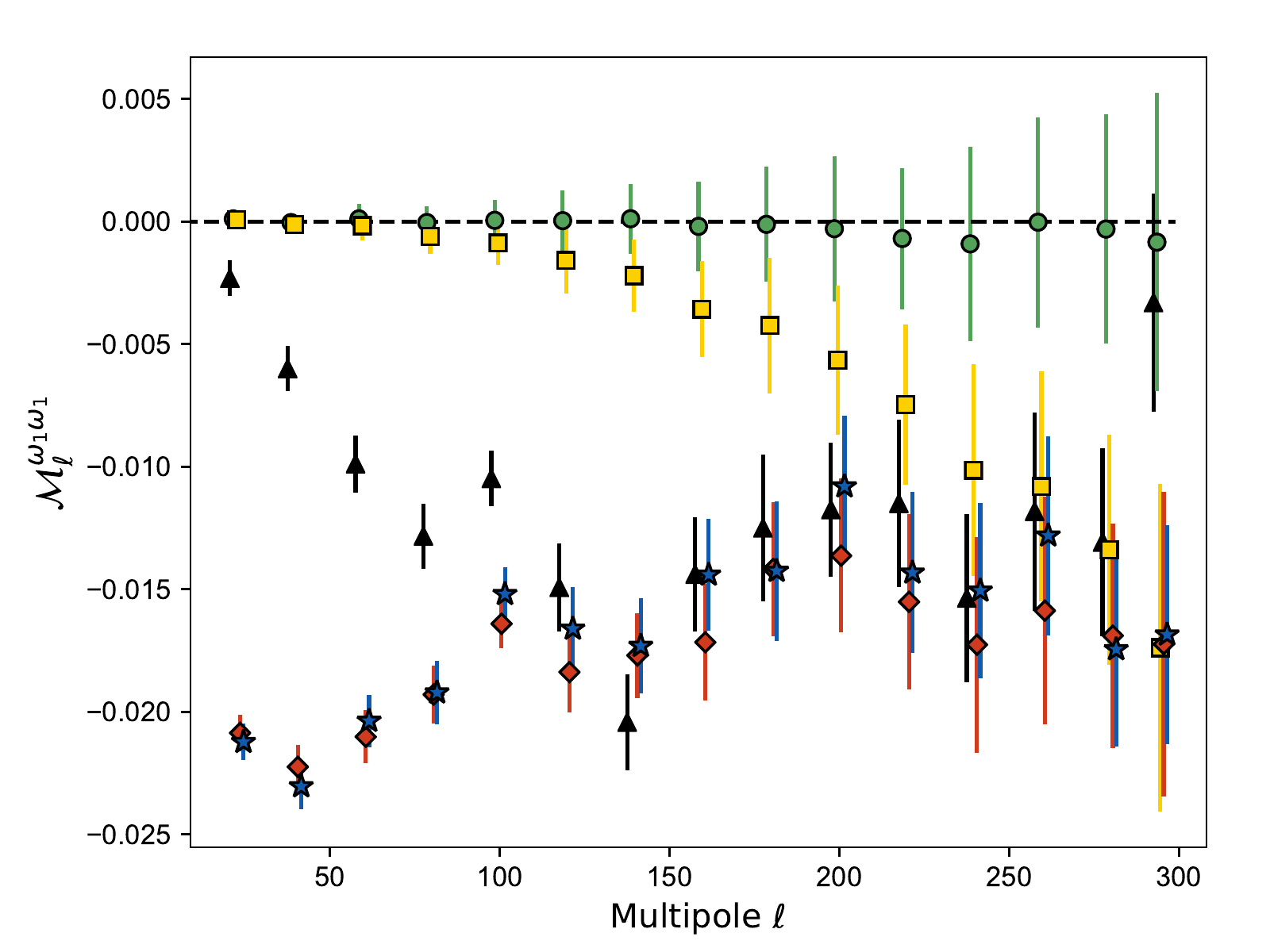}
\includegraphics[width=0.65\columnwidth]{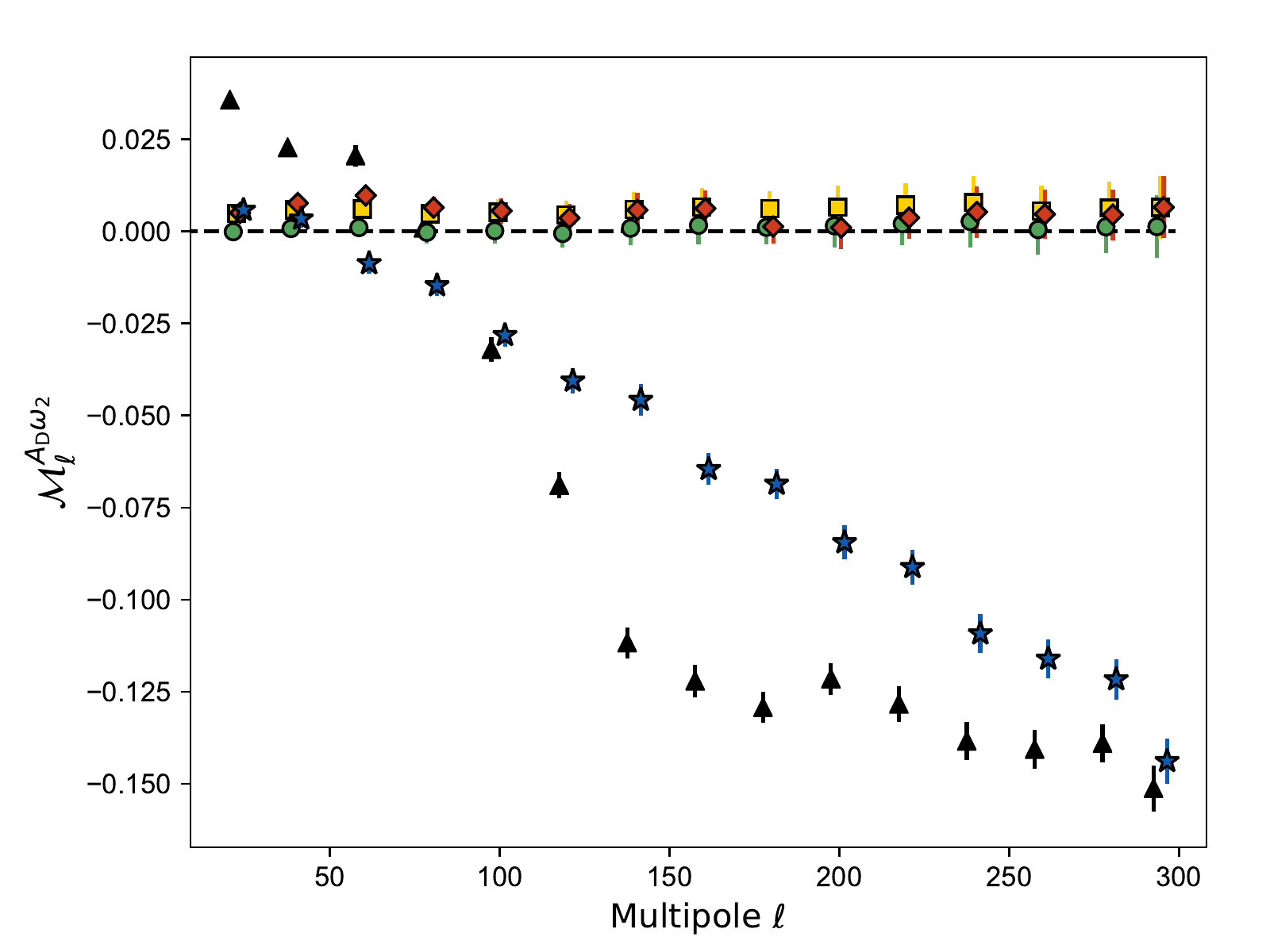}
\includegraphics[width=0.65\columnwidth]{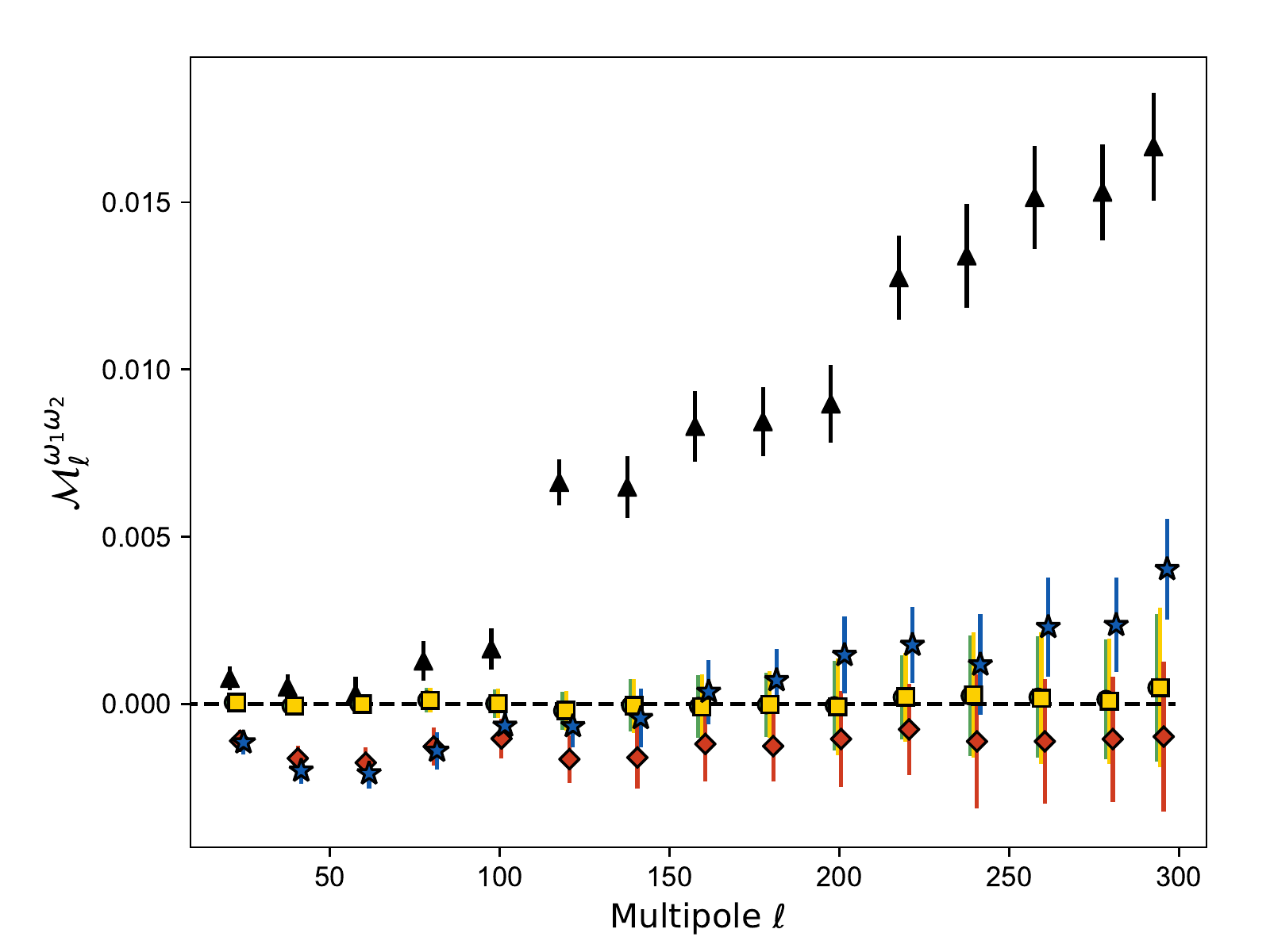}
\includegraphics[width=0.65\columnwidth]{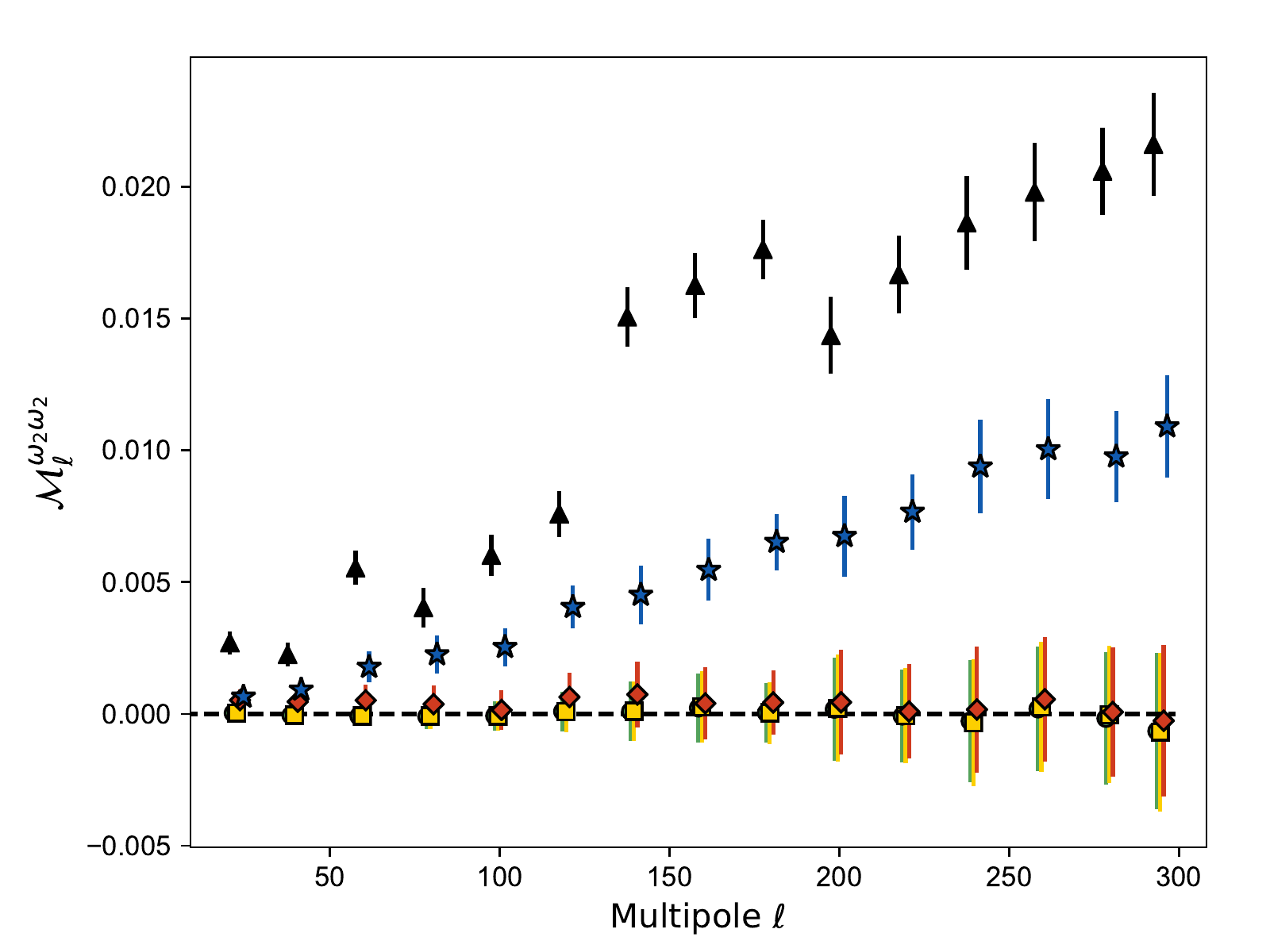}
\includegraphics[width=0.65\columnwidth]{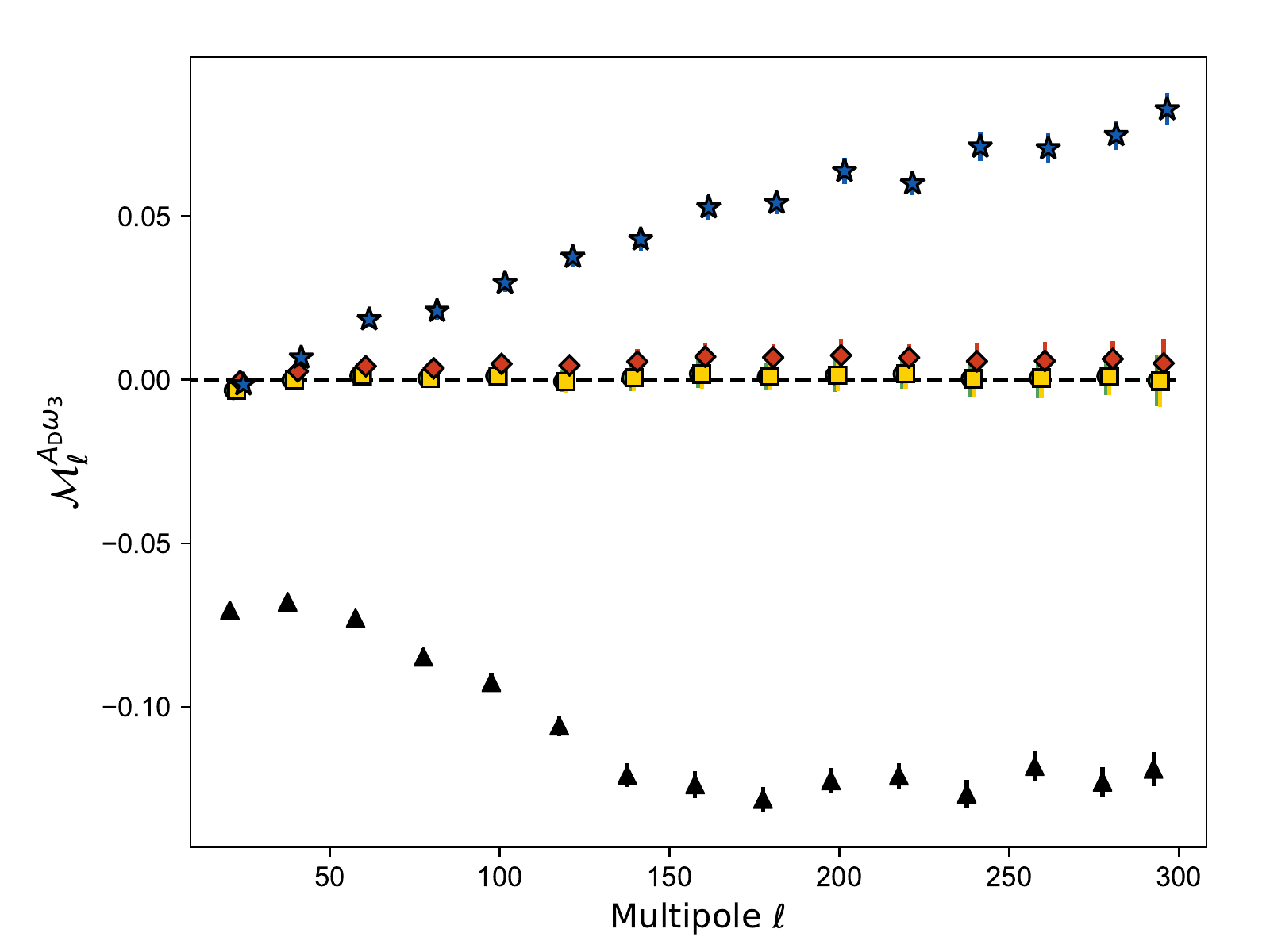}
\includegraphics[width=0.65\columnwidth]{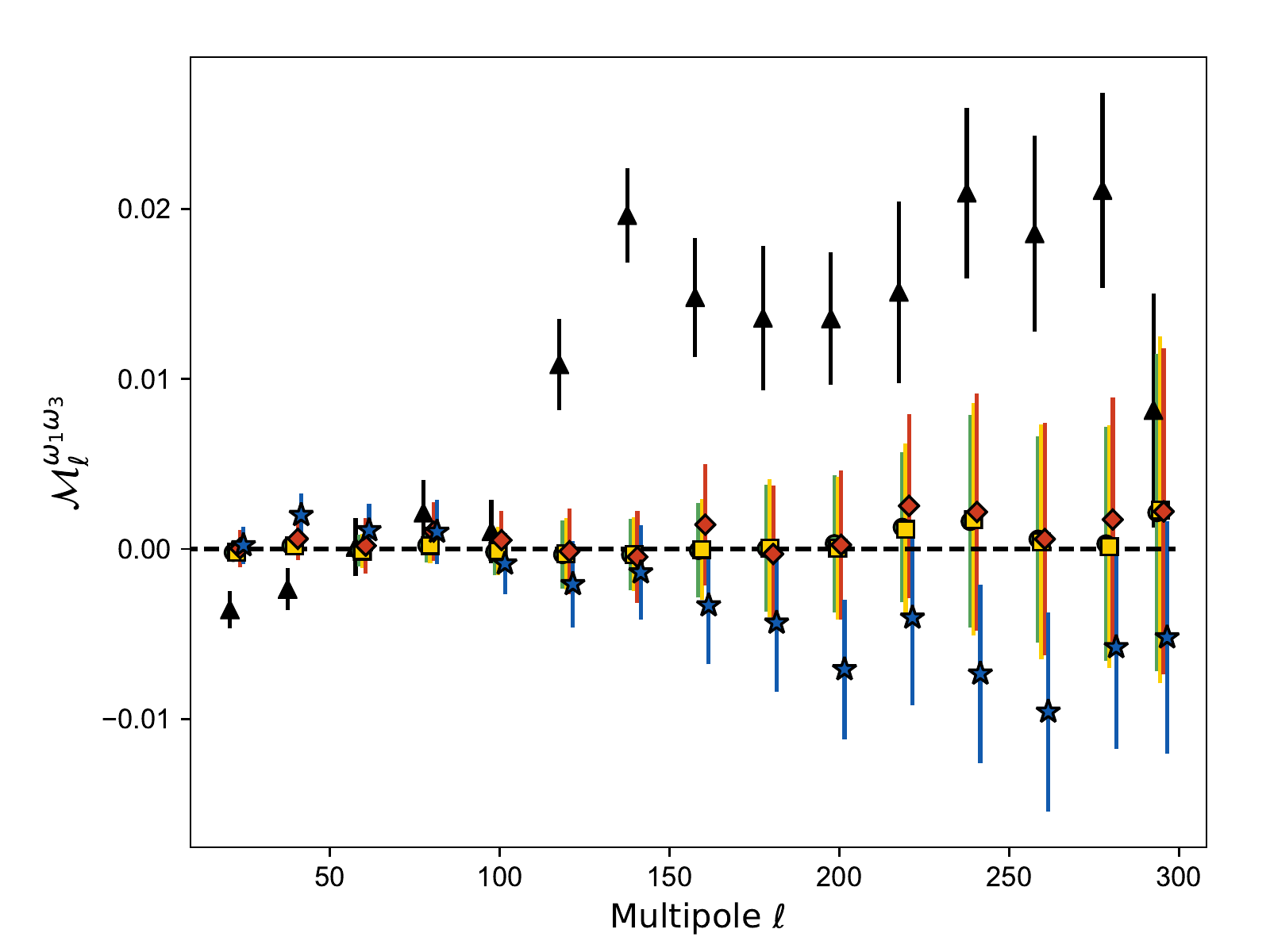}
\includegraphics[width=0.65\columnwidth]{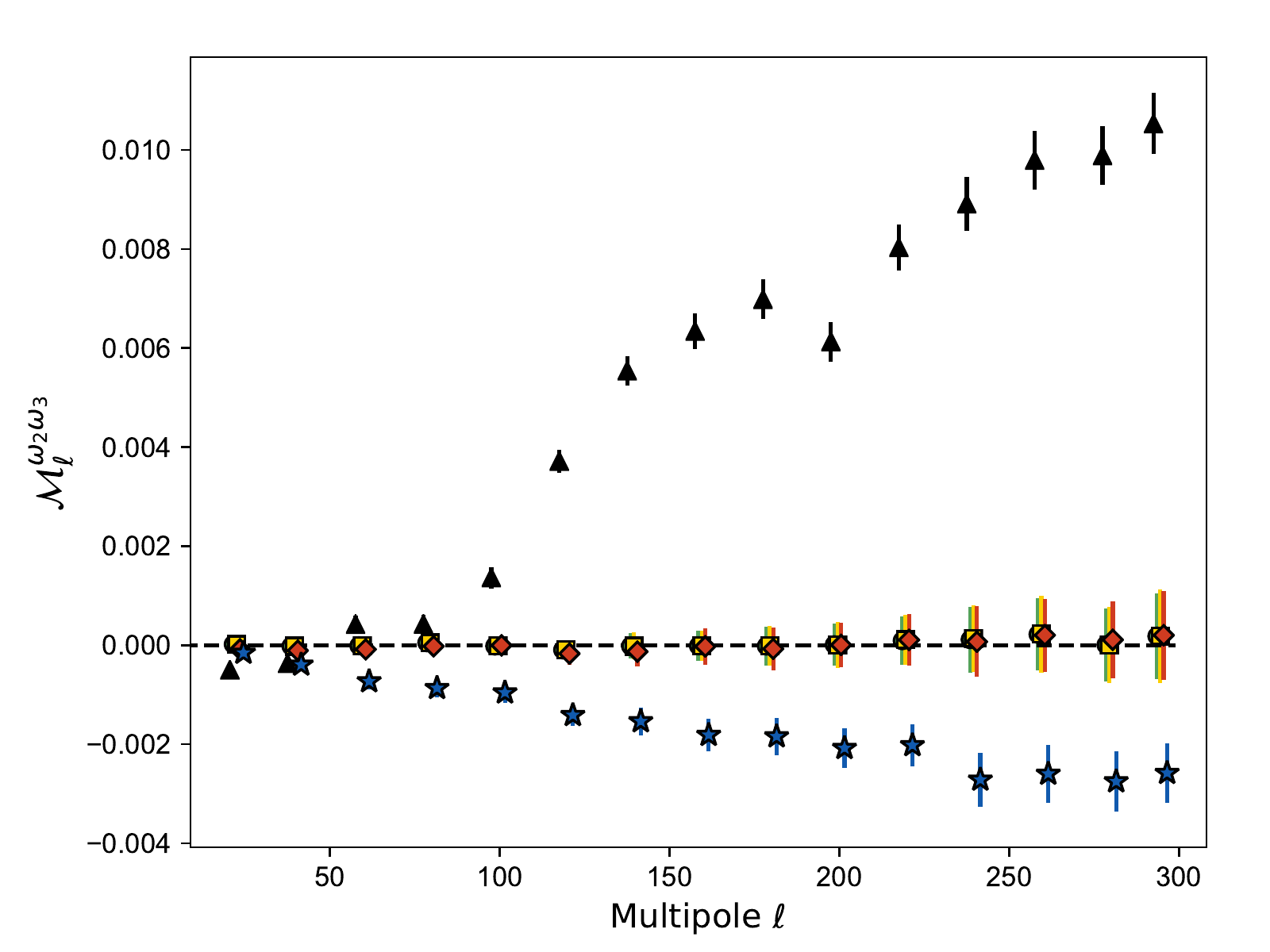}
\includegraphics[width=0.65\columnwidth]{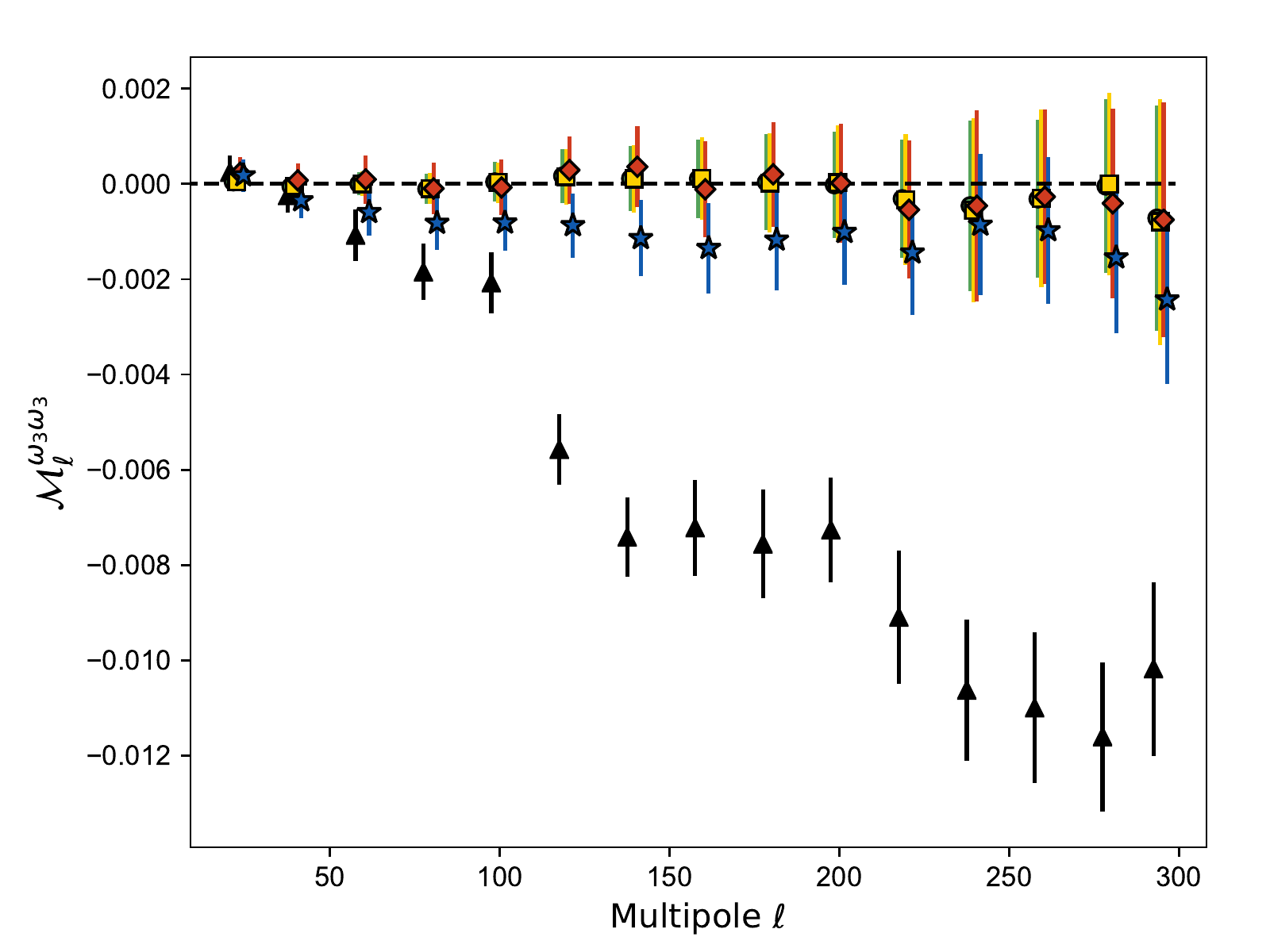}
\caption{Step 3 moment functions $\mathcal{M}_\ell^{ab}$ as a function of the multipole $\ell$ defined in \eq{eq:moment_vivj} for the $143 \times545$ cross-spectrum. 
The plots refer, from top left to bottom right, to $\mathcal{M}_\ell^{\AD\momi1}$, $\mathcal{M}_\ell^{\momi1\momi1}$, $\mathcal{M}_\ell^{\AD\momi2}$, $\mathcal{M}_\ell^{\momi1\momi2}$, $\mathcal{M}_\ell^{\momi2\momi2}$, $\mathcal{M}_\ell^{\AD\momi3}$, $\mathcal{M}_\ell^{\momi1\momi3}$, $\mathcal{M}_\ell^{\momi2\momi3}$ , and $\mathcal{M}_\ell^{\momi3\momi3}$. The symbols refer to the different data sets: SIM1 (green circles), SIM2 (yellow squares), SIM3 (red diamond), SIM4 (blue stars), and PR3 (black triangles).} 

\label{Fig:all_moments_LR42_143x545}
\end{figure*}

In this section, we discuss the results for the high-order (first- to third-order) moments in \eq{eq:dust_moment_cross_3}. We choose to show the \emph{moment functions} $\mathcal{M}_\ell^{ab}(\nu_i,\nu_j)$ defined in \eq{eq:moment_vivj} with $\nu_i=143$\,GHz and $\nu_j=545$\,GHz. We stress that this specific choice of frequencies only impacts the scaling pre-factor $c^{ab}(\nu_i,\nu_j,\nu_0)$ in \eq{eq:moment_vivj}, and not the $\ell$-dependence of the moments or the relative values between data sets for a given moment. 
The moment functions are presented in five plots 
in \fig{Fig:all_moments_LR42_143x545}, one for each data set. All moments in this figure were obtained from fits to third order.  
The top left panel of \fig{Fig:all_moments_LR42_143x545} represents the first-order moment function $\mathcal{M}_\ell^{\AD\momi1},$ which is
made compatible with zero through iteration on  $\Delta \betaellb$
(see \sect{sec:method_and_implementation}).

As expected, we do not detect any moment function for SIM1. This result gives us 
confidence that residual systematic errors included in our noise simulations (see \sect{sec:noise_component})
do not have a significant impact on high-order moments. 
For the SIM2 data set, the $\mathcal{M}_\ell^{\momi1 \momi1}$ function is detected with an amplitude increasing with $\ell$, while the other moments are consistent with zero. 
For SIM3, we find that the $\mathcal{M}_\ell^{\momi1 \momi1}$ moment is nearly scale-independent and 
is significantly larger than  that of SIM2 at low $\ell$. 
For this set, two additional moment functions are detected, namely  $\mathcal{M}_\ell^{\momi1 \momi2}$ and more marginally $\mathcal{M}_\ell^{\AD \momi2}$. 
Comparing moment functions for the SIM3 and SIM4 sets, we find that the CIB has a significant impact on several moments at $\ell\gtrsim100$, but not on $\mathcal{M}_\ell^{\momi1 \momi1}$. Nearly all of the other moment functions increase with $\ell$ and are close to zero in the lowest $\ell$-bins. 
 
For the PR3 data, all the moment functions (from first- to third-order) are detected with an absolute amplitude larger than that measured on the simulated data sets. The $\mathcal{M}_\ell^{\momi1 \momi1}$ moment function shows a similar amplitude as for SIM3 and SIM4 for $\ell\gtrsim100$, while at larger angular scales it does not. For the $\mathcal{M}_\ell^{\AD \momi2}$ and $\mathcal{M}_\ell^{\momi2 \momi2}$, the PR3 data set has an overall $\ell$ behavior close to that of SIM4 but with an increased absolute value. The $\mathcal{M}_\ell^{\momi1 \momi2}$, $\mathcal{M}_\ell^{\momi1 \momi3}$, $\mathcal{M}_\ell^{\momi2 \momi3}$ and $\mathcal{M}_\ell^{\momi3 \momi3}$ of PR3  match those of SIM4 for $\ell\lesssim70$ but progressively deviate from them for higher multipoles. Finally, we point out that the $\mathcal{M}_\ell^{\AD \momi3}$ moment function is the
one with highest absolute amplitude for the PR3 set. Surprisingly, it has the opposite sign, and a different $\ell$-dependence from the SIM4 moment.

\subsection{Discussion}
\label{sec:moments_discussion}

Here, we summarize the main results of the fits before briefly discussing our interpretation.

\begin{itemize}

\item The goodness of the fit obtained for the simulations demonstrates the ability of the moment expansion to account for spatial variations of the dust SED, even when the MBB law provides a very poor fit. Except for the simplest SIM1 simulations, with constant temperature and spectral index, high-order moments are significantly detected for the other simulations and the PR3 data. 

\item When spatial variations of the dust SED are present, the spectral index inferred from the MBB fit is biased. Fitting high-order moments, we obtain a significantly different value that cancels the first-order moment $A\momi1$

\item The comparison of the moments obtained for the SIM3 and SIM4 simulations, which only differ by the addition of the CIB, shows that the CIB has a significant impact on moments at $\ell\gtrsim 100$. To account for this additional emission component, we need to extend the moment expansion to third order.

\item The moments on the SIM4 simulations that include realistic spatial SED variations and the CIB are quantitatively different from those measured on the PR3 data.

\end{itemize}

The moment functions in \fig{Fig:all_moments_LR42_143x545} are difficult to interpret at $\ell>100$ due to the CIB contribution, but for lower multipoles one may relate the results to dust emission properties. 
For the PR3 data at $\ell < 100$, the three most significant moment functions in decreasing order are $\mathcal{M}_\ell^{\AD \momi3}$, $\mathcal{M}_\ell^{\AD \momi2}$ and 
$\mathcal{M}_\ell^{\momi1\momi1}$. 
We point out that the simulations do not match any of these three moment functions. The most immediate interpretation of this mismatch is the lack of variations of the dust MBB parameters along the line of sight in the simulations, but this may not be the sole explanation. The dust emission could also comprise two or more emission components, which are not
fully correlated on the sky \citep{Draine13,Guillet18,Hensley2018}. 

It is not straightforward to provide a specific interpretation of the amplitude, the scale dependence, or the hierarchy of the moments fits. One difficulty lies in the fact that the moments expansion does not decompose the data into independent components: the high-order moments depend on the expansion order as illustrated in \fig{fig:multi_order2} to \fig{fig:multi_order5} of \app{sec:multi_order}. Furthermore, 
the moments also quantify the SED averaging that occurs when going from pixel space to harmonic space. Without a model of the
sky emission, it is therefore difficult  to link a given high-order moment to a physical property of the dust emission. An iteration on dust and CIB simulations converging towards a moment decomposition in agreement with that of the 
\Planck{} data would be needed to quantify possible interpretations. We foresee that the moment decomposition could be used as a quantitative metric to obtain simulations that better match the data, but this is beyond the scope of the present work. 

\section{Implications for the tensor-to-scalar ratio measurement }\label{sec:impact_on_r}

We finally discuss the potential impact of our results on the measurement of the tensor-to-scalar ratio $r$. We performed the moment expansion of the dust SED from the \Planck{} intensity power spectra. There is no reason for the departures from the MBB SED that we observed and quantified to be absent in polarization. Here, we indeed assume that the SED of the dust $B$-modes  shows spectral departures from the mean MBB SED of the same relative order as the ones we observed in intensity and we derive the potential bias on $r$ that would result from neglecting them.

In order to be conservative in this determination, we use the SIM3 simulated data set: we know from our $\chi^2$ analysis of \sect{sec:chi2} that the actual dust intensity in the \Planck{} PR3 data contains {at least} the same level of departure from the MBB that is present in this simulated data set. We could have used SIM4 or PR3 data sets, but as we see in \sect{sec:results}, it is not trivial to decipher which part of the SED distortion is due to dust and CIB; the latter component being expected to contribute much less than the dust to the $B$-modes.

We consider the SIM3 cross-spectra fit with \eq{eq:dust_moment_cross_3} at different orders in the moment expansion. We look here at the relative difference between the SIM3 data set cross-spectra and the fitted model of \eq{eq:dust_moment_cross_3}, for every cross-spectra between frequencies $\nu_i$ and $\nu_j$. We focus on the multipole bin centered at $\ell_0=80$, as this scale corresponds to the CMB primordial $B$-modes peak. This relative difference reads:

\be
\label{eq:residuals_ell80}
\Delta \mathcal{D}_{\ell_0}\left({\nu_i}\times {\nu_j}\right)\equiv\frac{\mathcal{D}_{\ell_0}^{\rm SIM3}\left({\nu_i}\times{\nu_j}\right)-\mathcal{D}_{\ell_0}^{\rm fit}\left({\nu_i}\times{\nu_j}\right)}{\mathcal{D}_{\ell_0}^{\rm SIM3}\left({\nu_i}\times{\nu_j}\right)}.
\ee

\begin{figure}%
\centering
\subfigure{
\includegraphics[width=\columnwidth]{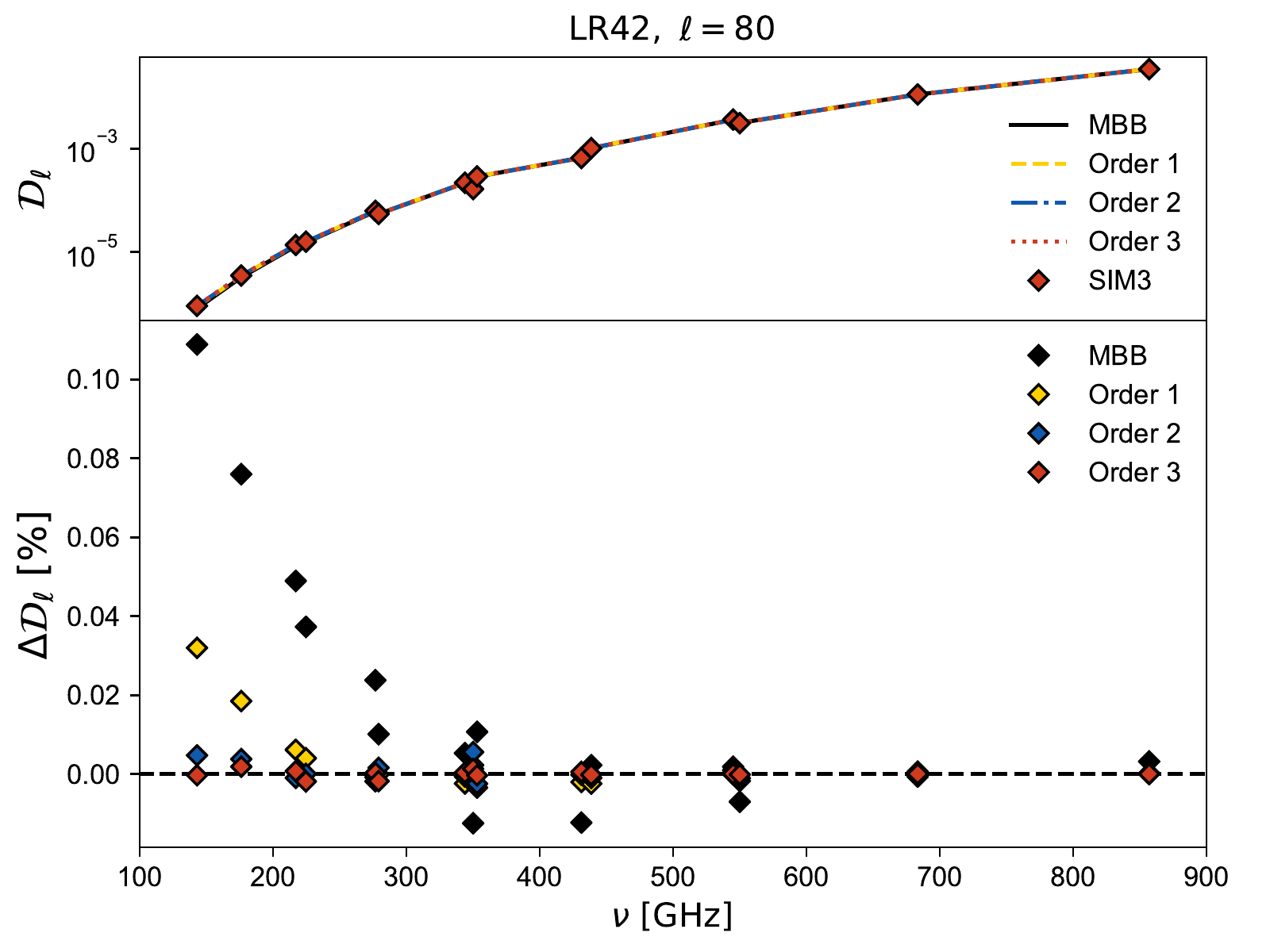}}
\caption{\emph{Upper panel}. Mean SED in MJy$^2\cdot$sr$^{-1}$ for the 100 SIM3 cross-spectra as a function of the effective frequency ($\sqrt{\nu_i\cdot\nu_j}$, in GHz) for the multipole bin centered at $\ell=80$. The MBB (solid black), first- (dashed yellow), second- (dashed-dotted blue), and third-order (dotted red) best fits are displayed and not distinguishable. \emph{Bottom panel}. Relative percentage difference between the SIM3 SED and the MBB (black diamonds) and the first- (yellow diamonds), second- (blue diamonds) and third-order (red diamonds) best fits.
}
\label{fig:dl_residuals_ell80}
\end{figure}

This relative difference is displayed in \fig{fig:dl_residuals_ell80}. 
Focusing on the $143\times143$ cross-spectrum, which is a frequency channel indicative of typical CMB $B$-mode experiments, the simple MBB fit leaves a $\Delta \mathcal{D}_{\ell_0}(143\times143)= 10.9\,\%$ residual, the first-order fit leaves $\Delta \mathcal{D}_{\ell_0}(143\times143)=3.2\,\%$, the second-order fit leaves $\Delta \mathcal{D}_{\ell_0}(143\times143)=0.5\,\%$ and the third-order fit leaves $\Delta \mathcal{D}_{\ell_0}(143\times143)=0.06\,\%$. 

Let us now consider a future CMB $B$-mode experiment that has the same frequency channels and a S/N for the dust $B$-modes that is similar to those of \PlanckHFI{} for the dust intensity \citep[e.g., LiteBIRD,][]{litebird}. We know from \citet{2016A&A...586A.133P} that the dust $B$-mode power spectrum computed on the LR42 mask has an amplitude of $\AD(\ell_0)=78.6\,\mu{\rm K}^2$ at 353\,GHz. If we convert this amplitude into the $r$-equivalent amplitude at 150\,GHz $r_{\rm D}$ \citep{2016A&A...586A.133P}, it corresponds to $r_{\rm D}=1.8$. Our results therefore suggest that a CMB $B$-mode experiment looking at half the sky could find an analysis bias of $\Delta r=0.109\cdot r_{\rm D}=0.20$ by assuming that the dust $B$-modes follow a MBB SED, $\Delta r=0.032\cdot r_{\rm D}=0.06$ by assuming an first-order moment expansion, $\Delta r=0.005\cdot r_{\rm D}=0.009$ by assuming a second-order moment expansion, and $\Delta r=0.0006\cdot r_{\rm D}=0.001$ by assuming a third-order moment expansion (as we see in the above analysis, parameterizing a dust decorrelation amounts to fitting the first order, if the parametrization is correct). 

The region of the sky observed by the BICEP2/Keck experiment has $r_{\rm D}=0.11$ \citep{bicep2018}. If we transpose our results to this region, we see that a MBB fit of the dust $B$-modes would lead to $\Delta r=0.109\cdot r_{\rm D}=0.01$, a decorrelation or first-order analysis would lead to $\Delta r=0.032\cdot r_{\rm D}=4\times10^{-3}$, a second-order analysis to $\Delta r=0.005\cdot r_{\rm D}=5\times10^{-4}$, and a third-order analysis to $\Delta r=0.0006\cdot r_{\rm D}=7\times10^{-5}$.

\begin{table}
\begin{center}
\begin{tabular}{ccc}
 & LR42 & BICEP2/Keck\\
 & ($f_{\rm sky}=0.42$)& ($f_{\rm sky}=0.01$)\\
\hline
MBB                   & 0.2&  0.01 \\
1$^{\rm st}$ order    & 0.06&  $4\times10^{-3}$ \\
2$^{\rm nd}$ order    & $9\times10^{-3}$& $5\times10^{-4}$ \\
3$^{\rm rd}$ order    & $1\times10^{-3}$& $7\times10^{-5}$ \\
\end{tabular}
\vspace{0.1cm}\caption{Estimates of the tensor-to-scalar ratio bias $\Delta r$ at 150\,GHz for the LR42 and the BICEP2/Keck regions, in the case of the SIM3 cross-spectra SED fitted assuming a MBB and a first-, second-, and or third-order SED moment expansion.\label{tab:rd_table}}
\end{center}
\end{table}

The $\Delta r$ values in the different cases are summarized in Table~\ref{tab:rd_table}. Although these $\Delta r$ values are rough estimates (that might be overestimated in the BICEP2/Keck case because fewer SED spatial variations could occur on this small region), they provide an insight into the order of magnitude of the potential bias. The values seen are in strong support of the need to take into account the spectral departures from the dust MBB in future CMB $B$-mode analyses, targeting $r$ values down to $10^{-3}$ and beyond. 

These conclusions are further supported by the moment decomposition of the \Planck{} PR3 data set at $\ell<100$, where the CIB contribution is likely to be negligible. As the moment functions $\mathcal{M}_\ell^{ab}$ measure a fractional departure from a simple MBB law, we can see that in order to reach an accuracy in the dust subtraction of 10$^{-3}$, we need to consider all moments with an absolute amplitude larger than 10$^{-3}$. As can be seen in \fig{Fig:all_moments_LR42_143x545}, most of the moments in the PR3 data set decomposition up to third order have an absolute amplitude that is greater than this threshold. In that sense, dust-dominated angular scales of the intensity PR3 data set additionally stress the need for a third-order expansion of the polarized dust SED in order to reach the accuracy targeted by future CMB $B$-mode experiments.


\section{Summary and conclusions}\label{sec:summary_and_conclusions}
In this paper, we present a model that describes the Galactic dust SED for total intensity at \PlanckHFI{} frequencies in terms of SED distortions with respect to the MBB
emission law. This model accounts for variations in the dust SED on the sky and along the line of sight in order to provide an astrophysically motivated description at the power spectrum level.  
The model formalism relies on expansion of the dust emission SED in moments around the MBB law related to derivatives with respect to the dust spectral index. These high-order moments lead to frequency decorrelation; departures from the MBB inevitably appear because of averaging effects along the line of sight and within the beam, and, as is most relevant here, because of the spherical harmonic expansion performed in the data processing.

We applied our analysis to total intensity cross-spectra computed from the combination of CMB-corrected PR3 \Planck{} data at the five HFI channels at 143, 217, 353, 545, and 857 GHz, and to four sets of foreground simulations of increasing complexity.
The main conclusions of our analysis can be summarized as follows. 

Our analysis quantifies the spectral complexity of the \textit{Planck} total intensity data at frequencies larger than $143\,$GHz. At $\ell \gtrsim 100$, the CIB is a significant component contributing to the complexity. At lower multipoles, the data are dust dominated but the dust simulations based on MBB parameters fitted on \Planck{} maps fail to  match the most significant moments. In future work, the moment decomposition could be used 
to obtain improved simulations of dust and CIB emission, which better match \Planck{} data and may
be used to test possible interpretations.

We extend our results to $B$-mode analyses within a simplified framework. We find that neglecting the dust SED distortions of the dust polarization with respect to the MBB, or trying to model them with a single ad-hoc parameter, could lead to biases larger than the accuracy of the component separation required to search primordial B-modes down to a tensor-to-scalar ratio $r=10^{-3}$.

If our results extend to polarized emission without any additional complexity, we anticipate that moment expansion up to third order would be required to model the dust polarization SED  to the accuracy of future CMB $B$-mode experiments. If this is a valid statement, it sets constraints on the number of frequency bands required to separate dust and CMB polarization. At least four and five dust-dedicated frequency channels are needed in order to perform second- and third-order moment expansion fits, respectively.

Additional difficulties for $B$-mode searches could arise from changes in polarization angles across frequencies, which would make the decomposition of polarized dust emission in $E$ and $B$-modes frequency-dependent. Further complexity may arise because of variation in dust temperature, which we did not include here. Similarly, synchrotron foregrounds at low frequencies will require an independent moment expansion.
Based on these findings, we conclude that the moment expansion of the dust SED is a new promising tool to model the dust component at the level of precision needed for the measurement of the CMB primordial $B$-modes. 
This paper presents a first step in this direction, providing the formalism and the first qualitative results based on the \Planck{} total intensity data and simulations. When dealing with the polarization, other details should be carefully considered and added, such as for instance the fact that the magnetic field direction will project variations of the SED differently in $Q$ and $U$, which implies that, generally, two independent moment expansions are needed. Studying the moment expansion method specifically for polarization will be another important step and the focus of a forthcoming publication.

It will also be important to study applications of the power spectrum moment expansion for extractions of primordial CMB spectral distortions. The expected signals are small \citep[e.g.,][]{Chluba2016} and heavily obscured by foregrounds. Most extraction methods mainly use information based on the SED shapes and neglect spatial information \citep{Mayuri2015, Vince2015, Abitbol2017}, but a more  recent study explores the benefits of using spatial information \citep{rotti2020}. Spatial information could be further exploited using the techniques described here, which warrants further investigation.


\begin{acknowledgements} 
This research was supported by the Agence Nationale de la Recherche (project BxB: ANR-17-CE31-0022).
AM acknowledges the support of the DIA-ASTRO fellowship from the French space agency (Centre National d'Etudes Spatiales, CNES).
AR is supported by the ERC Consolidator Grant {\it CMBSPEC} (No.~725456) as part of the European Union's Horizon 2020 research and innovation program.
JC was supported by the Royal Society as a Royal Society University Research Fellow at the University of Manchester, UK.
 \end{acknowledgements}

\bibliographystyle{aa}

\bibliography{Moments}

\begin{appendix}

\section{Mask and foreground templates}
\label{sec:templates}
In this section we provide details on the mask and on the foreground templates that we used to generate the simulations.
For both the data and the simulations, we used a mask that combines a 50\% apodized galactic cut and a point source mask. This mask, referred to as LR42 and defined in \cite{2016A&A...586A.133P},  has been extensively used in the \Planck{} analyses, has $f_{sky}=0.42,$ and is shown in \fig{Fig:maskLR42}.

As described in \sect{sec:data_and_sims}, we generated different sets of dust and multi-component simulations. We provide an illustrative example of the foreground templates at 353\,GHz in ${\rm MJy\, sr^{-1}}$ units in \fig{Fig:templates}. From top to bottom, the figure shows the dust template defined in \eq{eq:dust_353}, the CIB template, and the synchrotron template.

\begin{figure}%
\centering
\subfigure{%
\includegraphics[width=\columnwidth, height=6cm]{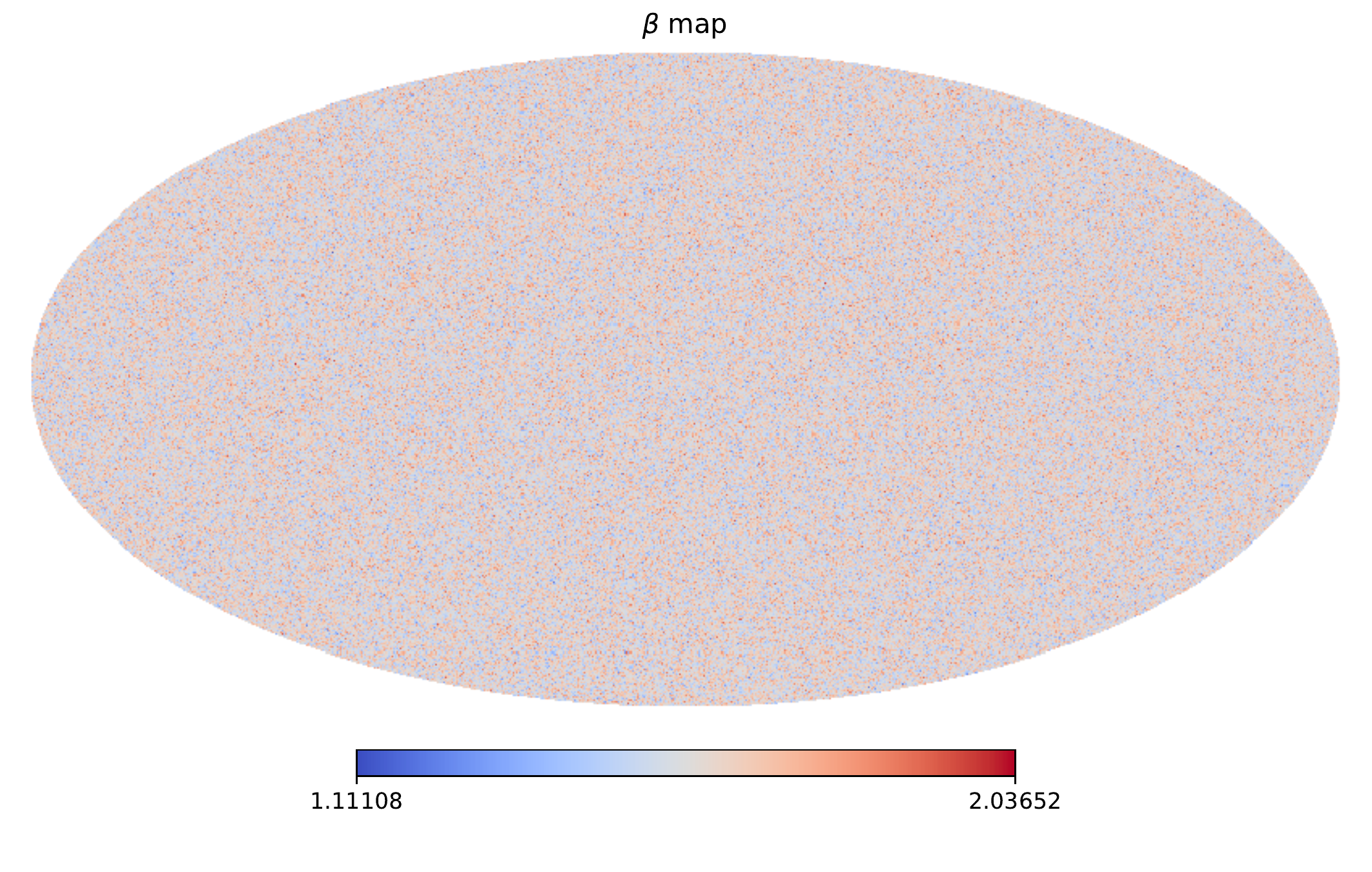}}
\subfigure{%
\includegraphics[width=\columnwidth, height=6cm]{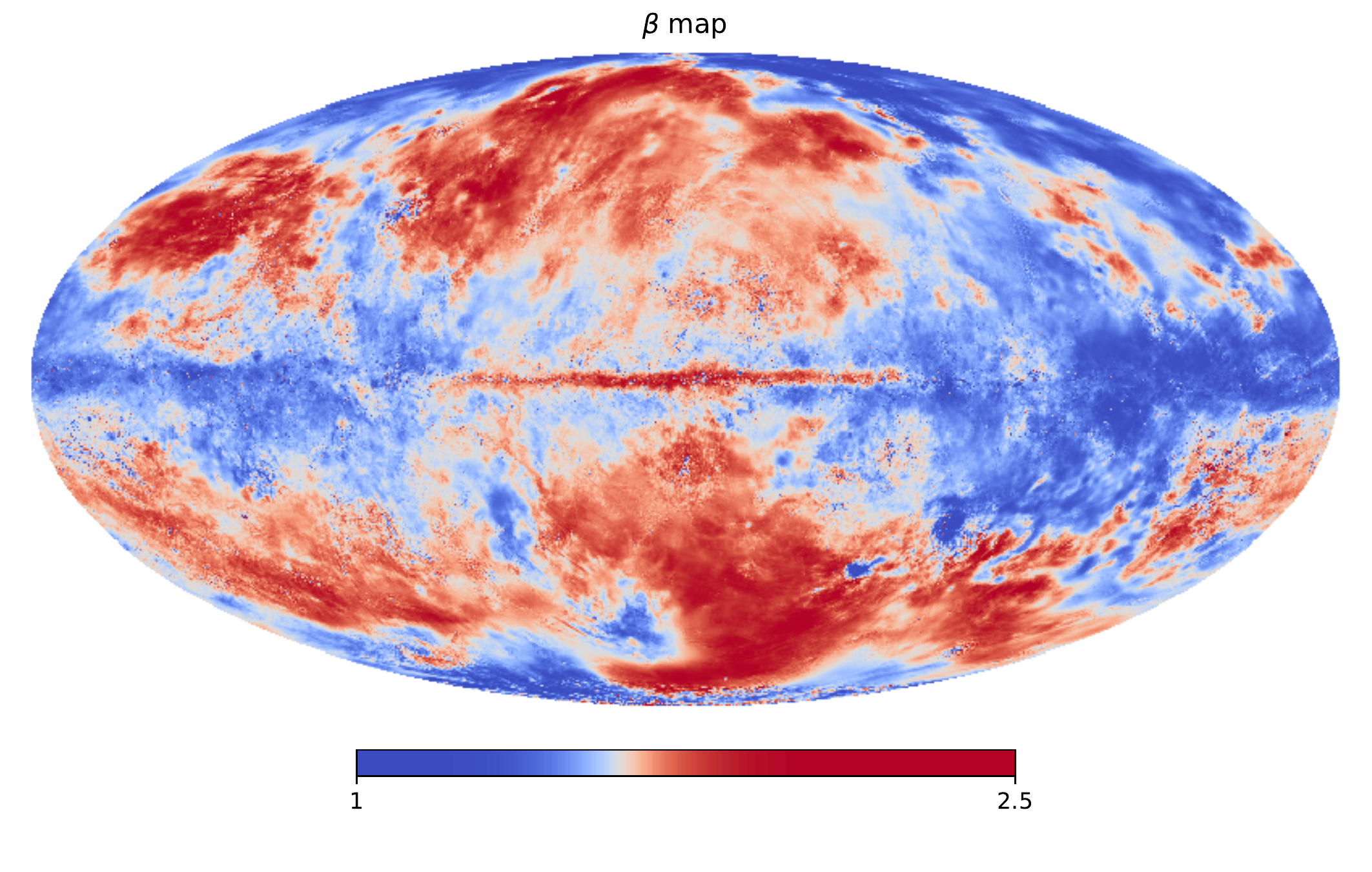}}
\caption{Dust spectral index map $\beta({\bf \hat n})$ used for the dust simulations SIM2 with Gaussian $\beta$ variations with $\Delta \beta=0.1$ (top panel) and for the dust simulations SIM3 with $\beta$ variations estimated from the data with the GNILC component separation method (bottom panel). }
\label{Fig:beta_maps}
\end{figure}

Given that investigating the impact of the spatial variations of the dust spectral index is a key part of our analysis, we show in \fig{Fig:beta_maps} the $\beta({\bf \hat{n}})$ maps used in the simulations. The top panel of \fig{Fig:beta_maps} shows the $\beta({\bf \hat{n}})$ map used to generate the SIM2 dust simulations with Gaussian $\beta({\bf \hat{n}})$ variations with $\Delta \beta=0.1$ around $\beta_0=1.59$. The bottom panel of \fig{Fig:beta_maps} shows the GNILC $\beta({\bf \hat{n}})$ map used to generate SIM3 simulations \citep{2016A&A...596A.109P}.

\section{Cross-spectra definition, covariance, and correlation matrices}
\label{sec:correlations}

Here we provide more details on how we compute the angular power spectra of the data sets we consider in the paper, as introduced in \sect{sec:crosscomputation}, and on how this impacts the cross-power spectra correlations and their statistical independence.

\subsection{Definitions}

In order to avoid the noise auto-correlation bias and to reduce the level of correlated systematic errors, we compute the cross-spectra from data split maps (the 2 \Planck{} half-mission maps, HM, namely HM1 and HM2). The general philosophy would be to reproduce \citet{2016A&A...586A.133P} to construct the cross-power spectra from two frequency maps $M_{\nu_i}$ and $M_{\nu_j}$:

\begin{figure}%
\centering
\subfigure{%
\includegraphics[width=\columnwidth, height=6cm]{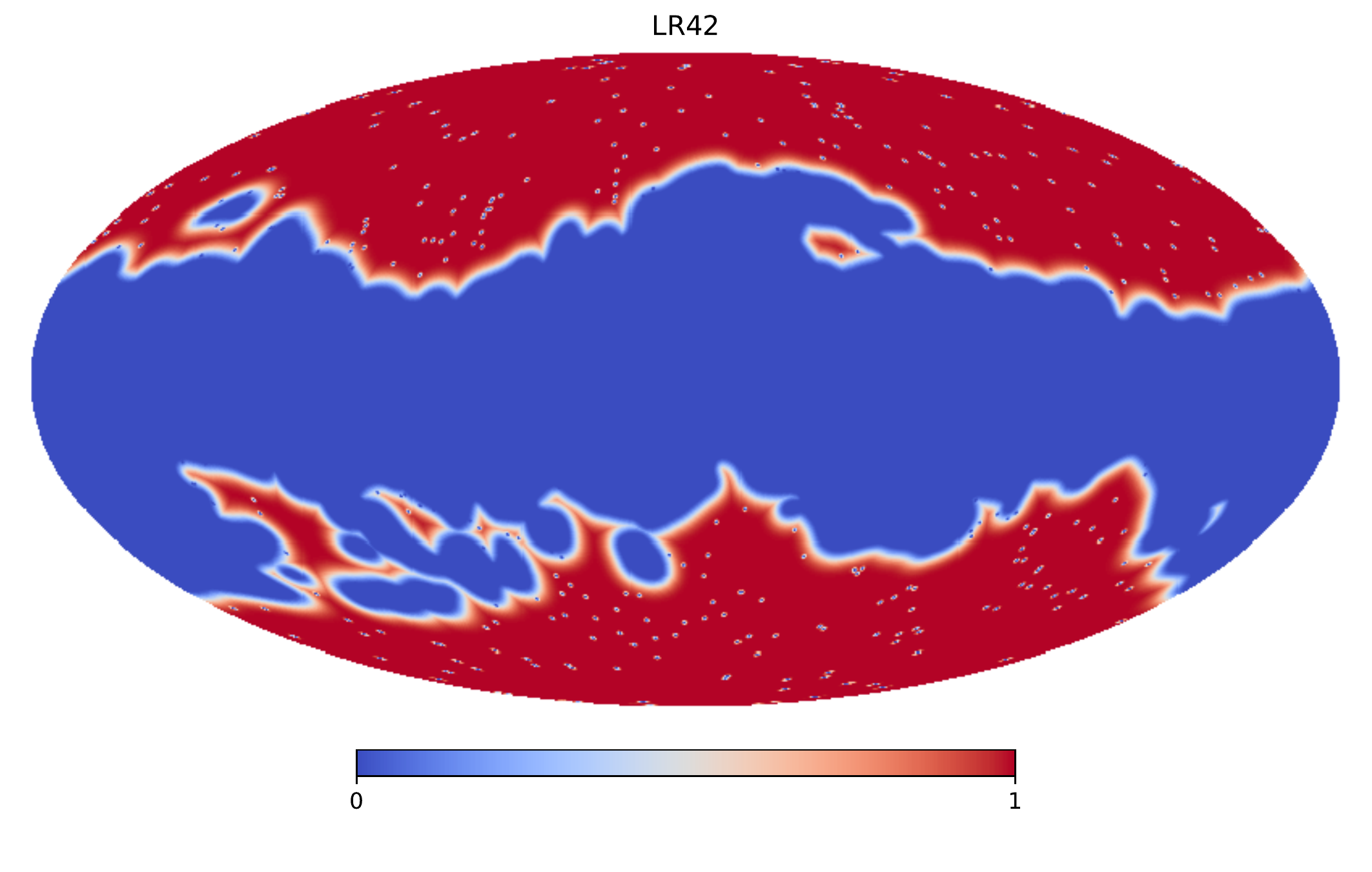}}
\caption{LR42 mask used in the analysis.}
\label{Fig:maskLR42}
\end{figure}

\begin{figure}%
\centering
\subfigure{%
\includegraphics[width=\columnwidth, height=6cm]{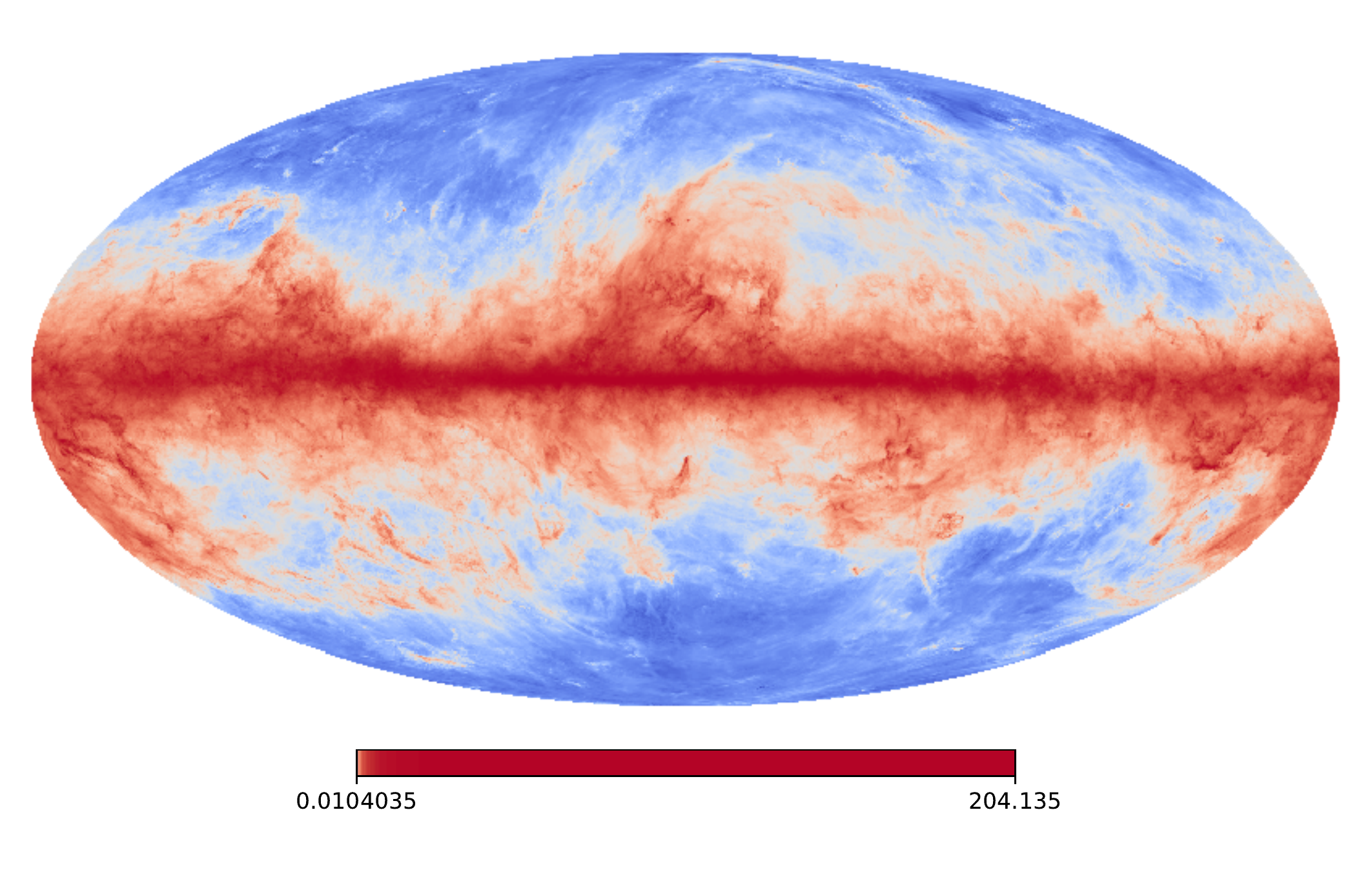}}
\subfigure{%
\includegraphics[width=\columnwidth, height=6cm]{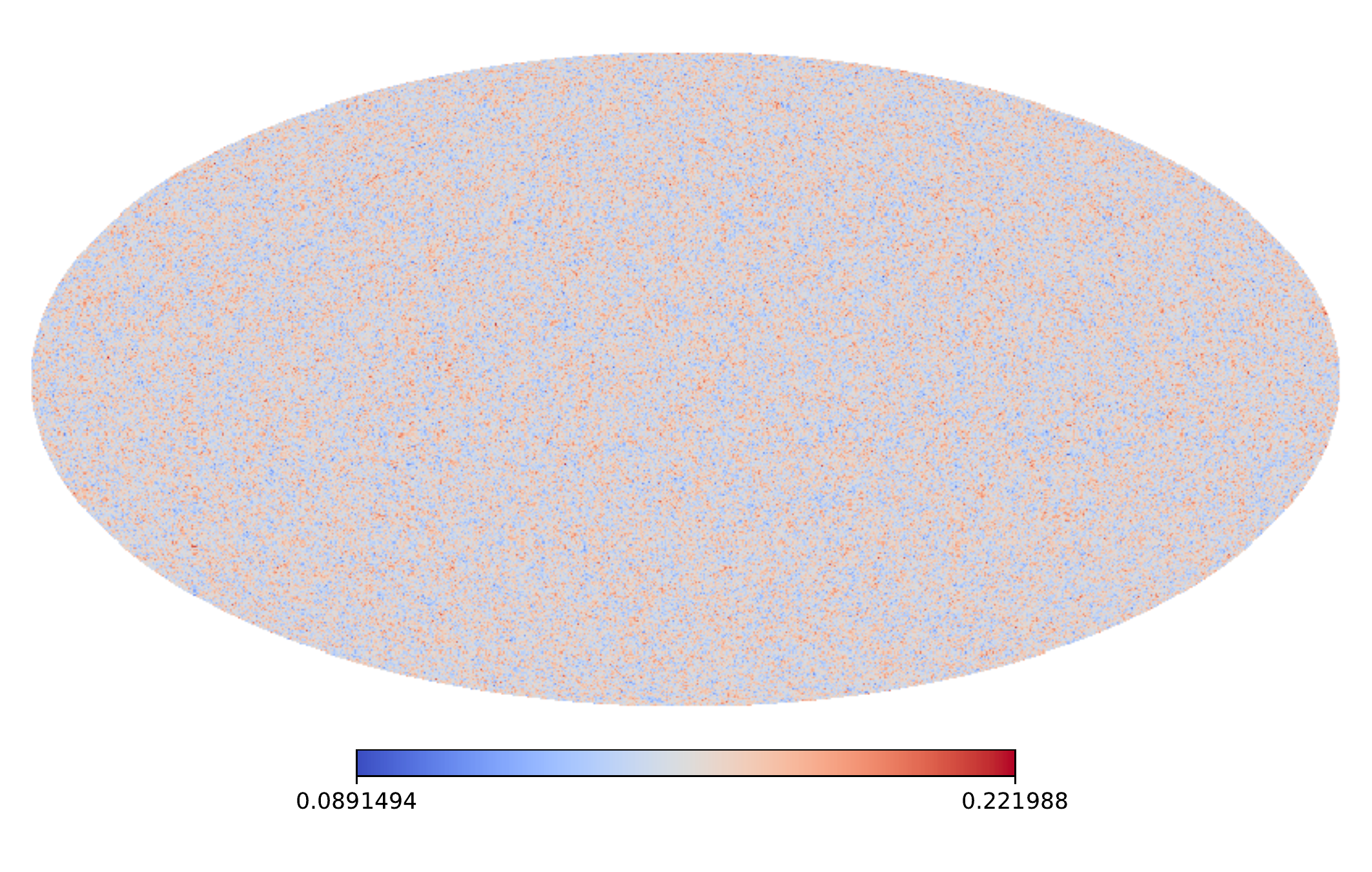}}
\subfigure{%
\includegraphics[width=\columnwidth, height=6cm]{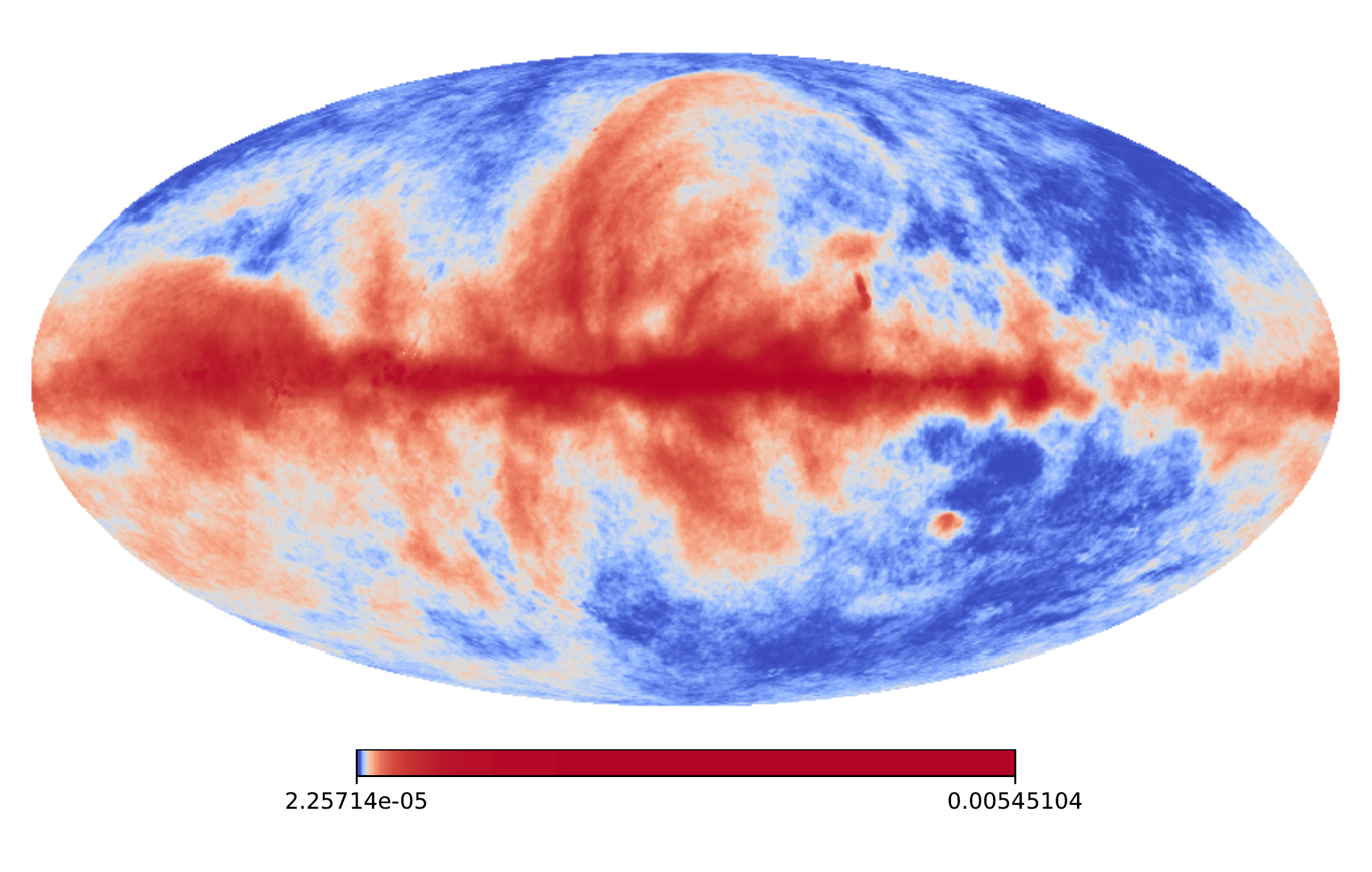}}
\caption{Foreground templates at 353 GHz in ${\rm MJy\, sr^{-1}}$ units. {\it Top panel:} Dust template defined in \eq{eq:dust_353}. {\it Middle panel:} CIB template. {\it Bottom panel:} Synchrotron template.}
\label{Fig:templates}
\end{figure}

\begin{equation}
\label{eq:def_cross_spectra}
\left\{
\begin{array}{ll}
\left.\mathcal{D}_\ell\left(M_{\nu_i}\times M_{\nu_j}\right)\right|_{i=j}=&\mathcal{D}_\ell\left(M_{\nu_i}^{\rm HM1}\times M_{\nu_i}^{\rm HM2}\right) 
\\  
\left.\mathcal{D}_\ell\left(M_{\nu_i}\times M_{\nu_i}\right)\right|_{i\neq j}=&\frac{1}{4}\left[\mathcal{D}_\ell\left(M_{\nu_i}^{\rm HM1}\times M_{\nu_i}^{\rm HM1}\right)\right.\\
&+\mathcal{D}_\ell\left(M_{\nu_i}^{\rm HM1}\times M_{\nu_i}^{\rm HM2}\right)\\
&+\mathcal{D}_\ell\left(M_{\nu_i}^{\rm HM2}\times M_{\nu_i}^{\rm HM1}\right)\\
&+\left.\mathcal{D}_\ell\left(M_{\nu_i}^{\rm HM2}\times M_{\nu_i}^{\rm HM2}\right)\right]
\end{array}
\right.
\end{equation}

The idea of doing the sum of the four cross-spectra in \eq{eq:def_cross_spectra} is to increase the S/N of $\left.D_\ell\left(M_{\nu_i}\times M_{\nu_i}\right)\right|_{i\neq j}$, presumably at the cost of the statistical independence between distinct cross-spectra, as we see in the following. To preserve the statistical independence between the cross-spectra, we do not adopt the definition of \eq{eq:def_cross_spectra} but that of \eq{eq:def_cross_spectra2}, in \app{sec:minimized_correlations}, for the reasons that are presented in \app{sec:toy_model}.

The covariance matrix $\mathbb C$ from \eq{eq:covariance}, which is used for the fits we perform throughout the paper, is computed from 100 pairs of half-mission maps of a given data set (e.g., SIM1 simulations are used to compute $\mathbb C$ when dealing with SIM1). As they are the closest to the data in terms of physical components and component complexity, the covariance matrix is inferred from the SIM4 simulations when dealing with the actual \Planck{} data set PR3. 

In order to assess the statistical independence between the cross-spectra of our data sets, we build the correlation matrix from the covariance matrix of \eq{eq:covariance}:

\begin{align}
\label{eq:correlation_matrix}
\mathbb{R}&\equiv\mathcal{R}_{ijkl}\left(\ell\right)\equiv{\rm corr}_{ijkl}\left(\ell\right)
\equiv\frac{\mathcal{C}_{ijkl}\left(\ell\right)}{\sqrt{\mathcal{C}_{ijij}\left(\ell\right)\mathcal{C}_{klkl}\left(\ell\right)}}\nonumber\\
&=\frac{\mathcal{C}_{ijkl}\left(\ell\right)}{\sqrt{{\rm var}_{ij}\left(\ell\right){\rm var}_{kl}\left(\ell\right)}}=\frac{\mathcal{C}_{ijkl}\left(\ell\right)}{\sigma_{ij}\left(\ell\right)\sigma_{kl}\left(\ell\right)}.
\end{align}

This correlation matrix is displayed for the SIM1 data set in \fig{Fig:corr_matrix_sims_donly}, for the multipole bin centered at $\ell=100$ (the shape of $\mathbb{R}$ is qualitatively the same in each multipole bin). It is significantly nondiagonal, showing large correlations between numerous cross-spectra. This highlights that the cross-spectra, as they are defined in \ref{eq:def_cross_spectra}, are not all independent. In the following section we see why this happens. Below, we change the definition of the cross-spectra in \ref{eq:def_cross_spectra} to minimize these correlations.

\subsection{Expected correlations from a toy model}
\label{sec:toy_model}

In a high-S/N mixture of interstellar dust and instrumental noise, the correlation between frequency channels cross-spectra can become significant and needs to be taken into account in minimization processes.

Let us suppose that a map of the sky $M_i\equiv M_{\nu_i}$ observed at a frequency channel $\nu_i$ can be written as the sum of a dust component $D$ and an instrumental noise term $N$ so that $M_i\simeq D_i+N_i$. The cross angular power spectra then read:

\begin{equation}
\label{eq:def_b1}
M_i\times M_j = D_i\times D_j + D_i\times N_j + D_j\times N_i + N_i\times N_j
.\end{equation} 

As the dust templates are the same in every simulation, the $D\times D$ term does not contribute to the covariance (nor to the variance) and in the high-S/N regime the $N\times N$ term can be neglected with respect to the $D\times N$ terms. Thus, the correlation coefficient between the power spectra computed from these maps, as defined in \ref{eq:correlation_matrix}, reads\footnote{We drop the $\ell$ dependence, as the reasoning below does not depend on the multipole.}:

\begin{align}
\label{eq:def_b2}
&\mathcal{R}_{ijkl}\simeq\nonumber\\
&\frac{{\rm cov}\left(D_i\times N_j+D_j\times N_i,D_k\times N_l+D_l\times N_k\right)}{\sqrt{{\rm var}\left(D_i\times N_j+D_j\times N_i\right){\rm var}\left(D_k\times N_l+D_l\times N_k\right)}}. 
\end{align}

Let us now consider as an example the specific correlation between 2 \Planck{} cross-spectra, namely $143\times217$ and $143\times353$ ($\nu_i=143$, $\nu_j=217$, $\nu_k=143$ and $\nu_l=353$\, GHz). As ${\rm cov}(X+Y,Z)={\rm cov}(X,Z)+{\rm cov}(Y,Z)$, the most significant among the developed terms in the numerator of \eq{eq:def_b2} is ${\rm cov}(D_{217}\times N_{143},D_{353}\times N_{143})$ as it is the only one to involve twice the same noise map and hence would be the most "covariant". If we make the assumption that the dust component is spatially the same at each frequency (i.e., the same dust spatial template $D$), scaling as $D_i=A_i\cdot D$ and that the noise basically scales as $N_i=B_i\cdot N$ (where $A_i$ and $B_i$ are scalars), we find:

\begin{align}
\label{eq:def_b3}
&\mathcal{R}_{143\times217,143\times353}\simeq\nonumber\\
&\frac{A_{217}A_{353}B_{143}^2}{\sqrt{\left(A_{217}^2B_{143}^2+A_{143}^2B_{217}^2\right)\cdot\left(A_{353}^2B_{143}^2+A_{143}^2B_{353}^2\right)}}.
\end{align}

For a typical dust MBB SED with $\beta_0=1.59$ and $T_0=19.6$\,K and the \PlanckHFI{} noise levels, we find that $\mathcal{R}_{143\times217,143\times353}=0.90$, which is a significant correlation. We note that it would be markedly decreased if the noise was independent between the 2 143\,GHz maps used in the two different cross-spectra. This can happen using two different data splits (as the \Planck{} HM maps) as we see in the following.

\begin{figure}[t!]%
\includegraphics[width=\columnwidth]{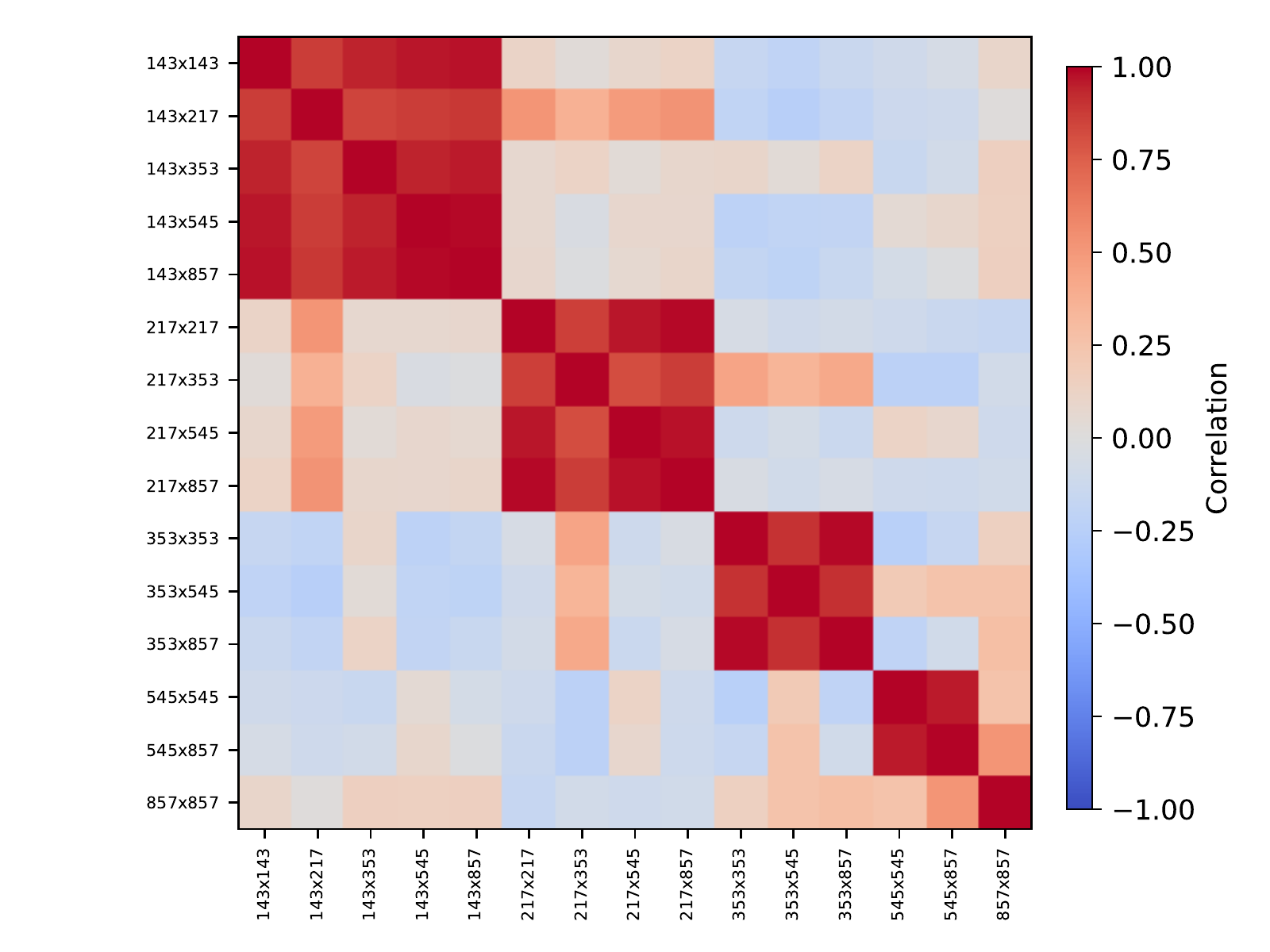}
\centering
\includegraphics[width=\columnwidth]{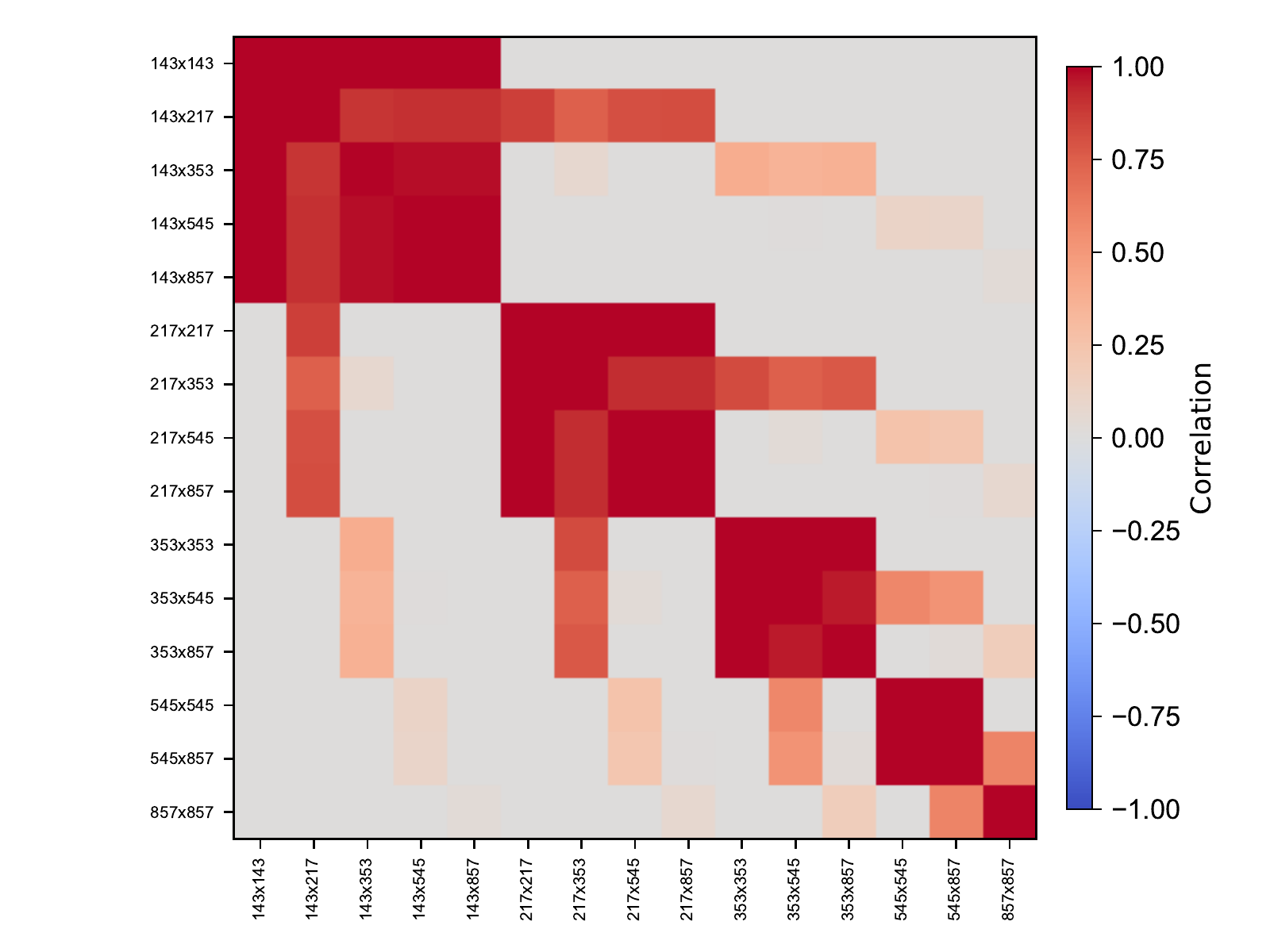}
\includegraphics[width=\columnwidth]{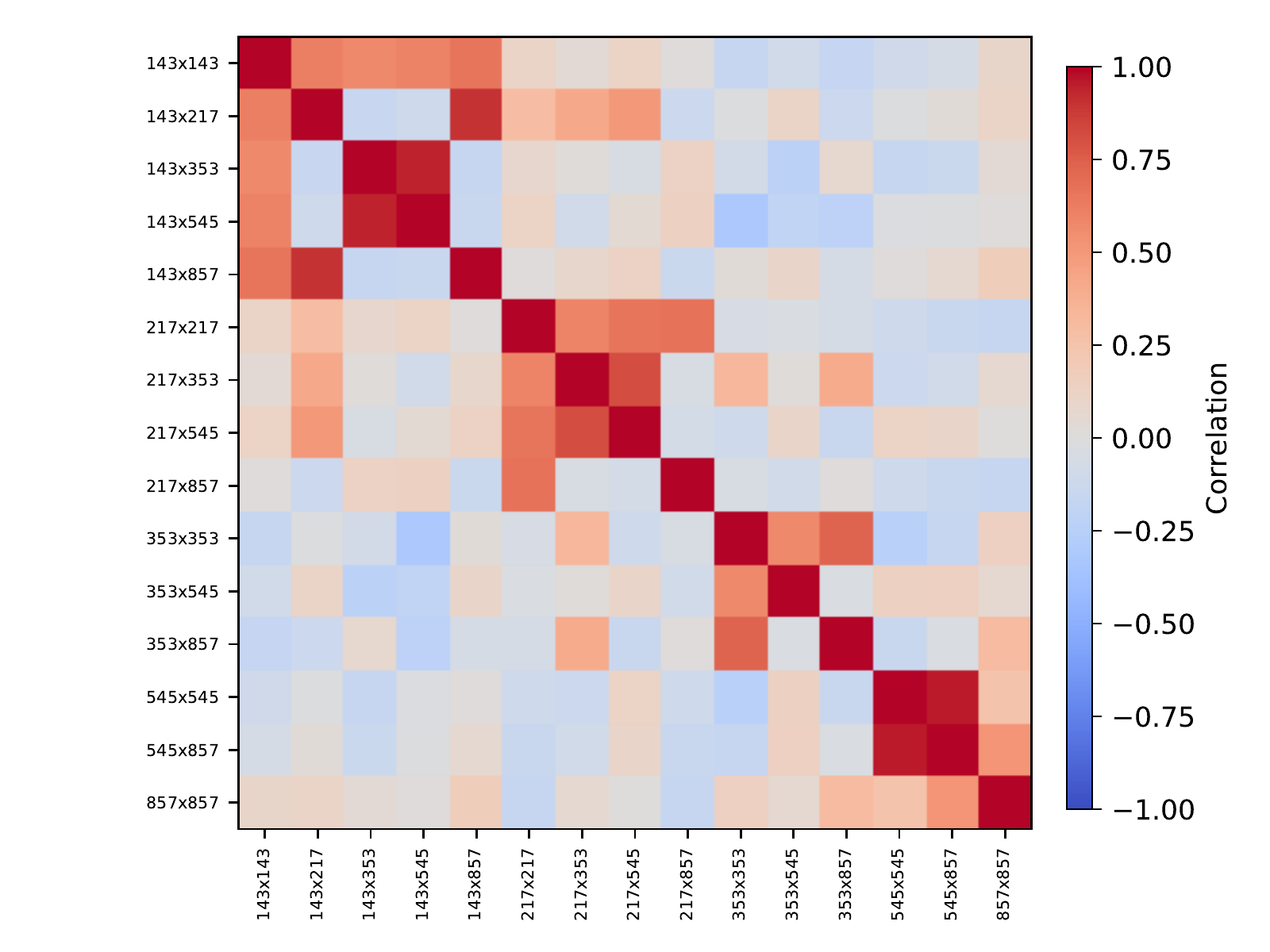}
\caption{Cross-spectra correlation matrices computed from \eq{eq:correlation_matrix} for the SIM1 data set in the multipole bin centered at $\ell=100$. The upper panel shows the correlation matrix computed from the cross-spectra as defined in \eq{eq:def_cross_spectra}. The middle panel shows the correlation matrix computed for our toy-model of the correlation presented in \app{sec:toy_model}. The \textbf{bottom} panel shows the correlation matrix computed from the cross-spectra as defined in \eq{eq:def_cross_spectra2}. }
\label{Fig:corr_matrix_sims_donly}
\end{figure}

By applying a reasoning equivalent to that of the example above to all the cross-spectra covariances in the correlation matrix $\mathbb{R}$, we can build the full toy-model correlation matrix for the data sets we use in this paper. It is displayed in the middle panel of \fig{Fig:corr_matrix_sims_donly}. We can see that despite the simplicity of our assumptions, this toy-model correlation matrix qualitatively reproduces that of our data sets (top panel of the same figure).

\subsection{Minimizing the correlations and effective degrees of freedom}
\label{sec:minimized_correlations}

In order to minimize the correlations between the cross-spectra of our data sets, we change the definition of \eq{eq:def_cross_spectra} in the case $i\neq j$:

\begin{equation}
\label{eq:def_cross_spectra2}
\left\{
\begin{array}{l}
\left.\mathcal{D}_\ell\left(M_{\nu_i}\times M_{\nu_j}\right)\right|_{i=j}=\mathcal{D}_\ell\left(M_{\nu_i}^{\rm HM1}\times M_{\nu_i}^{\rm HM2}\right) 
\\  
\mathcal{D}_\ell\left(M_{143}\times M_{217}\right)\equiv \mathcal{D}_\ell\left(M_{143}^{\rm HM2}\times M_{217}^{\rm HM1}\right)\\
\mathcal{D}_\ell\left(M_{143}\times M_{353}\right)\equiv \mathcal{D}_\ell\left(M_{143}^{\rm HM1}\times M_{353}^{\rm HM2}\right)\\
\mathcal{D}_\ell\left(M_{143}\times M_{545}\right)\equiv \mathcal{D}_\ell\left(M_{143}^{\rm HM1}\times M_{545}^{\rm HM1}\right)\\
\mathcal{D}_\ell\left(M_{143}\times M_{857}\right)\equiv \mathcal{D}_\ell\left(M_{143}^{\rm HM2}\times M_{857}^{\rm HM2}\right)\\
\mathcal{D}_\ell\left(M_{217}\times M_{353}\right)\equiv \mathcal{D}_\ell\left(M_{217}^{\rm HM2}\times M_{353}^{\rm HM1}\right)\\
\mathcal{D}_\ell\left(M_{217}\times M_{545}\right)\equiv \mathcal{D}_\ell\left(M_{217}^{\rm HM1}\times M_{545}^{\rm HM1}\right)\\
\mathcal{D}_\ell\left(M_{217}\times M_{857}\right)\equiv \mathcal{D}_\ell\left(M_{217}^{\rm HM2}\times M_{857}^{\rm HM2}\right)\\
\mathcal{D}_\ell\left(M_{353}\times M_{545}\right)\equiv \mathcal{D}_\ell\left(M_{353}^{\rm HM1}\times M_{545}^{\rm HM1}\right)\\
\mathcal{D}_\ell\left(M_{353}\times M_{857}\right)\equiv \mathcal{D}_\ell\left(M_{353}^{\rm HM2}\times M_{857}^{\rm HM2}\right)\\
\mathcal{D}_\ell\left(M_{545}\times M_{857}\right)\equiv \mathcal{D}_\ell\left(M_{545}^{\rm HM2}\times M_{857}^{\rm HM1}\right).
\end{array}
\right.
\end{equation}

This is the definition of the data sets cross-spectra we adopt in this paper. The correlation matrix built for the SIM1 data set from this definition of the cross-spectra is displayed, for the multipole bin centered at $\ell=100$, in the bottom panel of \fig{Fig:corr_matrix_sims_donly}. We can see that the correlations have been significantly decreased with respect to those of the cross-spectra defined in \eq{eq:def_cross_spectra} and that the correlation matrix is closer from being diagonal.

An eigenvalue analysis of the correlation matrix as computed in \eq{eq:def_cross_spectra} indicates that the effective degrees of freedom from the 15 cross-spectra is $N_{\rm dof}=5$. When defining the cross-spectra as in \eq{eq:def_cross_spectra2}, it becomes $N_{\rm dof}=11$, allowing us to fit the SED moment expansion of the cross-spectra up to third order (see \sect{sec:results}).

\section{Truncating the dust moment expansion at difference orders}\label{sec:multi_order}

In \sect{sec:results} we present results from the third-order dust SED moment expansion. Here, we present the corrected dust amplitude $\mathcal{D}_\ell^{A_{\rm D}A_{\rm D}}$, the corrected spectral index $\betaellbcorr,$ and the relevant moment functions $\mathcal{M}_\ell$ in the case where \eq{eq:dust_moment_cross_3} is truncated at first and second order in our three-step fitting procedure.

\begin{figure}[ht!]
\centering
\includegraphics[width=\columnwidth]{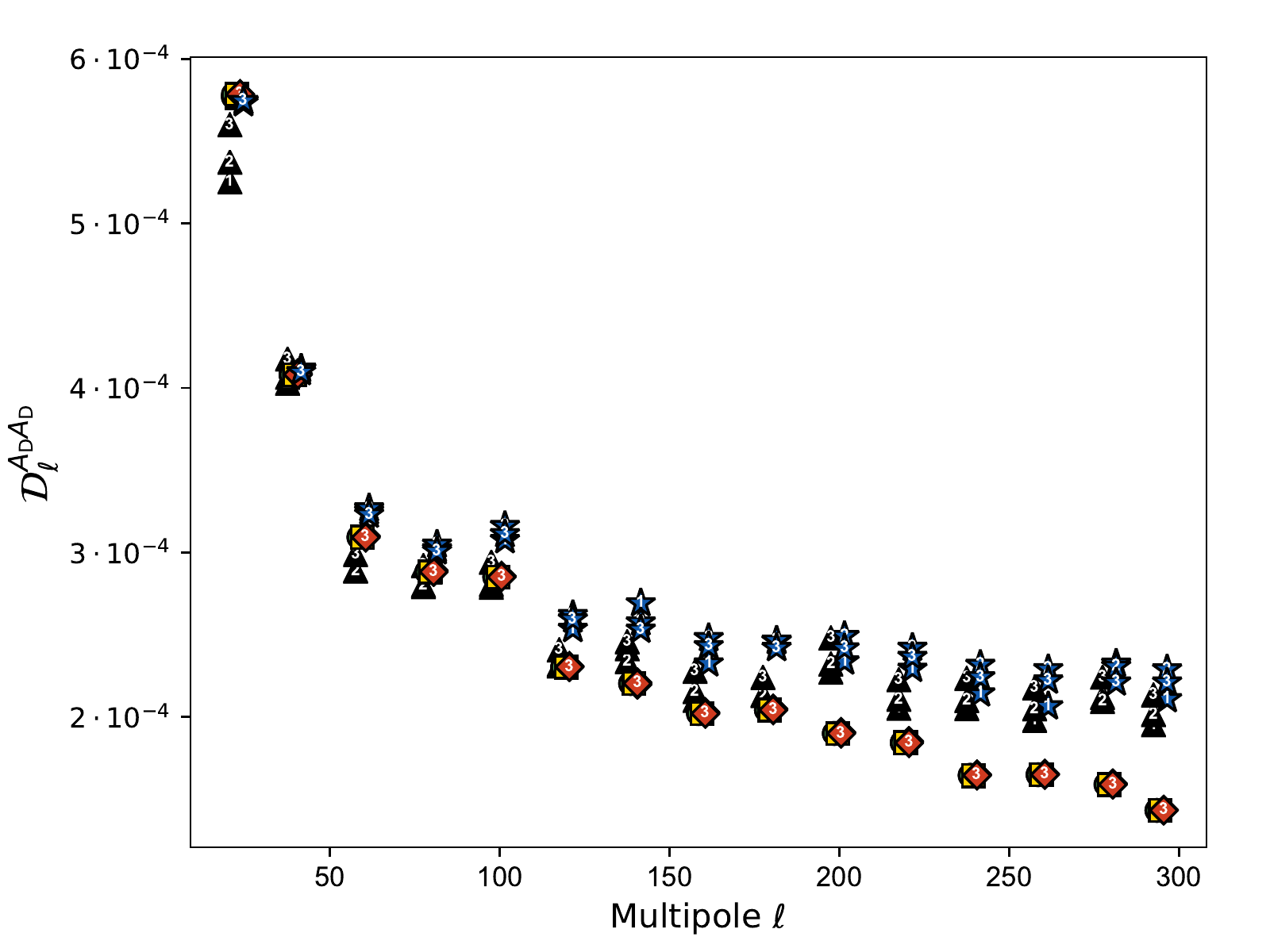}
\caption{Corrected dust amplitude $\mathcal{D}_\ell^{A_{\rm D}A_{\rm D}}$ after step 3 of the fit, when truncating \eq{eq:dust_moment_cross_3} at first order ("1" label on the plot markers), second order ("2" label), and third order ("3" label, same values as those in \sect{sec:results}). The symbols refer to the different data sets: SIM1 (green circles), SIM2 (yellow squares), SIM3 (red diamond), SIM4 (blue stars), and PR3 (black triangles).}
\label{fig:multi_order0}
\end{figure}

\begin{figure}[ht!]
\centering
\includegraphics[width=\columnwidth]{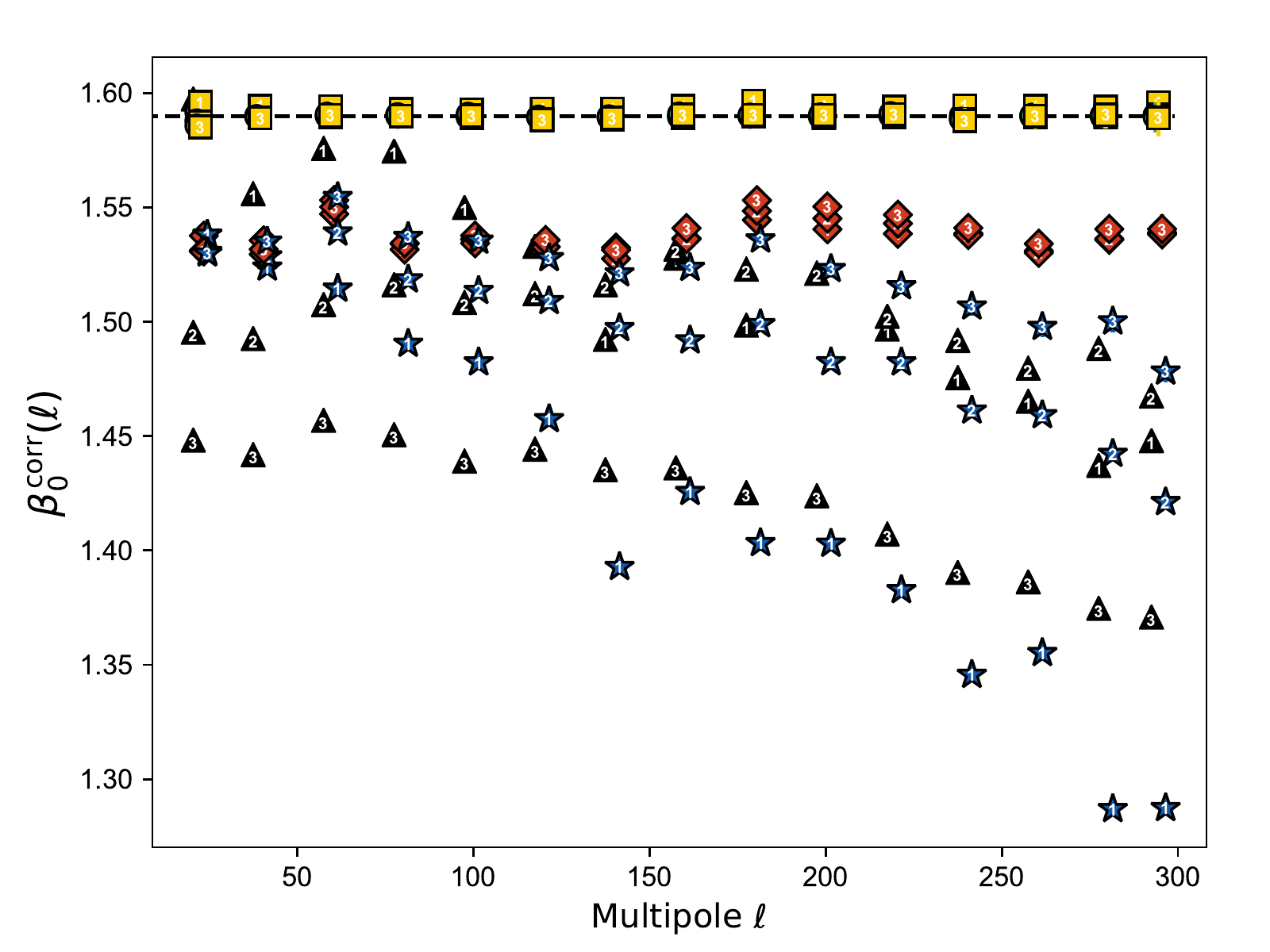}
\caption{Same as \fig{fig:multi_order0} but for $\betaellbcorr$.}
\label{fig:multi_order1}
\end{figure}

\begin{figure}[ht!]
\centering
\includegraphics[width=\columnwidth]{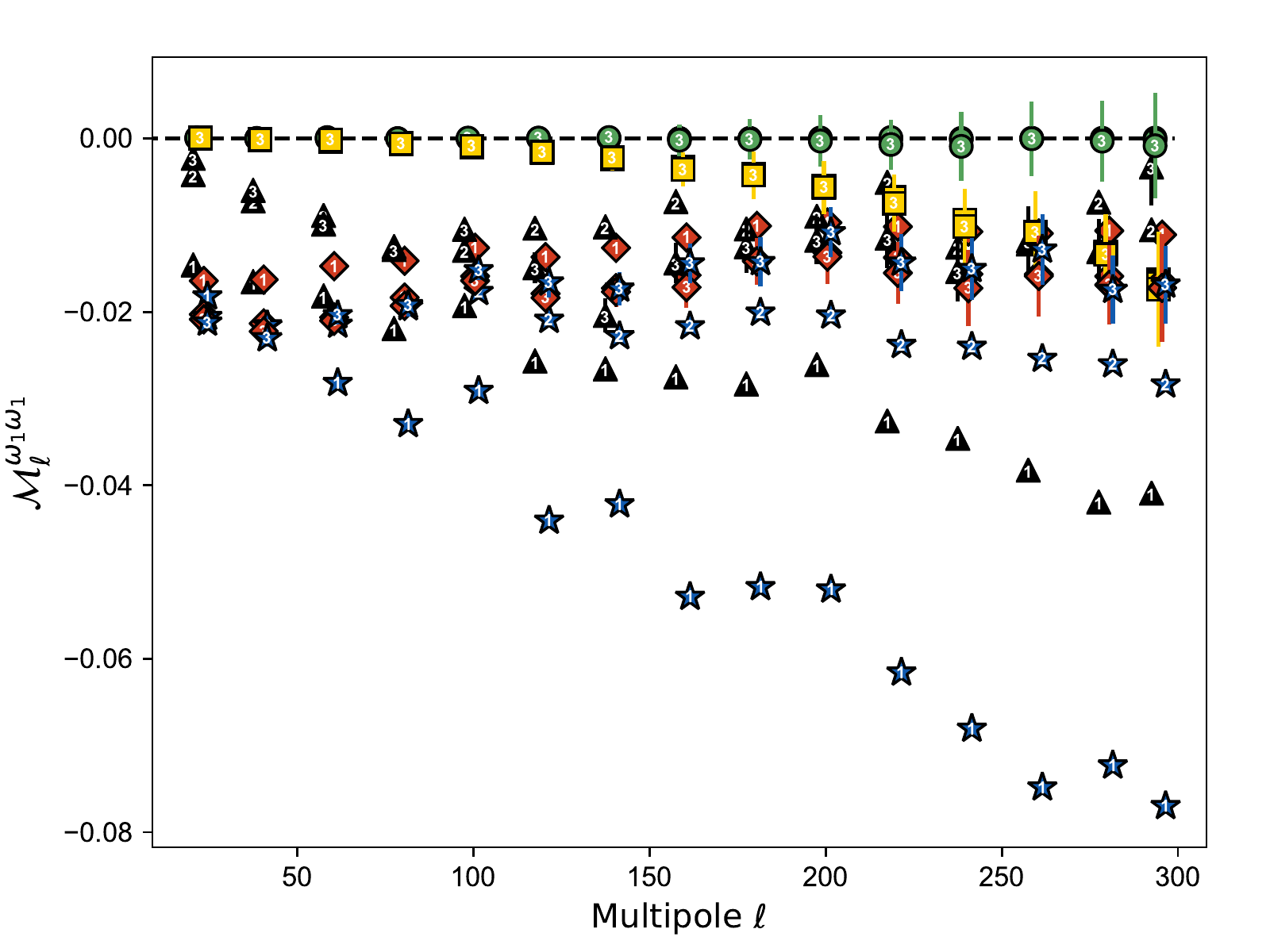}
\caption{Same as \fig{fig:multi_order0} but for $\mathcal{M}_\ell^{\omega_1\omega_1}$, normalized for the $143\times545$\,GHz cross-spectrum.}
\label{fig:multi_order2}
\end{figure}

\begin{figure}[ht!]
\centering
\includegraphics[width=\columnwidth]{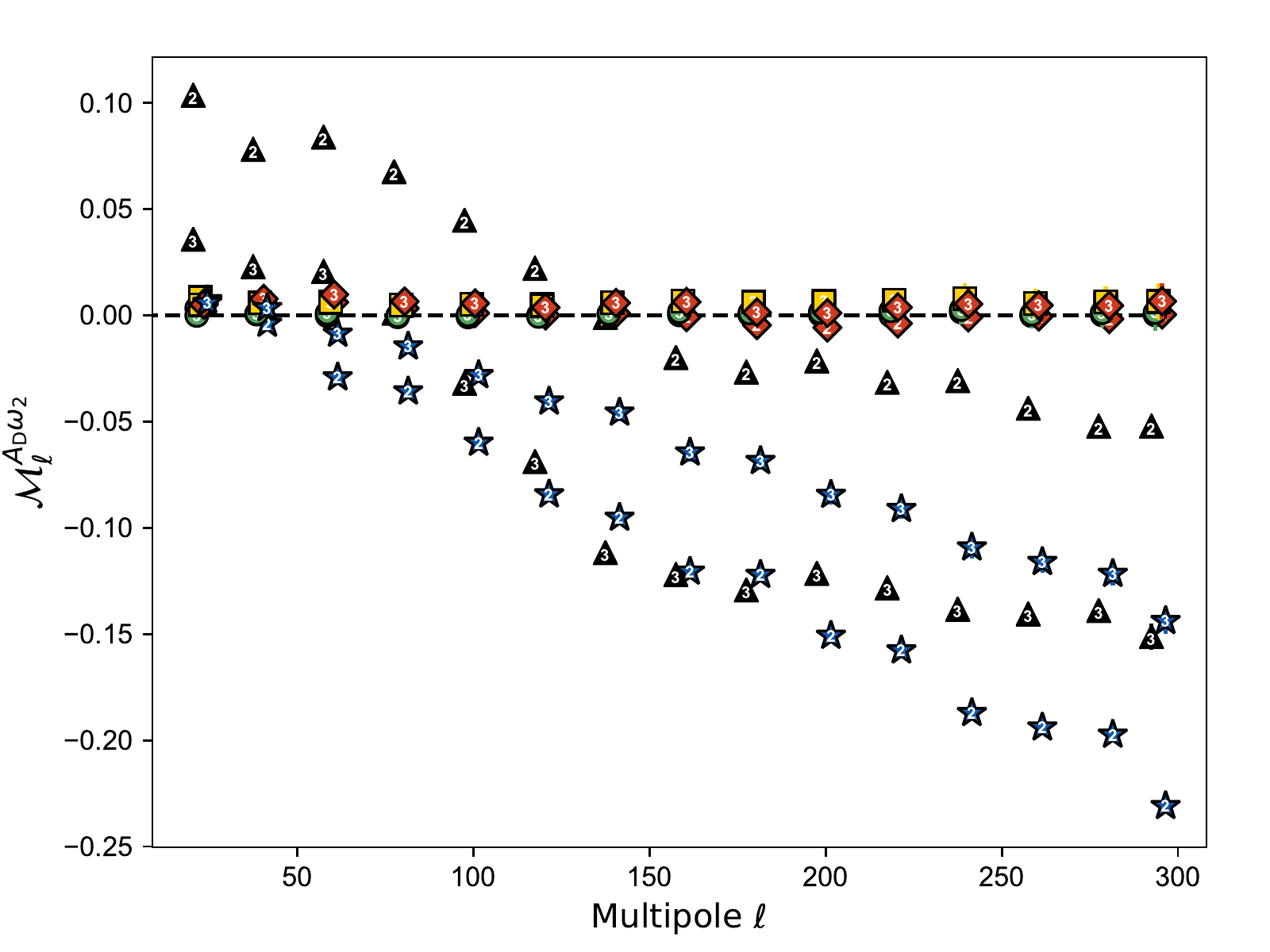}
\caption{Same as \fig{fig:multi_order0} but for $\mathcal{M}_\ell^{A_{\rm D}\omega_2}$, normalized for the $143\times545$\,GHz cross-spectrum.}
\label{fig:multi_order3}
\end{figure}

\begin{figure}[ht!]
\centering
\includegraphics[width=\columnwidth]{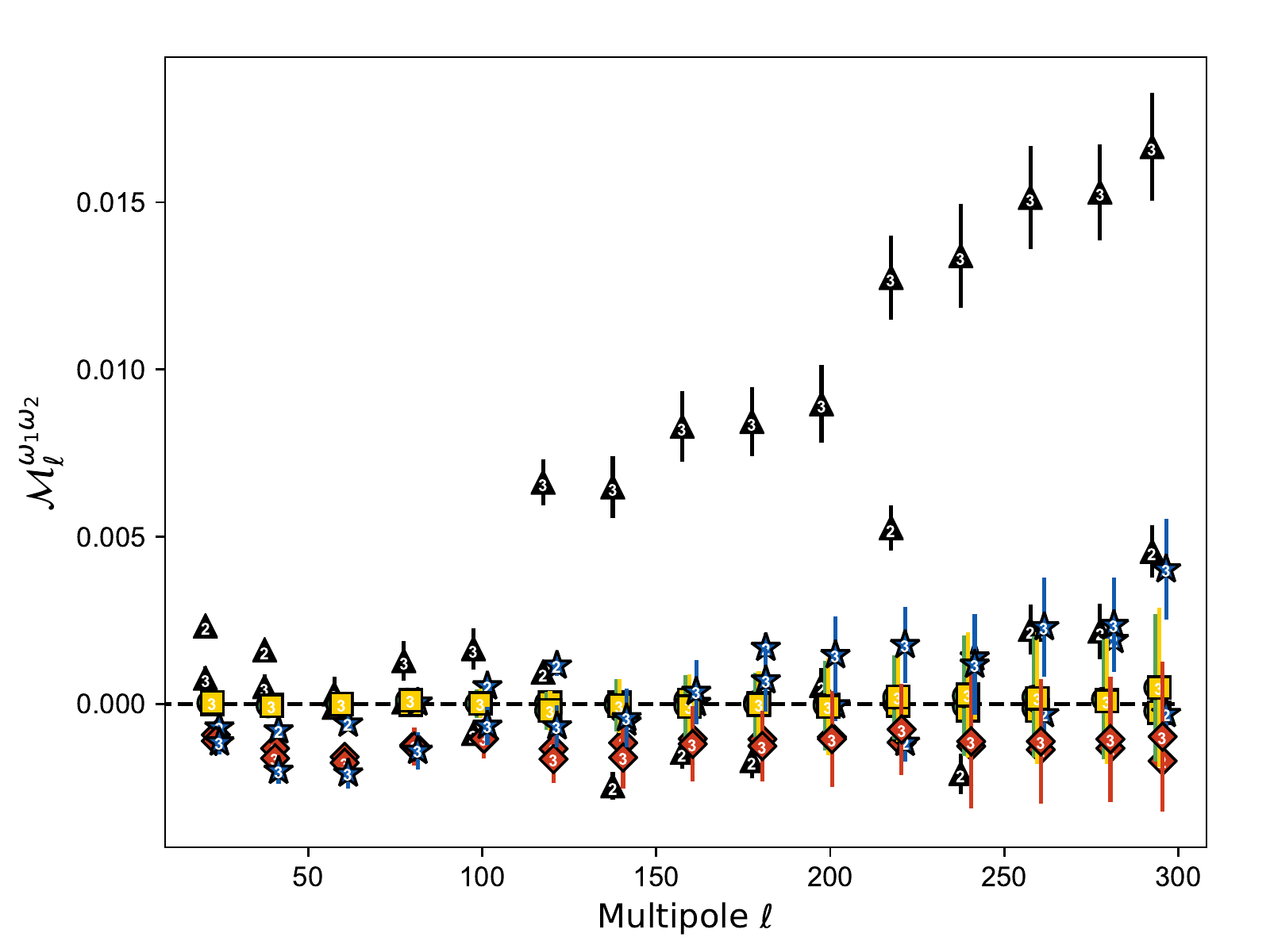}
\caption{Same as \fig{fig:multi_order0} but for $\mathcal{M}_\ell^{\omega_1\omega_2}$, normalized for the $143\times545$\,GHz cross-spectrum.}
\label{fig:multi_order4}
\end{figure}

\begin{figure}[ht!]
\centering
\includegraphics[width=\columnwidth]{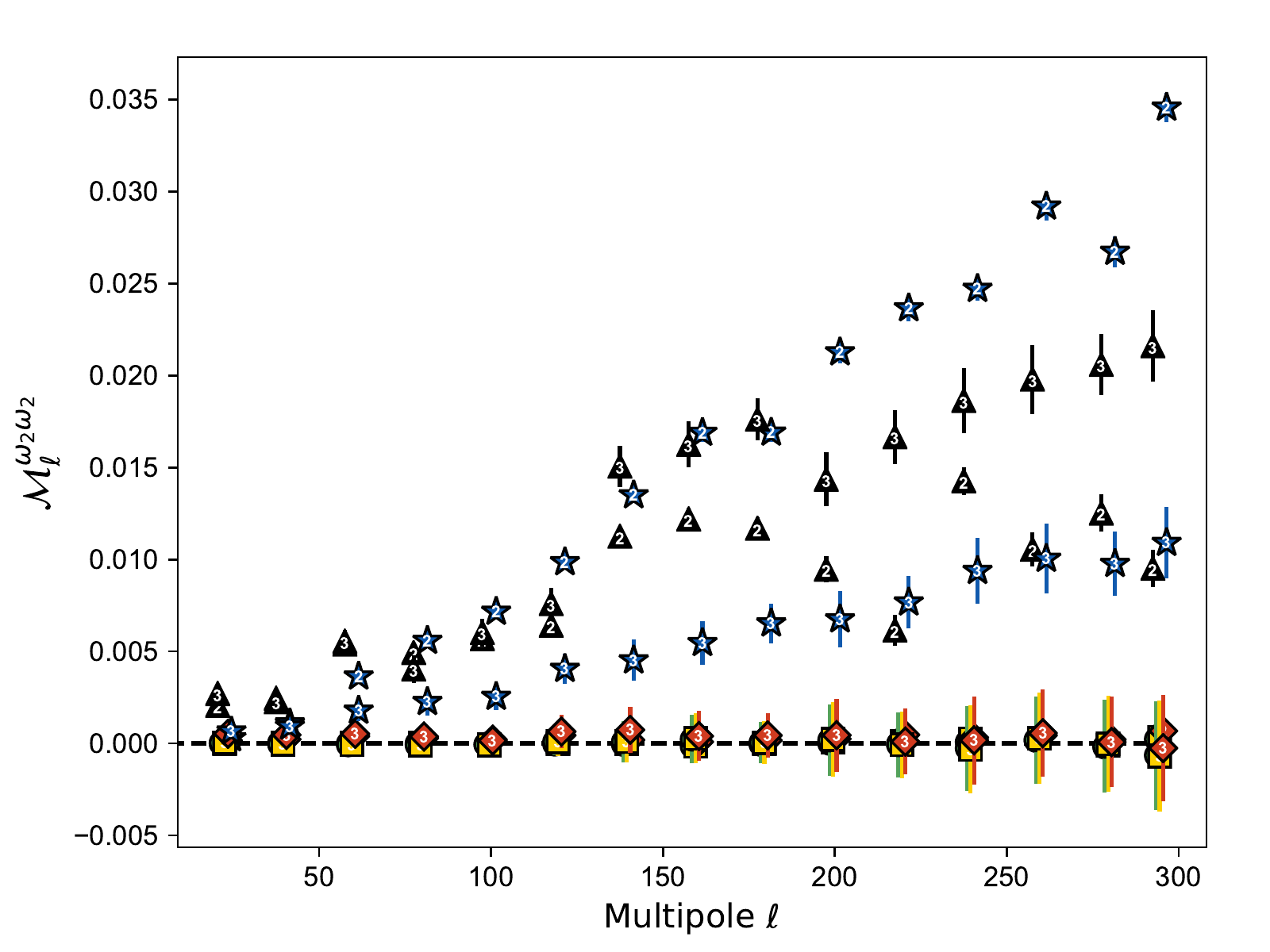}
\caption{Same as \fig{fig:multi_order0} but for $\mathcal{M}_\ell^{\omega_2\omega_2}$, normalized for the $143\times545$\,GHz cross-spectrum.}
\label{fig:multi_order5}
\end{figure}

These other truncating order results are displayed in Figs.~\ref{fig:multi_order0}, \ref{fig:multi_order1}, \ref{fig:multi_order2}, \ref{fig:multi_order3}, \ref{fig:multi_order4} and \ref{fig:multi_order5} for $\mathcal{D}_\ell^{A_{\rm D}A_{\rm D}}$, $\betaellbcorr$, $\mathcal{M}_\ell^{\omega_1\omega_1}$, $\mathcal{M}_\ell^{A_{\rm D}\omega_2}$, $\mathcal{M}_\ell^{\omega_1\omega_2}$ and $\mathcal{M}_\ell^{\omega_2\omega_2}$, respectively. 

SIM1, SIM2, and SIM3 data set results are stable with the truncating order, while SIM4 and PR3 change significantly. Nevertheless, SIM4 and PR3 results are affected in a very different way. For example, $\betaellbcorr$ values increase with the truncating order for SIM4, while they evolve in the opposite way for SIM3. However, this behavior is not observed for the moment functions.


\section{Impact of the synchrotron emission }\label{sec:sync_vs_CIB}

\begin{figure}%
\centering
\subfigure{%
\includegraphics[width=\columnwidth]{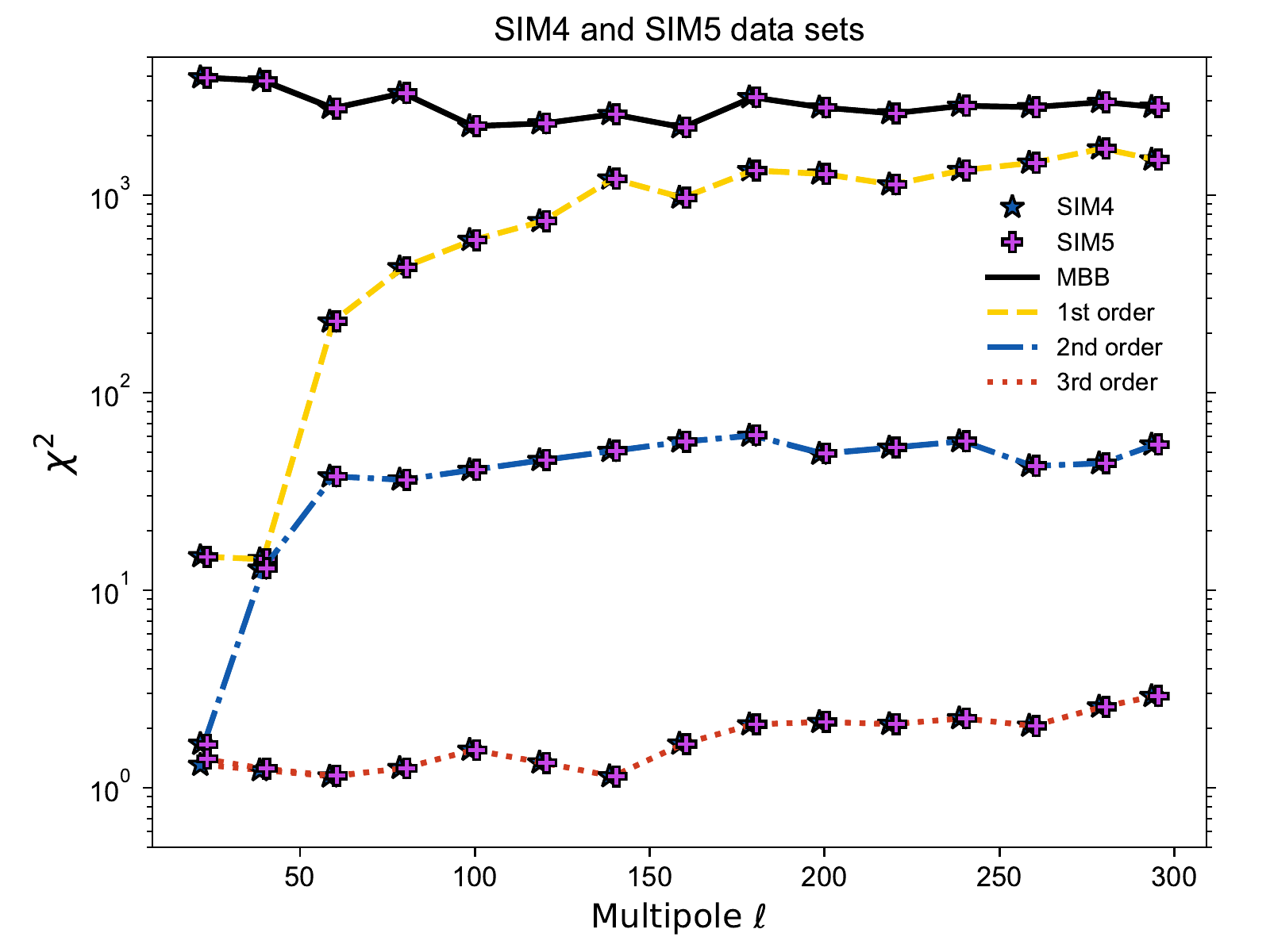}}
\caption{Reduced $\chi^2$ of the fit of \eq{eq:dust_moment_cross_3} at zero (solid black), first (dashed yellow), second (dashed-dotted blue), and third order (dotted red) for the SIM5 (including synchrotron, pink plus signs) and SIM4 (blue stars) data sets.}
\label{Fig:chi2_CIB_SYNC}
\end{figure}

\begin{figure}%
\centering
\subfigure{%
\includegraphics[width=\columnwidth]{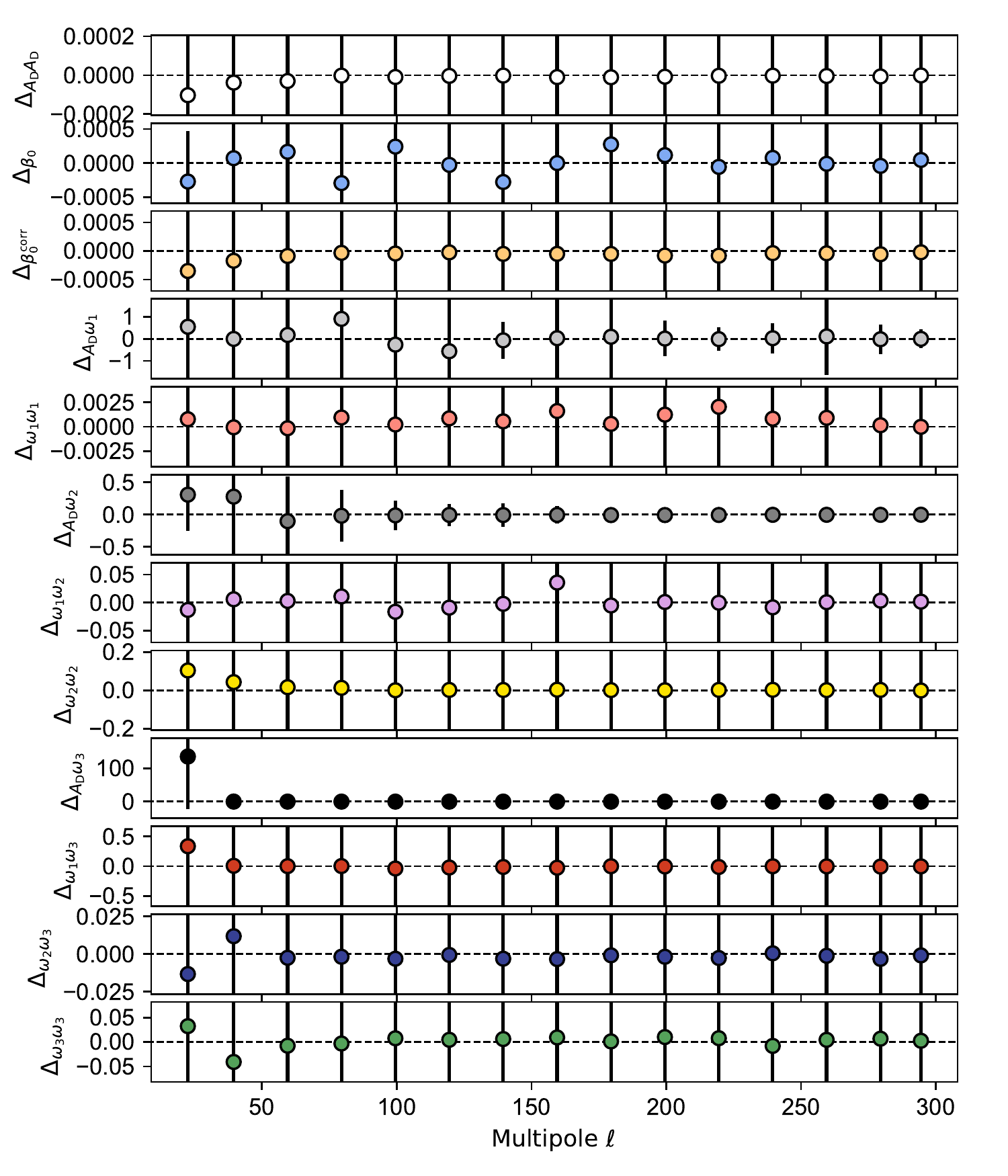}}
\caption{Relative difference between SIM5 and SIM4 fitted parameters. $\Delta_X=(X^{\rm SIM5}-X^{\rm SIM4})/X^{\rm SIM4}$ with $X\in\{\mathcal{D}_\ell^{A_{\rm D}A_{\rm D}},\beta_0,\beta_0^{\rm corr},\mathcal{M}_\ell^{A_{\rm D}\omega_1},\,{\rm etc}\dots\}$ as a function of the multipole $\ell$.}
\label{Fig:moments_CIB_SYNC}
\end{figure}

In this section we quantify the impact of the synchrotron emission on the moment analysis by comparing the results between two types of simulations: the SIM4 simulations, described in \sect{sec:data_and_sims}, and the SIM5 simulations that in addition to SIM4 also include a synchrotron component:\\

{\bf SIM5}: $M^{\rm SIM5,\eta}_{\nu_i} =  N^\eta_{\nu_i}+S^{{\rm D}_3}_{\nu_i}+S^{\rm CIB}_{\nu_i}+S^{\rm Sync}_{\nu_i}$,\\ 

\noindent where $S^{\rm Sync}_{\nu_i}$ is the synchrotron template at the frequency $\nu_i$. This component is generated, as for the CIB component, using the Planck Sky Model (PSM) version 1.9 described in \cite{2013A&A...553A..96D}. Examples of the CIB and synchrotron templates at 353\,GHz are shown in the middle and bottom panels of \fig{Fig:templates}, respectively. We can already note that, at this frequency, the synchrotron is sub-dominant with respect to the CIB by at least three orders of magnitude. 

\fig{Fig:chi2_CIB_SYNC} shows the reduced $\chi^2$ results of the MBB and the dust moment fits up to third order for the SIM4 and SIM5 simulations. The lines are barely distinguishable. The relative difference of the dust amplitude spectrum $\mathcal{D}^{\rm \AD \AD}_\ell$, the dust spectral index $\betaellb$, the corrected spectral index $\betaellbcorr,$ and the dust moments functions $\mathcal{M}_\ell^{ab}$ up to third order between SIM4 and SIM5 are shown in \fig{Fig:moments_CIB_SYNC}. For most of these quantities, this relative difference is very small. It is bigger for some of them, for example for the $\mathcal{M}_\ell^{A_{\rm D}\omega_i}$ moment functions, but still well within the propagated error bars (due to division by small numbers). 

According to these results we can therefore conclude that the synchrotron emission has a negligible impact on the dust moment analysis for the \PlanckHFI{} channels from 143 to 857\,GHz.

\section{Dust simulations with varying spectral index and constant temperature}\label{sec:dust_beta_tconst}

In the present work, the moment expansion is performed by derivation with respect to the dust spectral index $\beta$. Here, we present the results from an additional simulated data set in order to shed light on the capacity of the moments expansion on $\beta$ to capture the complexity arising specifically from the spatial variations of the dust temperature $T(\hat{\bf n})$. Our SIM1 data set has constant $\beta(\hat{\bf n})=\beta_0$  and $T(\hat{\bf n})=T_0$ over the sky, while our SIM3 data set has spatial variations of both these parameters; here we introduce SIM6, a data set simulated with a varying spectral index $\beta(\hat{\bf n})$ and a constant temperature $T_0$:\\

{\bf SIM6}: $M^{\rm SIM6,\eta}_{\nu_i} =  N^\eta_{\nu_i}+S^{{\rm D}_4}_{\nu_i},$\\ 

\noindent where $S^{{\rm D}_4}$ is the dust model with varying spectral index $\beta(\hat{\bf n})$ and a constant temperature $T_0$, presented in Sect.~\ref{sec:dust_component}. The SIM6 simulated data set underwent the same fits as for the other sets presented in the main body of this article. 

\begin{figure}%
\centering
\subfigure{%
\includegraphics[width=\columnwidth]{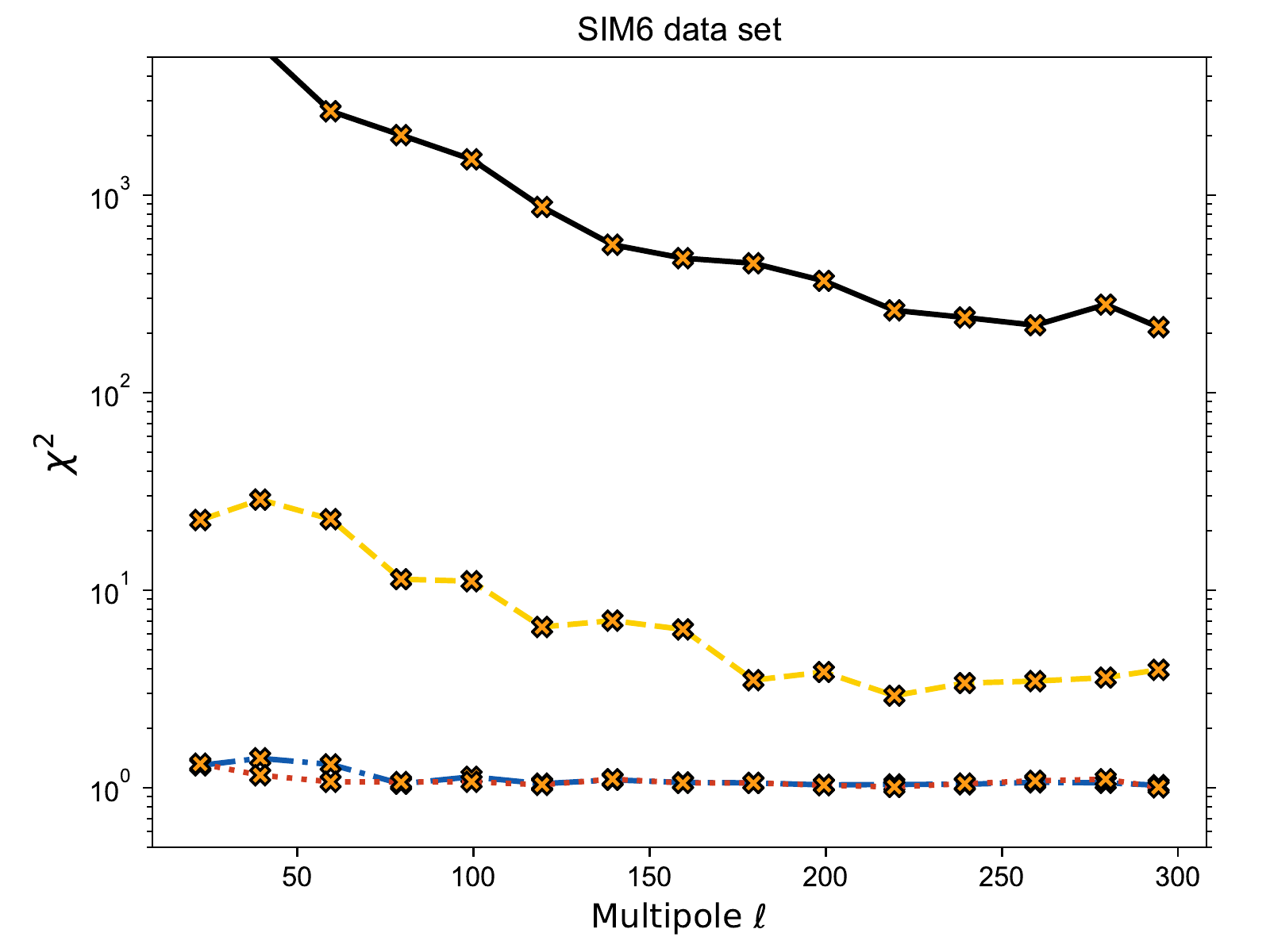}}
\caption{Reduced $\chi^2$ of the fit of \eq{eq:dust_moment_cross_3} at zero
(solid black), first (dashed yellow), second (dashed-dotted blue), and third
order (dotted red) for the SIM6 ($\beta({\hat{\bf n}}),T_0$, orange crosses).}
\label{Fig:chi2_SIM6}
\end{figure}

\fig{Fig:chi2_SIM6} shows the reduced $\chi^2$ results of the MBB and the dust moment fits up to third order for the SIM6 simulations. They globally show similar trends as the ones computed for SIM3 (see Fig.~\ref{Fig:dust_moment_fit_vs_MBB_chi2_vs_ell}). The reduced $\chi^2$ values for the MBB are significantly higher than that of SIM3, while higher order fits give slightly better $\chi^2$. 

\begin{figure}%
\centering
\subfigure{%
\includegraphics[width=\columnwidth]{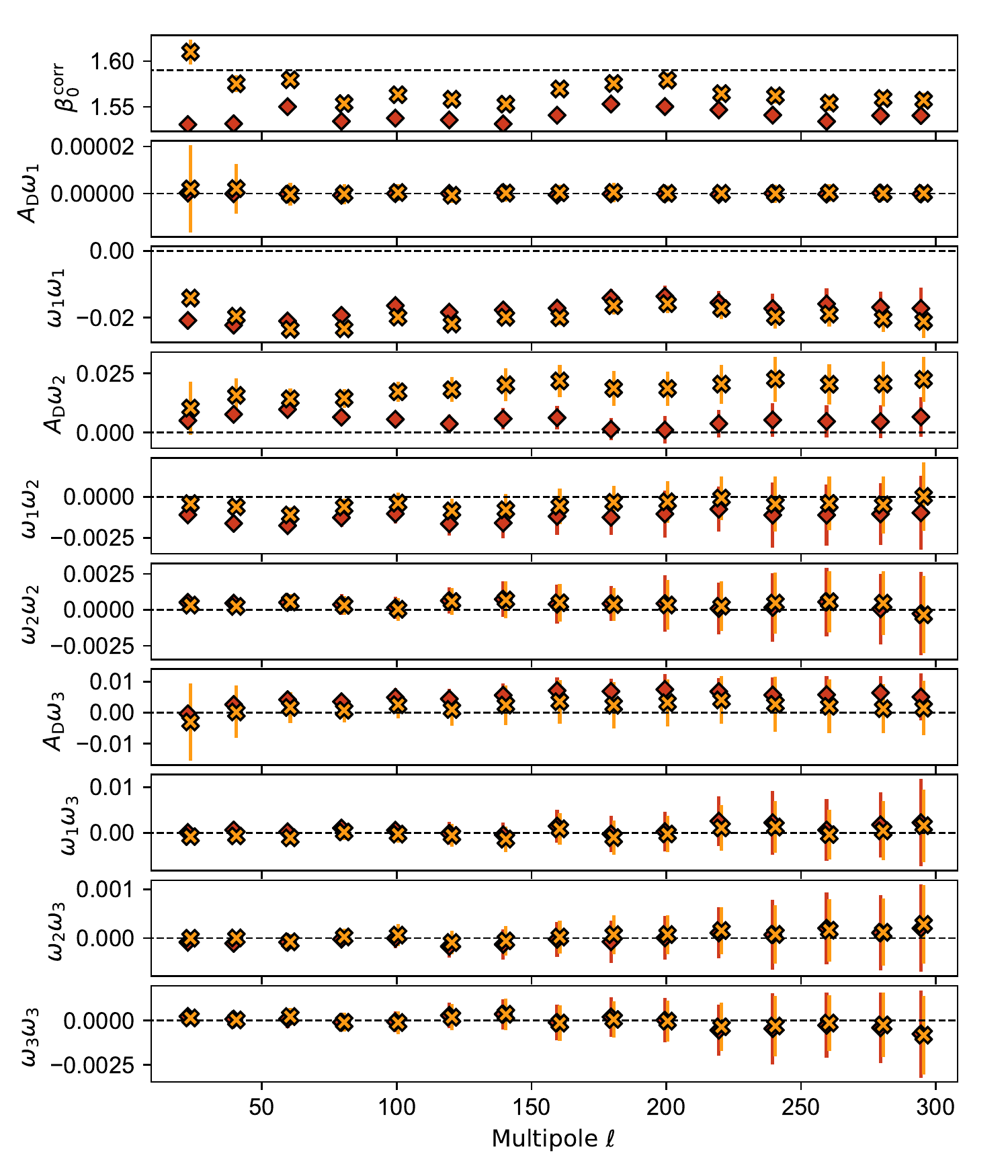}}
\caption{Comparison between SIM3 (red diamonds) and SIM6 (orange crosses) fitted parameters for the corrected spectral index $\beta_0^{\rm corr}$ and the higher order moments ($\mathcal{M}_\ell^{ab}\ (143\times545)\in\{A_{\rm D},\omega_1,\omega_2,\omega_3\}$) as a function of the multipole $\ell$.}
\label{Fig:moments_SIM6}
\end{figure}

Pushing further the comparison between SIM3 and SIM6, we can see from Figure~\ref{Fig:moments_SIM6} that most of the parameters from the fits are similar, except for $\beta_0^{\rm corr}$ and $\mathcal{M}_\ell^{A_{\rm D}\omega_2}$. SIM6 shows significantly higher values than SIM3 for $\beta_0^{\rm corr}$ and $\mathcal{M}_\ell^{A_{\rm D}\omega_2}$.  

The known dust emission $\beta$--$T$ anti-correlation \citep[e.g.,][]{juvela2013} could be an explanation for the differences between results on SIM6 and SIM3. SIM6, with a constant $T_0$ on the sky, shows a larger variability in SED due to the $\beta$ spatial variations, whereas in SIM6 the spatial variations of $T$ ---anti-correlated to those of $\beta$--- tend to compensate the SED variability, hence the larger $\chi^2$ when fitting the MBB.


\end{appendix}

\end{document}